\documentclass[a4paper,fleqn,usenatbib]{mnras}
\usepackage[T1]{fontenc}
\usepackage{ae,aecompl}
\usepackage{graphicx}	% Including figure files
\usepackage{amsmath}	% Advanced maths commands
\usepackage{amssymb}	% Extra maths symbols
\usepackage{makecell}
\usepackage{multirow}
\usepackage{ulem}
\usepackage{xcolor}

\newcommand{\kms}{\mbox{km s$^{-1}$}}
\newcommand{\dego}{$^\circ$}
\title[The morpho-kinematics of the circumstellar envelope
around the AGB star EP Aqr]{The morpho-kinematics of the circumstellar envelope around the AGB star EP Aqr}
\author[D.T. Hoai et al.]{{D.T. Hoai$^{1,2}$\thanks{E-mail: dthoai@vnsc.org.vn}, P.T. Nhung$^1$, P. Tuan-Anh$^1$, P. Darriulat$^1$, P.N. Diep$^1$,}
\newauthor{T. Le Bertre$^3$, N.T. Phuong$^1$, T.T. Thai$^1$, J.M. Winters$^4$} 
\\
$^1$Department of Astrophysics, Vietnam National Space Center (VNSC), Vietnam Academy of Science and Technology (VAST),\\
18 Hoang Quoc Viet, Cau Giay, Ha Noi, Viet Nam\\
$^2$Graduate University of Science and Technology (GUST), Vietnam Academy of Science and Technology (VAST), \\
18 Hoang Quoc Viet, Cau Giay, Ha Noi, Vietnam\\
$^3$LERMA, UMR 8112, CNRS and Observatoire de Paris, PSL Research University, 61 av. de l'Observatoire, F-75014 Paris, France\\
$^4$IRAM, 300 rue de la Piscine, Domaine Universitaire, F-38406 St. Martin d'H\`{e}res, France\\
}
\date{Accepted XXX. Received YYY; in original form ZZZ}

\pubyear{2018}
\begin{document}
\label{firstpage}
\pagerange{\pageref{firstpage}--\pageref{lastpage}}
\maketitle

\begin{abstract}
  ALMA observations of \mbox{CO(1-0)} and \mbox{CO(2-1)} emissions of the circumstellar envelope of EP Aqr, an oxygen-rich AGB star, are reported. A thorough analysis of their properties is presented using an original method based on the separation of the data-cube into a low velocity component associated with an equatorial outflow and a faster component associated with a bipolar outflow. A number of important and new results are obtained concerning the distribution in space of the effective emissivity, the temperature, the density and the flux of matter. A mass loss rate of $(1.6\pm0.4) 10^{-7}$ solar masses per year is measured. The main parameters defining the morphology and kinematics of the envelope are evaluated and uncertainties inherent to de-projection are critically discussed. Detailed properties of the equatorial region of the envelope are presented including a measurement of the line width and a precise description of the observed inhomogeneity of both morphology and kinematics. In particular, in addition to the presence of a previously observed spiral enhancement of the morphology at very small Doppler velocities, a similarly significant but uncorrelated circular enhancement of the expansion velocity is revealed, both close to the limit of sensitivity. The results of the analysis place significant constraints on the parameters of models proposing descriptions of the mass loss mechanism, but cannot choose among them with confidence.
  
\end{abstract}

%%%%%%%%%%%%%%%%%%%%%%%%%%%%%%%%%%%%%%%%%%%%%%%%%%

%%%%%%%%%%%%%%%%% BODY OF PAPER %%%%%%%%%%%%%%%%%%

\section{Introduction}\label{sec1}

EP Aqr is an oxygen-rich M type AGB star at a distance of only 119$\pm$6 pc from the solar system \citep[]{vanLeeuwen2007,Gaia2018} with a luminosity of 3450 solar luminosities. There is a silicate feature at 10 $\mu$m \citep[]{Speck2000}, which shows that the star is presently undergoing mass loss. Several observations suggest a mass loss episode at the scale of 10$^4$ to 10$^5$ years: the Herschel 70 $\mu$m observation \citep[]{Cox2012} of a large trailing circumstellar shell and the HI observation \citep[]{LeBertre2004} of its interaction with the interstellar medium. In CO rotational lines, EP Aqr is characterized by a composite profile with narrow ($\sim 2$ \kms) and broad ($\sim 10$ \kms) components \citep[]{Knapp1998, Winters2003}. Few oxygen-rich AGB stars have been observed with sufficient sensitivity and spectral resolution to reveal such composite profiles in CO (or SiO) rotational lines, making it difficult to reliably evaluate which fraction of the total population they represent and to state whether it corresponds to an intrinsic property of some stars, or to a common stage in the evolution of oxygen-rich AGB stars. Detailed studies of the kinematics and 3D-structure of the inner circumstellar shells ($<$ 100 AU) are needed. We are making efforts to obtain and model high spectral-resolution interferometric data on such objects \cite[]{LeBertre2016}. EP Aqr is particularly interesting because, from the absence of technetium in the spectrum \citep[]{Lebzelter1999}, and from its low $^{12}$C/$^{13}$C ratio \citep[]{Cami2000}, it appears to be in an early stage of evolution on the AGB.
  
Spatially resolved observations obtained at IRAM \citep[]{Winters2003, Nhung2015a} on EP Aqr, and on other sources such as RS Cnc \citep[]{Libert2010, Hoai2014, Nhung2015b} and X Her \citep[]{Kahane1996} are generally interpreted as the superposition of a slowly expanding wind with a bipolar outflow. The presence of irregularities in the morpho-kinematics has been interpreted as suggesting an episode of increased mass loss rate \citep[]{Knapp1998, Nakashima2006, Winters2007}. De-projection \citep[]{Diep2016, Nhung2015a} has shown that morphology and kinematics of EP Aqr display approximate axi-symmetry about an axis making a small angle with the line of sight (typically smaller than $\sim20$\dego) and projecting on the sky plane some 20\dego\ to 30\dego\ west of north. This was corroborated using recent ALMA observations having four times better spatial resolution than the preceding IRAM observations by \citet{Nhung2018a,Nhung2018b} and \citet{Homan2018}. The latter authors have also observed SO$_2$ and SiO line emissions exploring shorter distances to the star.
  
The present work presents a detailed study of the morphology and kinematics of the circumstellar envelope of EP Aqr, taking due account of what has been learned from previous studies and using observations of \mbox{CO(1-0)} and \mbox{CO(2-1)} emissions obtained at ALMA. The two components of the wind, bipolar and equatorial, are studied separately, an approach that is used for the first time and that helps greatly with identifying and describing properties of the morpho-kinematics more thoroughly than could be done previously. The method is described in Section \ref{sec3}; Section \ref{sec4} studies the equatorial component and Section \ref{sec5} studies the bipolar component. The results are then used in Section \ref{sec6} to present a joint description independent of the separation into two components and accounting for absorption and temperature effects. The under-determination inherent to the interpretation of radio observations and their de-projection in space plays a particularly important role in the case of EP Aqr as has been amply demonstrated by earlier studies. The emphasis of the present work is accordingly to attempt a critical and thorough study of the morpho-kinematics with the aim to identify those features that can be reliably ascertained. The results provide important new information on the properties of the circumstellar envelope at distances from the star exceeding $\sim250$ au, where CO molecules are particularly efficient tracers. Understanding the morpho-kinematics at shorter distances requires in addition different tracers, such as SiO and SO$_2$, a study that is postponed to a forthcoming publication (\citet{TuanAnh}, in preparation). The results are summarized and discussed in Section \ref{sec7}.

\section{Observations and data reduction}\label{sec2}

The observations used in the present article were made in cycle 4 of ALMA operation (2016.1.00026.S) between October 30$^{\rm th}$ 2016 and April 5$^{\rm th}$ 2017 and were presented in detail in \citet{Nhung2018a}. \mbox{CO(1-0)} emission was observed in 3 execution blocks in mosaic mode (3 pointings) with the number of antennas varying between 41 and 45. \mbox{CO(2-1)} emission was observed in 2 execution blocks in mosaic mode (10 pointings) with the number of antennas varying between 38 and 40.

Both lines were observed in two different configurations, C40-2 and C40-5 and the data were then merged in the \textit{uv} plane. The main parameters attached to these observations are listed in Table \ref{table1}. The \mbox{CO(1-0)} and \mbox{CO(2-1)} data have been calibrated using CASA\footnote{http//casa.nrao.edu} and the mapping was done with GILDAS\footnote{https://www.iram.fr/IRAMFR/GILDAS}.

We also included IRAM 30-m observations of \mbox{CO(1-0)} emission that have been reported in \citet{Winters2007}. We used GILDAS to re-centre the map and re-sample velocities in the same way as for ALMA \mbox{CO(1-0)} data. We then produced a list of pseudo visibilities containing short-spacing information and merged these with the 12-m ALMA data.

\begin{table}
  \centering
  \caption{Main observation parameters. $R$ is the distance from the star projected on the sky plane.}
  \begin{tabular}{|c|c|c|}
    \hline
    &CO(1-0)&\mbox{CO(2-1)}\\
    \hline
    Beam FWHM (arcsec$^2$) &0.78$\times$0.70 & 0.33$\times$0.30 \\
    \hline
    Beam PA (\dego)  & $-$56 & $-$80\\
    \hline
    Maximum baseline (m) & 1124 & 1400 \\
    \hline
    Minimum baseline (m) & 19 & 15 \\
    \hline
    Time on source (min) & 98.3  & 52.4  \\
    \hline
    Zero spacing & Yes & No \\
    \hline
    \makecell{Area covered by mosaic\\(arcsec$^2$)} & $\sim 60\times 45$ & $\sim 60 \times 45$ \\
    \hline
    \makecell{Noise \\ (mJy\,beam$^{-1}$/(0.2 \kms))} & 7 & 6 \\
    \hline
    \makecell{Flux, $R<$13 arcsec \\(Jy\,\kms)} & 110 & 549 \\
    \hline
    \end{tabular}
\label{table1}
\end{table}

\begin{figure*}
\centering
\includegraphics[height=4.5cm,trim=0.cm 0.cm 1.cm 9.cm,clip]{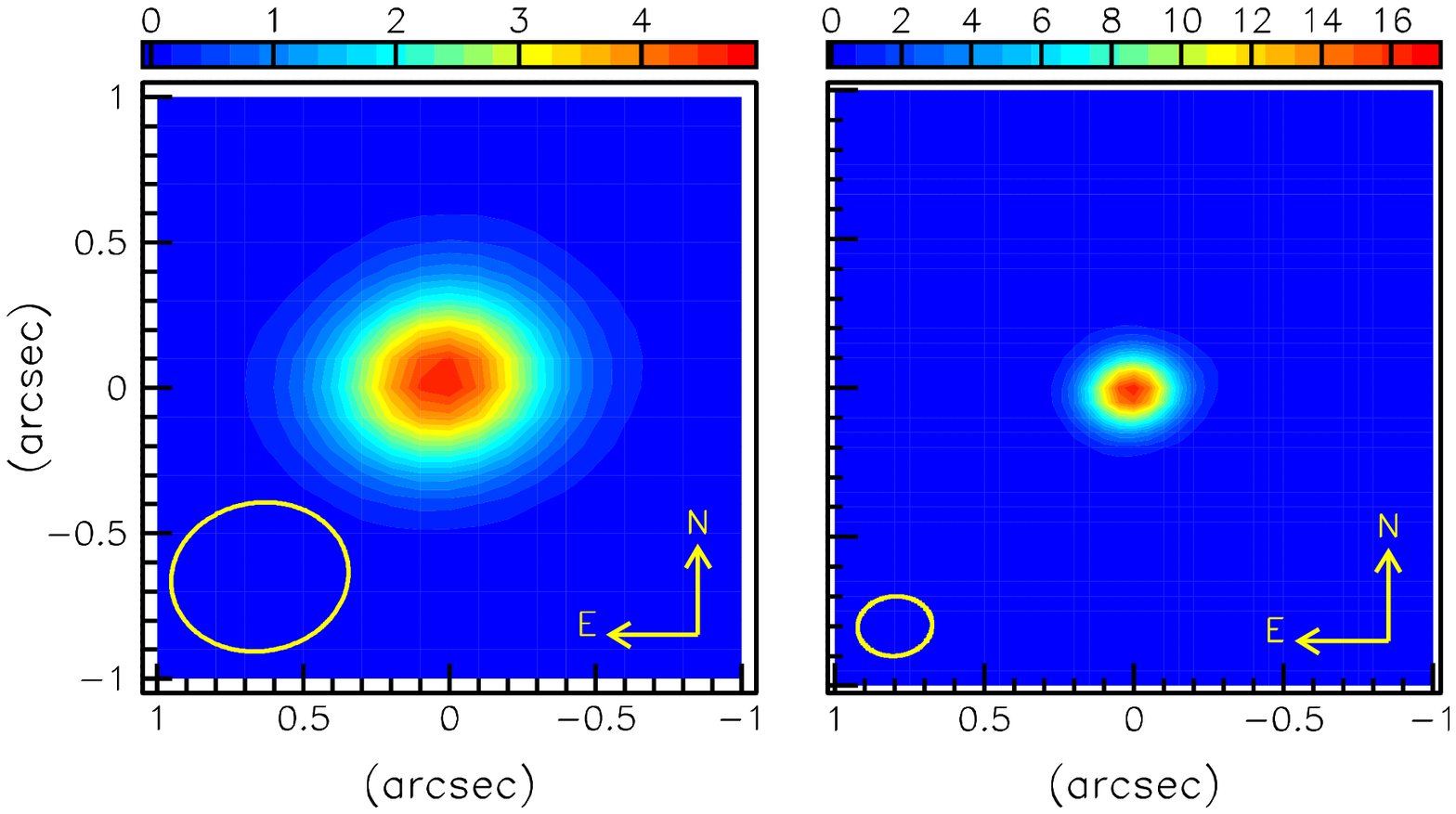}
\includegraphics[height=4.5cm,trim=0.cm .6cm 1.cm 0.cm,clip]{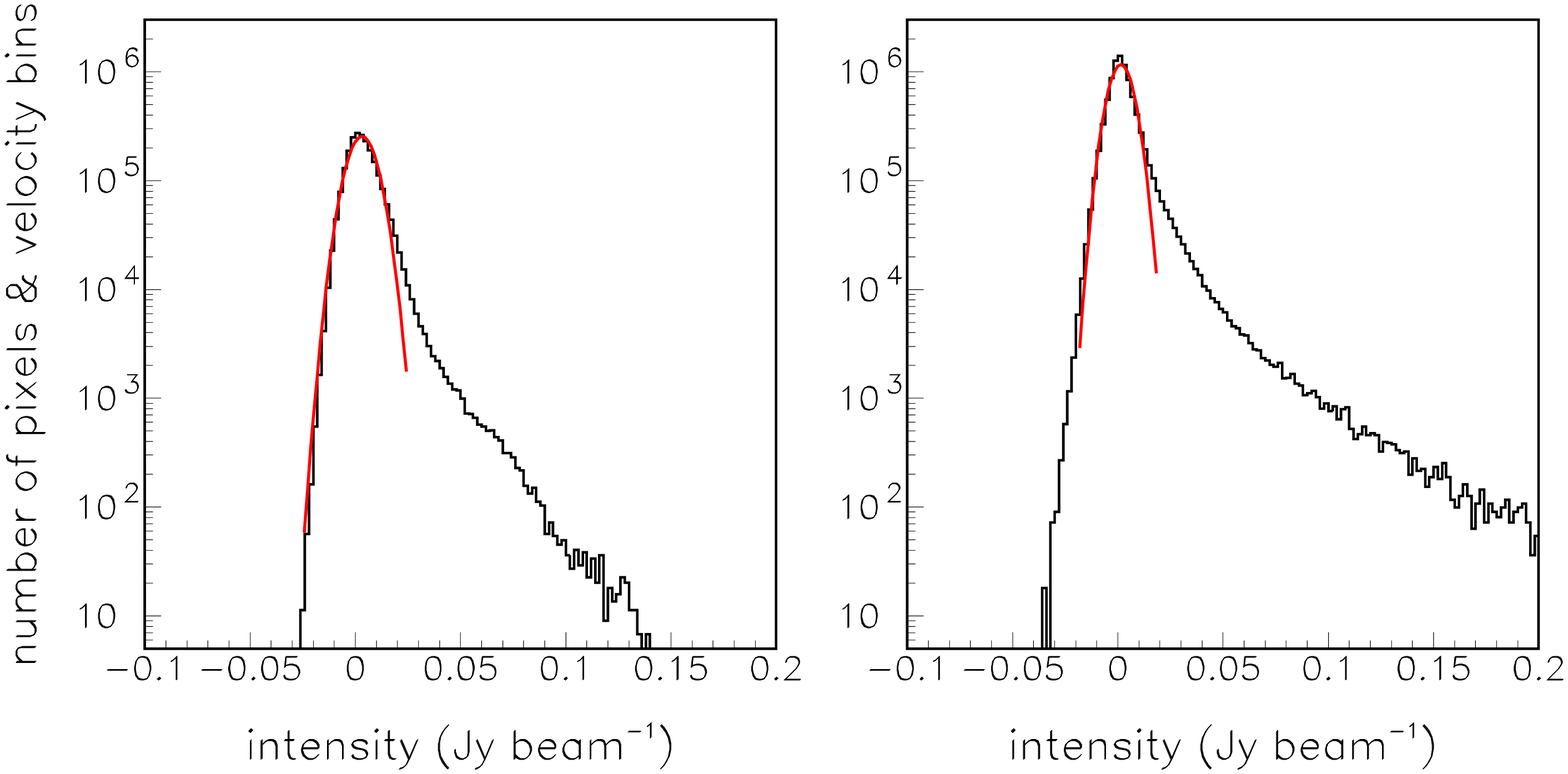}
\caption{Left panels: Sky maps of the continuum intensity (mJy\,beam$^{-1}$) in band 3 (left) and band 6 (middle left) respectively. Beam sizes are shown in the lower left corner. Right panels: distributions of the brightness measured in \mbox{CO(1-0)} (middle right) and \mbox{CO(2-1)} (right) emissions at a distance projected on the sky plane, $R$, smaller than 12 arcsec from the star. The curves are Gaussian fits to the noise peaks with (mean, $\sigma$)=(3.2, 6.7) and (1.4, 5.8) mJy\,beam$^{-1}$ for \mbox{CO(1-0)} and \mbox{CO(2-1)} respectively.}
\label{fig1}
\end{figure*}

The origin of coordinates at RA=21h46m31.848s and DEC=$-$02\dego12'45.93'' corresponds to year 2000. Between 2000 and the time of observation, the source has moved by 0.40 arcsec east and 0.31 arcsec north (proper motion of (24.98, 19.54) mas/year \citep[]{vanLeeuwen2007}); the data have been corrected accordingly. The spectral resolution (channel) has been smoothed to 0.2 \kms\ and the Doppler velocity covers between $-20$ and 20 \kms. We use as origin the Doppler velocity of $-33.6$ \kms\ (LSR) about which the profile is well symmetric.

In addition to line emission, we have also observed the continuum emission in the same frequency bands (2.6 mm and 1.3 mm). As illustrated in the left panels of Figure \ref{fig1}, the source is unresolved, with no hint at the presence of a possible companion. It is centred at the same location as CO line emission to within 30 mas. The noise is at the level of 23 $\mu$Jy for band 3 and 75 $\mu$Jy for band 6.  The continuum fluxes are measured as 4.9$\pm$0.1 mJy at 2.6 mm and 17.8$\pm$0.2 mJy at 1.3 mm, consistent with the result of \citet{Winters2007} and with expectation from the black-body emission of the central star. We note that \citet{Homan2018} erroneously quote a 75 times larger flux, implying surprisingly slow winds and unusual amounts of central dust.

\section{Separation of the data cubes into broad and narrow components} \label{sec3}

\subsection{General properties}\label{sec3.1}

In order to ease the analysis and interpretation of the observations, we use a system of coordinates rotated clockwise by 20\dego\ about the line of sight, meaning that the $y$ axis points 20\dego\ west of north and the $x$ axis 20\dego\ north of east, the $z$ axis pointing away from us along the line of sight. Indeed, earlier analyses \citep[see in particular][]{Nhung2018b,Homan2018} have given evidence for an approximate symmetry of the data cubes with respect to the plane containing the line of sight and projecting on the sky plane some 20\dego\ to 30\dego\ west of north. A summary of the geometry and of the variables used in the study is presented in Appendix \ref{app1}. Under the hypothesis that the morphology and kinematics of the star obey axi-symmetry, the star axis would project on the sky plane along the $y$ axis. The arguments that are developed below confirm this result. Note that strictly speaking, what we call ``star axis'' for brevity should be called ``axis of the circumstellar envelope'' as no evidence is presented in the present work for these to be identical. Moreover, we apply some very small shifts to the Doppler velocity scales, by adding 0.12 \kms\ to the \mbox{CO(1-0)} velocities and subtracting 0.06 \kms\ from the \mbox{CO(2-1)} velocities; these are smaller than the spectral resolution (the velocity is given in bins of 0.2 \kms\ for both \mbox{CO(1-0)} and \mbox{CO(2-1)} emissions) but fine tune the mean position of the narrow central peak of the Doppler velocity spectrum to be at the origin ($V_z=0$). Finally, in order to ease the comparison between the \mbox{CO(1-0)} and \mbox{CO(2-1)} observations, we normally use a common pixel size, $0.3\times0.3$ arcsec$^2$ in spite of the difference between the associated beam widths. The grouping is done in the image plane after cleaning, the latter operation being made as usual with optimal pixel sizes, at the level of $\sim20$ to 25\% of the beam size. The present analysis requires a good spectral resolution but is less demanding on spatial resolution, most resolved features being significantly larger than the beam. The new data cubes are obtained from the original data cubes by sharing the content of an original pixel between the new pixels that overlap it, in proportion to the overlap fraction of the original pixel area.

Figure \ref{fig1} (right panels) displays the distributions of the brightness measured in each pixel over the circle of projected distances to the star not exceeding 12 arcsec. It gives evidence for an effective noise level ($1\sigma$) of 6.7 mJy\,beam$^{-1}$ for \mbox{CO(1-0)} and 5.8 mJy\,beam$^{-1}$ for \mbox{CO(2-1)}, consistent with thermal noise levels between 1 and 1.5 mJy per 0.2 \kms\ bin. As expected from the pixel/beam ratios, the grouping of pixels has only little effect on the effective noise. Before grouping, these values were respectively 7.2 and 6.7 mJy\,beam$^{-1}$, namely respectively $\sim7$\% and $\sim16$\% larger than after grouping.

\begin{figure*}
\centering
\includegraphics[width=0.75\textwidth,trim=.0cm 0.cm 0cm 0.cm,clip]{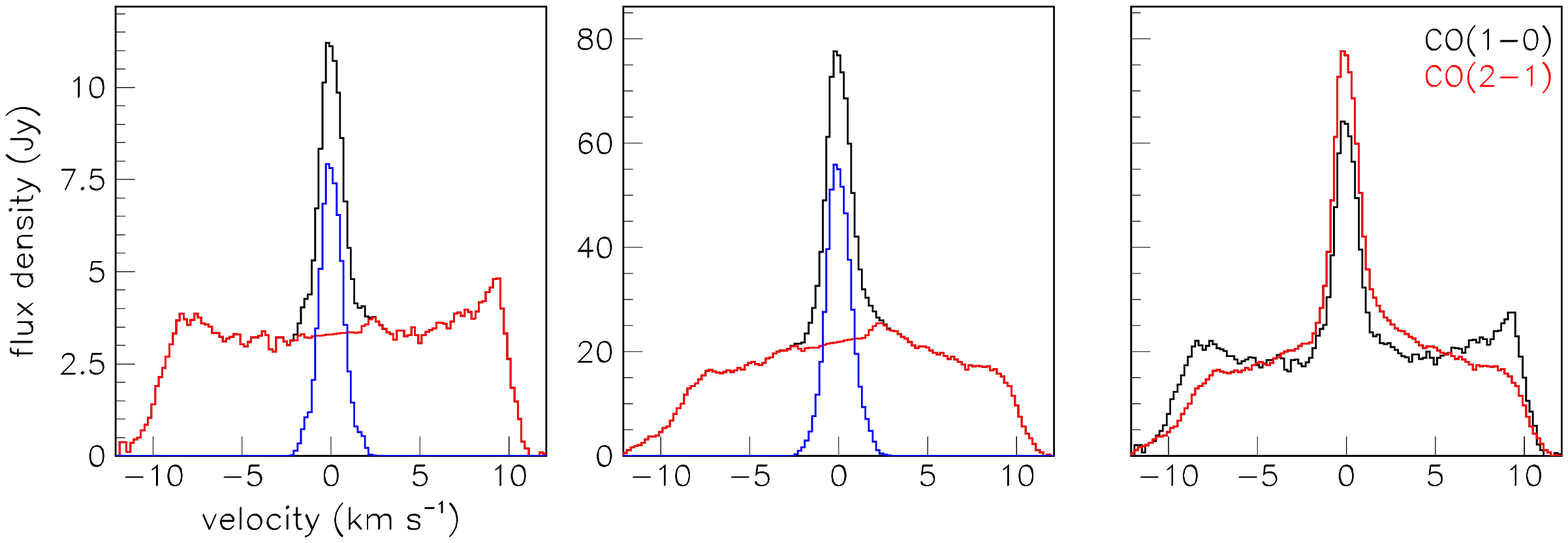}
\caption{Doppler velocity spectra of \mbox{CO(1-0)} (left) and \mbox{CO(2-1)} (middle) emissions. The broad component is shown in red and the narrow component is shown in blue. Right: comparison between the two spectra normalised to a common area (\mbox{CO(1-0)} black and \mbox{CO(2-1)} red).}
\label{fig2}
\end{figure*}

Figure \ref{fig2} displays the distributions of Doppler velocity, $V_z$. In this figure, like in the remainder of the present work, we limit our analysis to pixels distant by less than 8 arcsec from the central star. The rightmost panel of Figure \ref{fig2} compares the \mbox{CO(1-0)} and \mbox{CO(2-1)} spectra normalised to a same area, giving evidence for a significant difference of shape near the end-points. This feature has been discussed in \citet{Nhung2018a} and associated with a depression of the effective emissivity near the poles of the star. We discuss it in some detail in Section \ref{sec5.2}.

\subsection{Splitting the data cubes}\label{sec3.2}

Earlier analyses have associated, at least qualitatively, the narrow central peak of the Doppler velocity spectra with an equatorial outflow being seen nearly face-on, and the wings of the spectra with a bipolar outflow being seen nearly pole-on. We exploit this remark by separating the data cubes in two parts, referred to as narrow and broad components and associated with equatorial and bipolar outflows respectively. To do so, we use in each pixel two control-regions, each 2 \kms\ wide, located symmetrically with respect to the position of the central peak in the pixel and covering between 2 \kms\ and 4 \kms\ away from it on either side. We then interpolate linearly between the two control-regions what we attribute to the broad component, the narrow component being obtained by subtraction. We have checked thoroughly that the results are very robust and that the algorithm being used does not produce any significant bias. Indeed most of these results would not be strongly affected if the narrow component were simply defined as obeying the inequality $|V_z|<2$ \kms, the broad component being defined as its complement, $|V_z|>2$ \kms.

\begin{figure*}
\centering
\includegraphics[width=0.47\textwidth,trim=0.5cm 0.cm 0.5cm 0.cm,clip]{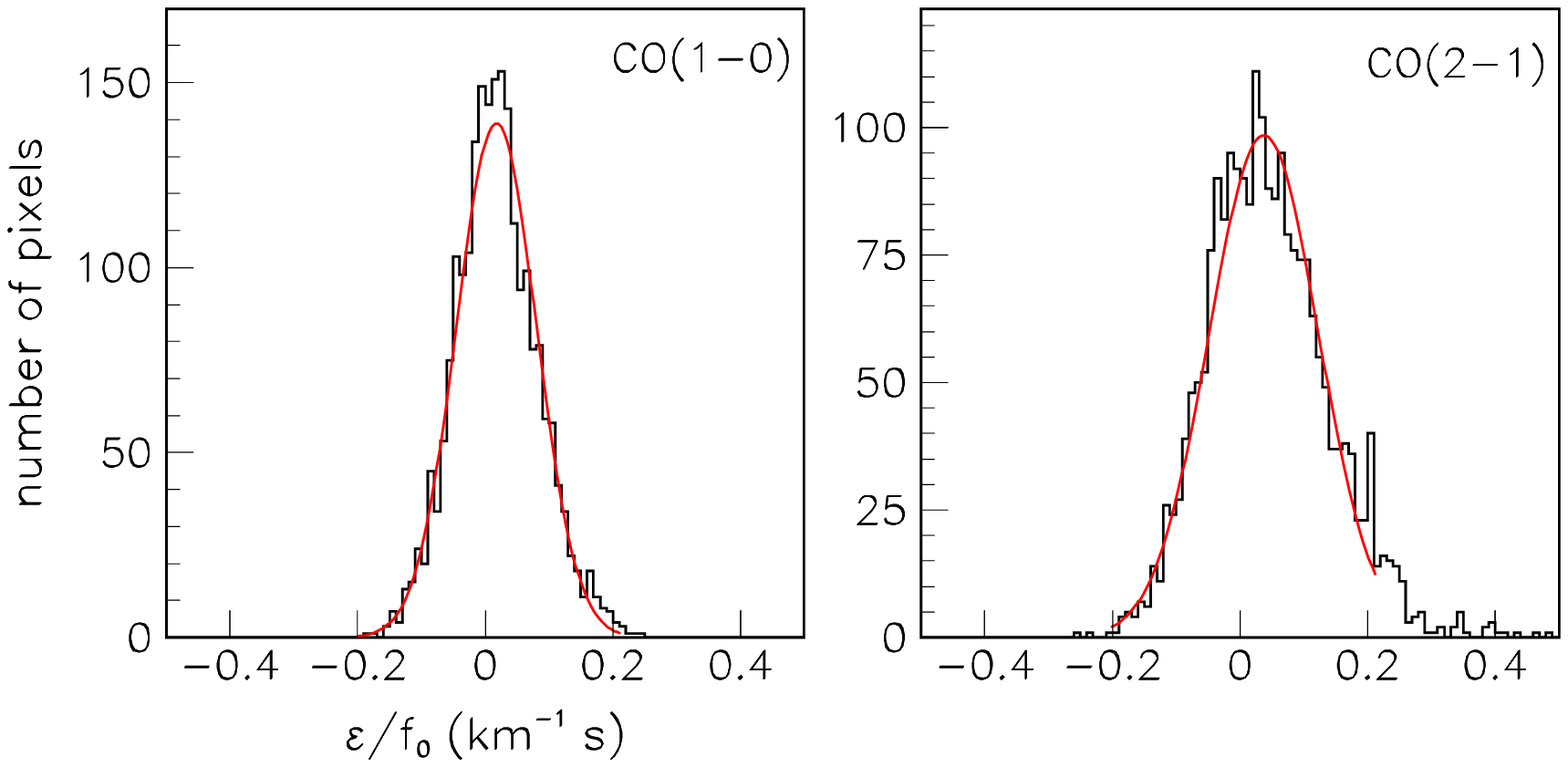}
\includegraphics[width=0.47\textwidth,trim=0.5cm 0.cm 0.5cm 0.cm,clip]{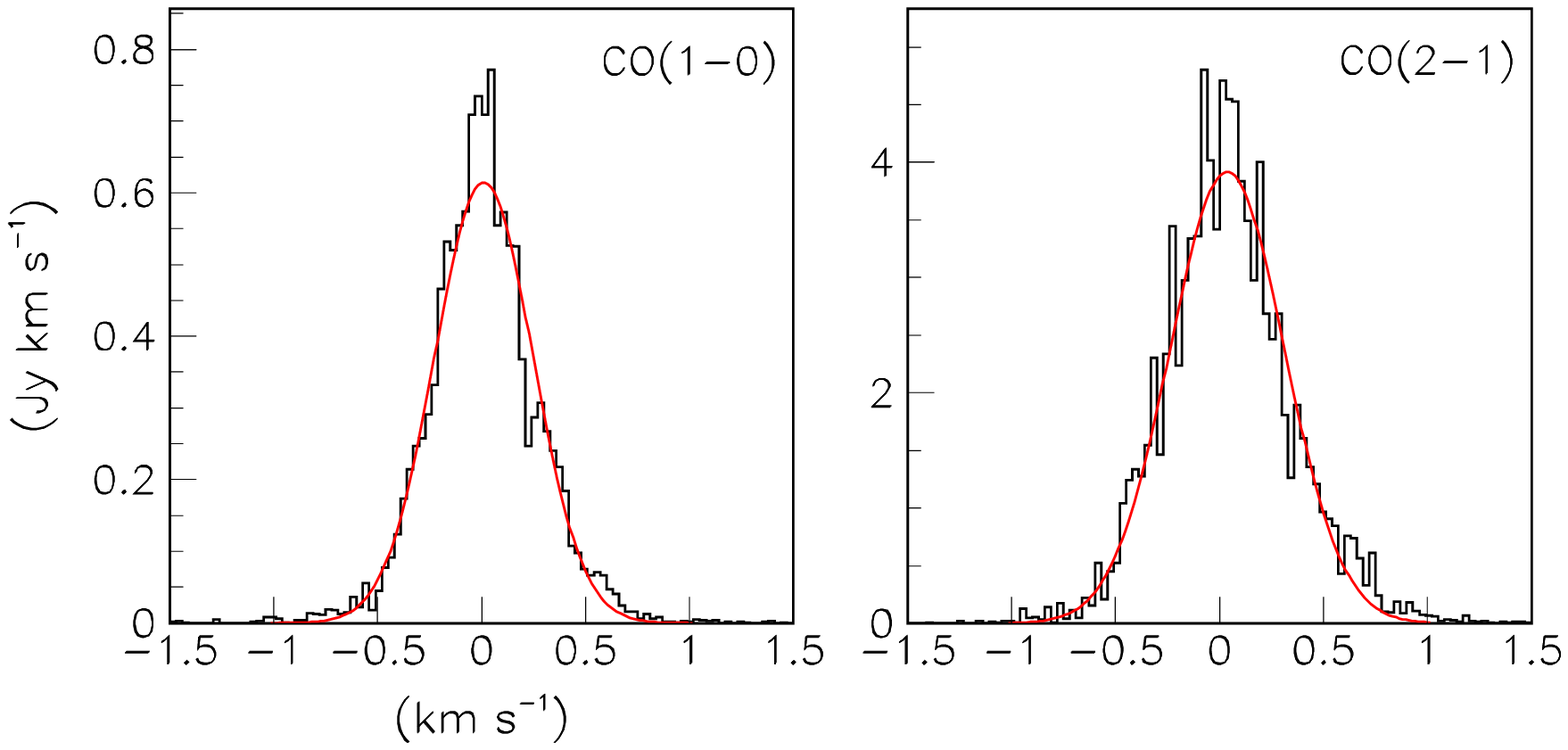}
\caption{Leftmost panels: distribution of the relative slope $\varepsilon/f_0$ of the linear interpolation below the peak for \mbox{CO(1-0)} (left) and \mbox{CO(2-1)} (middle left). Rightmost panels: shift in $V_z$ between the peak position estimated from the total spectrum used in the first iteration and the narrow component average used in the second iteration for \mbox{CO(1-0)} (middle right) and \mbox{CO(2-1)} (right).}
\label{fig3}
\end{figure*}

We show in Figure \ref{fig3} two distributions that are meant to illustrate the quality of the results: one is of the relative slope of the interpolated broad component below the peak, which is of the form $f=f_0+\varepsilon V_z$; the quantity plotted, $\varepsilon/f_0$, is well-behaved with an approximate Gaussian shape centred at 0.02 (0.04) km$^{-1}$\,s and having a $\sigma$ of 0.06 (0.09) km$^{-1}$\,s for \mbox{CO(1-0)} (\mbox{CO(2-1)}). The other distribution is of the difference between the Doppler velocity of the central peak estimated from the total spectrum in a first step and that obtained for the narrow component in a second iteration. Again, it is very well-behaved with an approximate Gaussian shape centred at 0.01 (0.04) \kms\ with a $\sigma$ of 0.24 (0.28) \kms\ for \mbox{CO(1-0)} (\mbox{CO(2-1)}). The separation between the narrow and broad components is indicated in Figure \ref{fig2} on each of the \mbox{CO(1-0)} and \mbox{CO(2-1)} Doppler velocity spectra.

\subsection{Dependence of the Doppler velocity on position angle: evidence for a tilt of the star axis}\label{sec3.3}

\begin{figure*}
\centering
\includegraphics[width=0.63\textwidth,trim=0cm 0.cm 0cm 0.cm,clip]{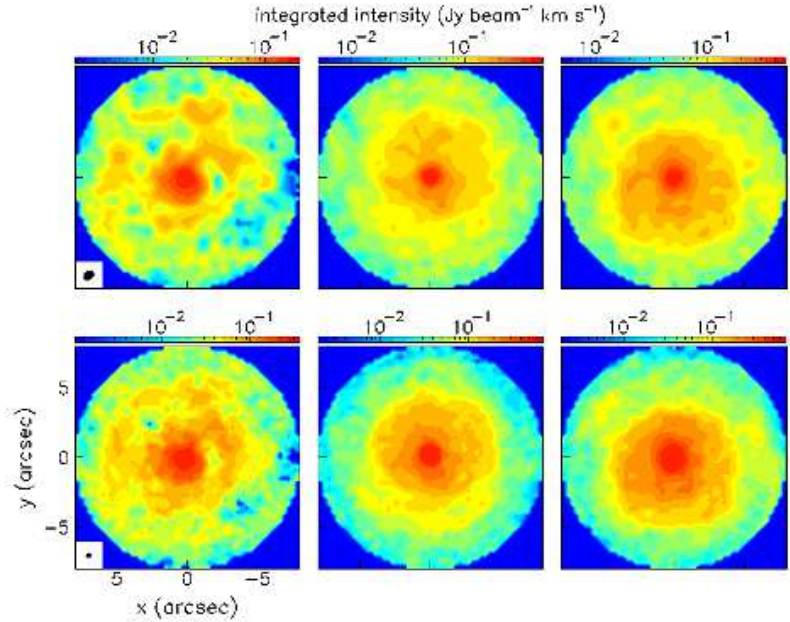}
\caption{Sky maps of the integrated intensity of the narrow (left), blue-shifted broad (middle) and red-shifted broad (right) components for \mbox{CO(1-0)} (upper panels) and \mbox{CO(2-1)} (lower panels). The beams are shown in the lower left corners of left panels.}
\label{fig4}
\end{figure*}

\begin{figure*}
  \centering
  \includegraphics[width=0.95\textwidth,trim=0cm .5cm 0cm 0.cm,clip]{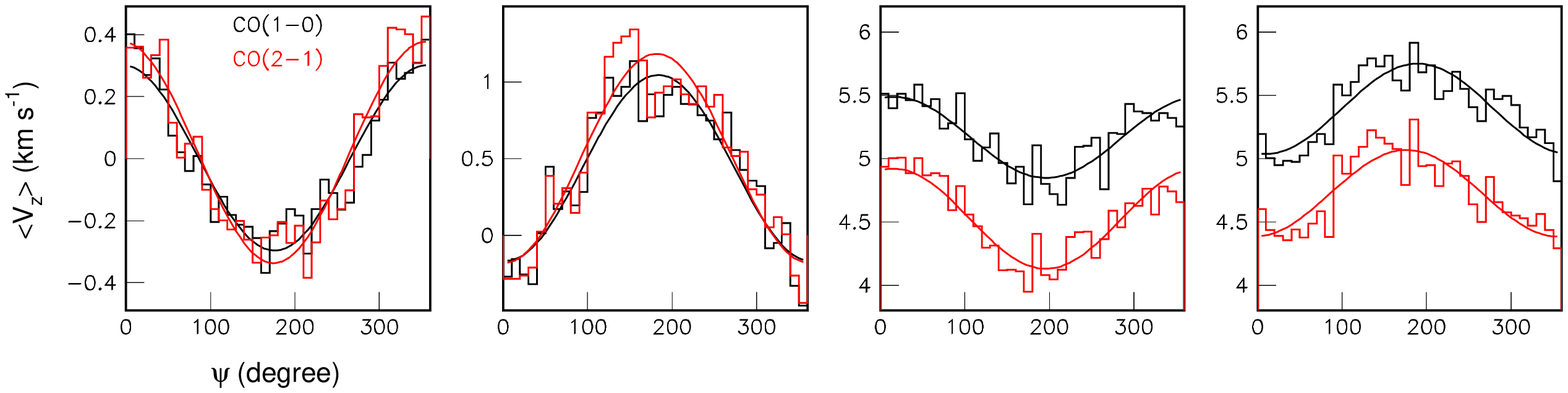}
  \caption{From left to right: Dependence on position angle $\psi$ of the mean Doppler velocity $<\!\!V_z\!\!>$ for each of the narrow, broad, blue-shifted broad and red-shifted broad components. For the blue-shifted broad component we use $<\!\!-V_z\!\!>$ instead of <$V_z$>. Fits of the form $<\!\!V_z\!\!>=a+b\cos(\psi-\psi_0)$ (see Table \ref{table2}) are illustrated in the figure. In all panels, black is for \mbox{CO(1-0)} and red for \mbox{CO(2-1)}.}
  \label{fig5}
\end{figure*}

Figure \ref{fig4} displays the sky maps of the integrated intensity of the broad and narrow components, the red-shifted and blue-shifted halves of the broad component being displayed separately. Figure \ref{fig5} displays the dependence on position angle $\psi$ (measured counter-clockwise from the $y$ direction) of the mean Doppler velocity for each of the narrow, broad, blue-shifted broad and red-shifted broad components separately. In the case of the blue-shifted broad component we plot $<\!\!-V_z\!\!>$ instead of $<\!\!V_z\!\!>$. Fits of the form $<\!\!V_z\!\!>=a+b\cos(\psi-\psi_0)$ are illustrated on the figure and their results are listed in Table \ref{table2}. The values of the offset $a$ and of the phase-shift $\psi_0$ are negligible for the narrow component; indeed excellent fits are obtained by simply writing $<\!\!V_z\!\!>=b\cos\!\psi$ with $b=0.30$ (0.36) \kms\ for \mbox{CO(1-0)} (\mbox{CO(2-1)}).

\begin{table*}
  \centering
  \caption{Result of the fits of the mean Doppler velocity of the narrow, broad, blue-shifted broad and red-shifted broad components of the form $<\!\!V_z\!\!>=a+b\cos(\psi-\psi_0)$. For the blue-shifted broad component we use $<\!\!-V_z\!\!>$ rather than $<\!\!V_z\!\!>$.}
  \begin{tabular}{|c|c|c|c|c|}
    \hline
    & & $a$ (\kms)&$b$ (\kms)&$\psi_0$ (degree) \\
    \hline
    \multirow{4}{*}{CO(1-0)} &Narrow component&0&0.30&$-$4.1\\
    \cline{2-5}
    &Broad component&0.44&$-$0.61&3.8\\
    \cline{2-5}
    &Blue-shifted broad component&5.2&0.32&16\\
    \cline{2-5}
    &Red-shifted broad component&5.4&$-$0.36&9\\
    \hline
    \multirow{4}{*}{\mbox{CO(2-1)}}&Narrow component&0.02&0.36&$-$5.0\\
    \cline{2-5}
    &Broad component&0.50&$-$0.68&1.8\\
    \cline{2-5}
    &Blue-shifted broad component&4.5&0.39&16\\
    \cline{2-5}
    &Red-shifted broad component&4.7&$-$0.34&$-$4\\
    \hline
  \end{tabular}
\label{table2}
\end{table*}

A radial expansion velocity $V$ produces a Doppler velocity $V_z=V\cos\!\psi\sin\!\varphi$ for the star equator and $V_z=V\cos\!\psi\cos\!\varphi$ for the star poles; $b$ is therefore a measure of some effective value of $V\sin\!\varphi$ for the narrow component, effective value meaning averaged over the relevant range of stellar latitudes: in particular, for $\varphi=10$\dego\ ($\sin\!\varphi=0.17$) $b=0.33\pm0.03$ \kms\ means an average expansion velocity of $\sim0.33/0.17=1.9\pm0.2$ \kms. 

\subsection{Effect of a possible rotation}\label{sec3.4}

The fact that $b$ takes values of opposite signs for the narrow and broad components shows that we are globally dealing with expansion rather than with rotation: rotation would be about an axis making a small angle with the line of sight and projecting on the $x$ axis; but the whole $y<0$ hemisphere would be shifted the same way, blue or red, and the whole $y>0$ hemisphere would be shifted the opposite way, red or blue, for both the narrow and broad components. The sign of $b$ implies that the axis of the star makes a small positive angle $\varphi$ with the line of sight, away from the $y$ axis, namely pointing to the $\cos\!\varphi<0$ hemisphere, as can be seen directly in the right panels of Figure \ref{fig4}.

The small values found for the phase-shift $\psi_0$ show that the rotation of 20\dego\ applied to the system of coordinates was correct.  We note however that $\psi_0\sim-4$\dego\ for the narrow component and $\sim+3$\dego\ for the broad component, a barely significant difference as we estimate the uncertainty on these numbers as $\sim10$\dego. Yet, the difference of $\sim7$\dego$\pm14$\dego\ may reveal a small rotation of the narrow component with respect to the broad component. A rotation of mean velocity $V_{rot}$ about the star axis contributes a term $V_{rot}\sin\!\psi\sin\!\varphi$ to $V_z$ instead of $V\cos\!\psi\sin\!\varphi$ for expansion, adding up to $\sqrt{V^2+V^2_{rot}}\cos[\psi-\tan^{-1}(V_{rot}/V)]\sin\!\varphi$. A difference of phase shift is therefore a direct measure of $V_{rot}/V$; namely the observed phase shift, 7\dego$\pm$14\dego, translates into $V_{rot}/V=0.12\pm0.26$ or, conservatively, in an upper limit of $\sim$0.38 for the ratio $|V_{rot}|/V$. To the extent that the broad component can reasonably be assumed to display no rotation, this is a direct estimate of the amount of rotation that the equatorial outflow can accommodate.

\subsection{Dependence on latitude of the outflow}\label{sec3.5}

The value of $<|V_z|>$ is larger for the \mbox{CO(1-0)} than for the \mbox{CO(2-1)} broad component, 5.3 \kms\ instead of 4.6 \kms\, as expected from the shape of the Doppler velocity spectra (Figure \ref{fig2} right). It is a factor $\sim15$ larger than the amplitude of the narrow component oscillation; for an inclination angle $\varphi\sim10$\dego, this implies a ratio between the average expansion velocity of the bipolar outflow and the equatorial expansion velocity of the order of $15\sin$(10\dego)$\sim2.6$. As the average expansion velocity of the equatorial outflow is $\sim1.9$ \kms\ for $\varphi\sim10$\dego, it means an average expansion velocity of the bipolar outflow of 2.6$\times$1.9 $\sim4.9$ \kms.  Moreover, the Doppler velocity spectra extend up to $|V_z|\sim11$ \kms, namely a factor $\sim2.2$ larger than the average expansion velocity of the bipolar outflow for $\varphi=10$\dego, implying an important increase of the expansion velocity when moving away from the equator to the poles. These estimates are in agreement with what found by \citet{Nhung2015a}.

\subsection{Radial dependence of the effective emissivity}\label{sec3.6}

\begin{figure*}
  \centering
  \includegraphics[height=5.cm,trim=0.5cm 0.5cm 1cm 1.2cm,clip]{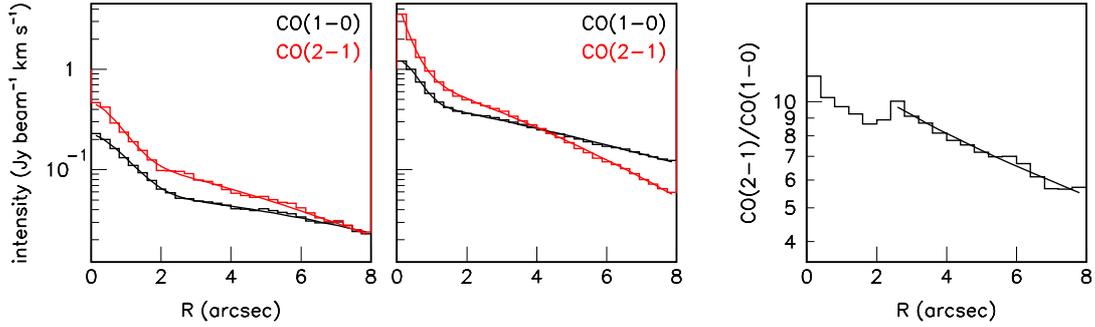}
  \caption{Left and middle: dependence on $R$ of the integrated intensity of the narrow (left) and broad (middle) components for \mbox{CO(1-0)} (black) and \mbox{CO(2-1)} (red) emissions. The lines show the result of fits listed in Table \ref{table4}. Right: ratio between \mbox{CO(2-1)} and \mbox{CO(1-0)} narrow component intensities corrected for beam sizes; the curve is the result of the temperature dependent prediction (see text). }
  \label{fig6}
\end{figure*}

The $R$-dependence of the integrated intensity of the narrow and broad components is shown separately for \mbox{CO(1-0)} and \mbox{CO(2-1)} emissions in the left panels of Figure \ref{fig6}. Beyond $R=2.5$ arcsec, fits of the form $I_0\times10^{-R/R_0}$ give results listed in Table \ref{table3}. The \mbox{CO(2-1)} distribution is steeper than the \mbox{CO(1-0)} distribution: on average, the ratio between the values of $R_0$ obtained for \mbox{CO(2-1)} and \mbox{CO(1-0)} is 0.55 (Figure \ref{fig6} right). The ratio between the values obtained for the broad and narrow components is 0.71. As $R$ is approximately equal to $r\cos\!\alpha$, where $\alpha$ is the stellar latitude, the latter ratio gives the scale of the mean angle made by the bipolar outflow with the equatorial plane: $\cos^{-1}(0.71)=45$\dego\ (an isotropic and uniform outflow gives 0.5, namely 60\dego). At least qualitatively, this implies that the broad component is associated with an outflow reaching  maximum emissivity at intermediate stellar latitudes.

\begin{table*}
  \centering
  \caption{Parameters of the fits $I_0\times10^{-R/R_0}$ to the $R$-dependence of the intensity ($R>2.5$ arcsec).}
  \begin{tabular}{|c|c|c|c|}
    \hline
    &Line&$I_0$ (Jy\,beam$^{-1}$\,\kms)&$R_0$ (arcsec)\\
    \hline
    \multirow{2}{*}{Narrow component}&CO(1-0)&0.076&16.2\\
    \cline{2-4}
    &\mbox{CO(2-1)}&0.177&9.0\\
    \hline
    \multirow{2}{*}{Broad component}&\mbox{CO(1-0)}&0.567&11.7\\
    \cline{2-4}
    &\mbox{CO(2-1)}&1.18&6.1\\
    \hline
  \end{tabular}
  \label{table3}
\end{table*}

The former ratio, between \mbox{CO(2-1)} and \mbox{CO(1-0)} emissions, was discussed by \citet{Nhung2018a} and interpreted as a temperature effect. The temperature-dependent factors $Q$ that scale the emissivity are of the form $Q=Q_0(2J+1)(2.8/T) \exp(-E_J/T)$; for \mbox{CO(1-0)} or respectively \mbox{\mbox{CO(2-1)}}, $J=1$ or 2, $E_J=5.5$ K or 16.6 K and $Q_0=7.4\ 10^{-8}$ s$^{-1}$ or $7.1\ 10^{-7}$ s$^{-1}$. This can be checked directly on the narrow component for which $r\sim R$: there is no need for de-projection. As illustrated in the right panel of Figure \ref{fig6}, a radial dependence of the temperature of the form $T=45\,r^{-0.72}$ K ($r$ in arcsec) reproduces the observed \mbox{\mbox{CO(2-1)}} to \mbox{CO(1-0)} ratio, $\sim16\exp(-11.1/T[K])$,  for $R>2.5$ arcsec. 

Below 2.5 arcsec, the observed intensity of the narrow component is steeper than what a simple scaling of $R_0$ between \mbox{CO(2-1)} and \mbox{CO(1-0)} on one hand and between broad and narrow components on the other would imply. This translates into a wiggle on the $R$-dependence of the intensity ratio, clearly seen in the right panel of Figure \ref{fig6}. However, in this region, significant absorption effects may be expected and the narrow and broad components should not be studied in isolation but jointly.

\section{The narrow component}\label{sec4}

\subsection{Doppler velocity and intensity fluctuations}\label{sec4.1}

\begin{figure*}
  \centering
  \includegraphics[height=5.5cm,trim=.5cm .4cm 1.5cm 0.9cm,clip]{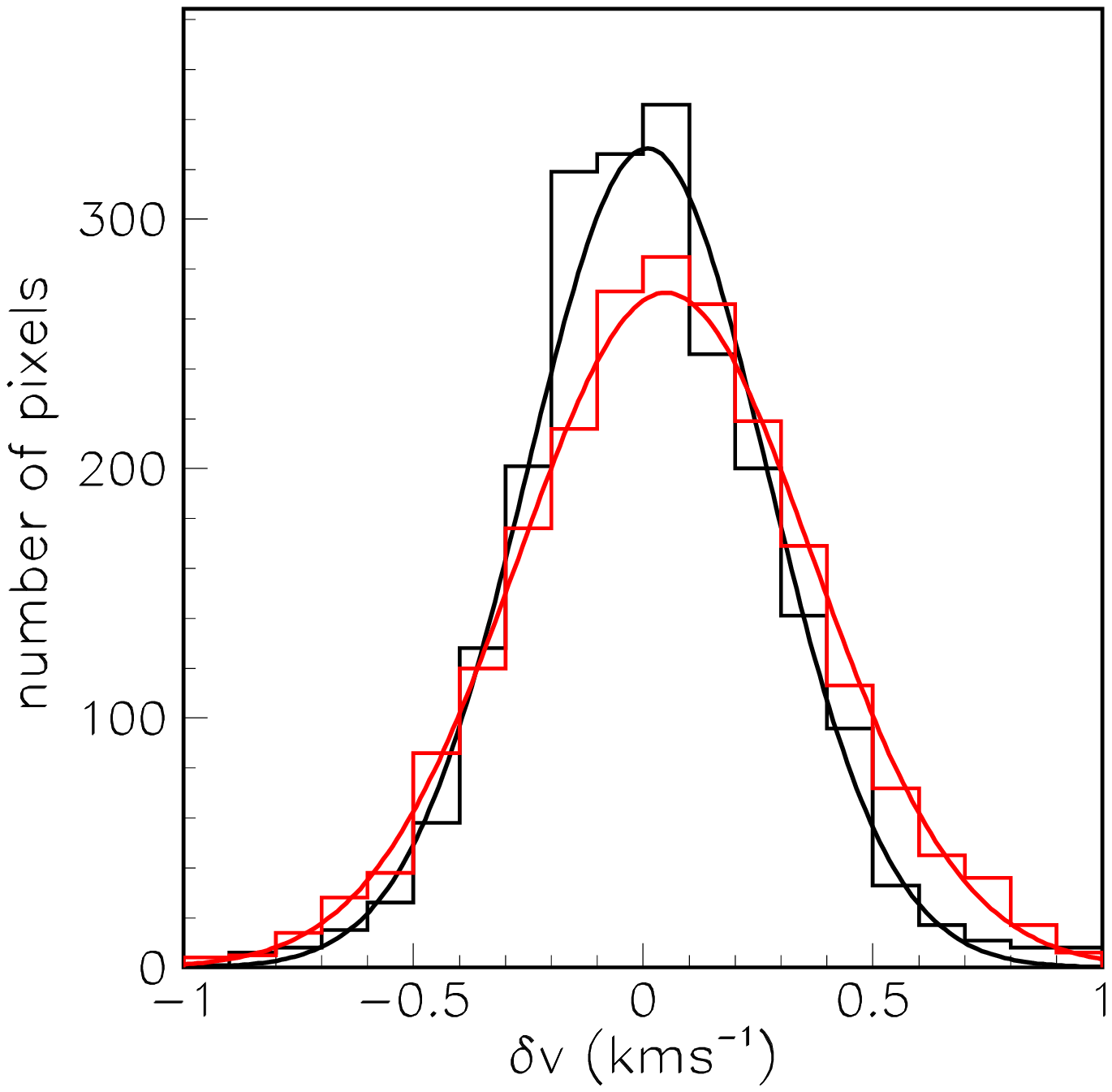}
  \includegraphics[height=5.5cm,trim=.5cm 1.cm 2.2cm 1.cm,clip]{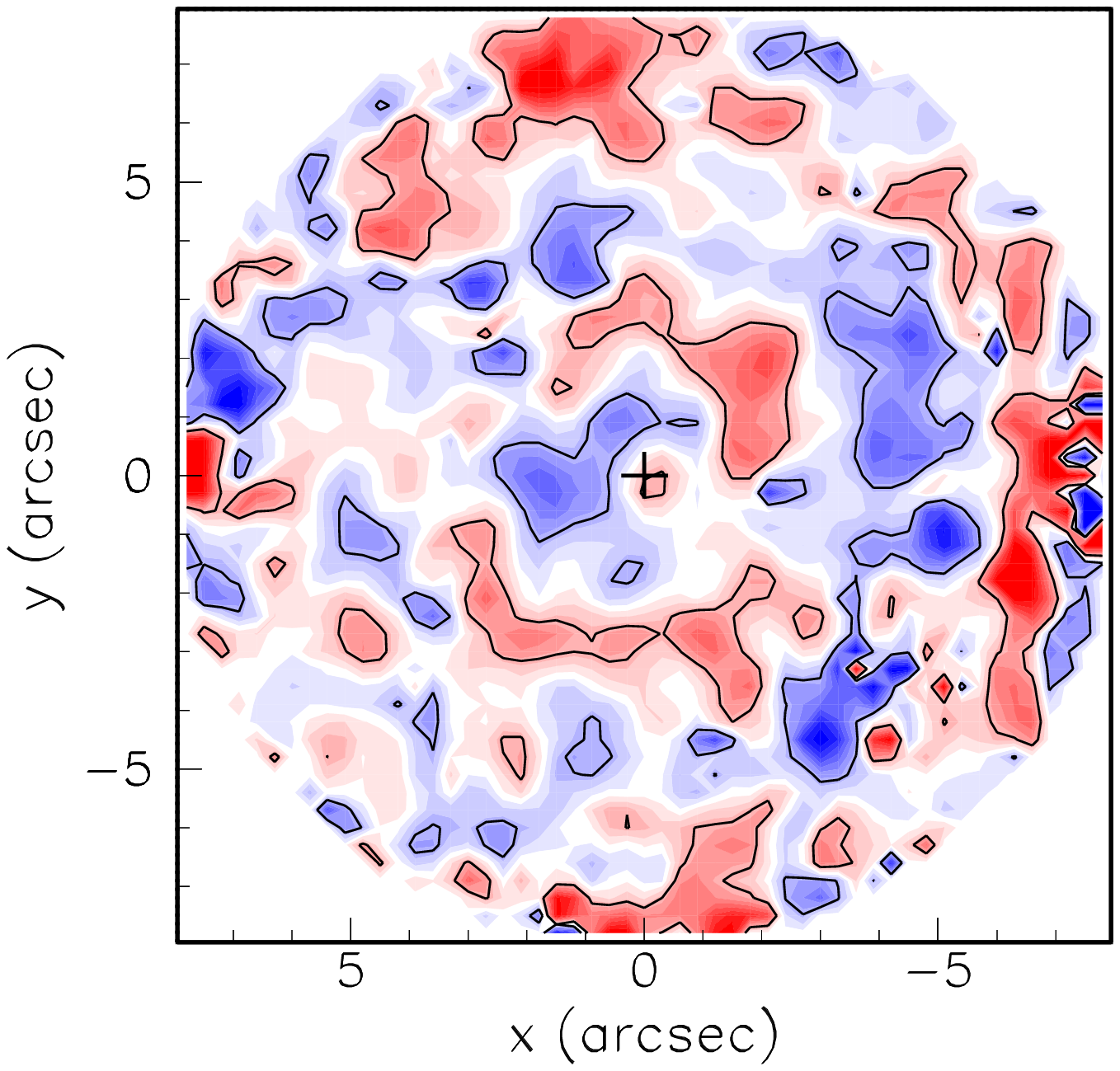}
  \includegraphics[height=5.5cm,trim=2.2cm 1.cm .0cm 1.cm,clip]{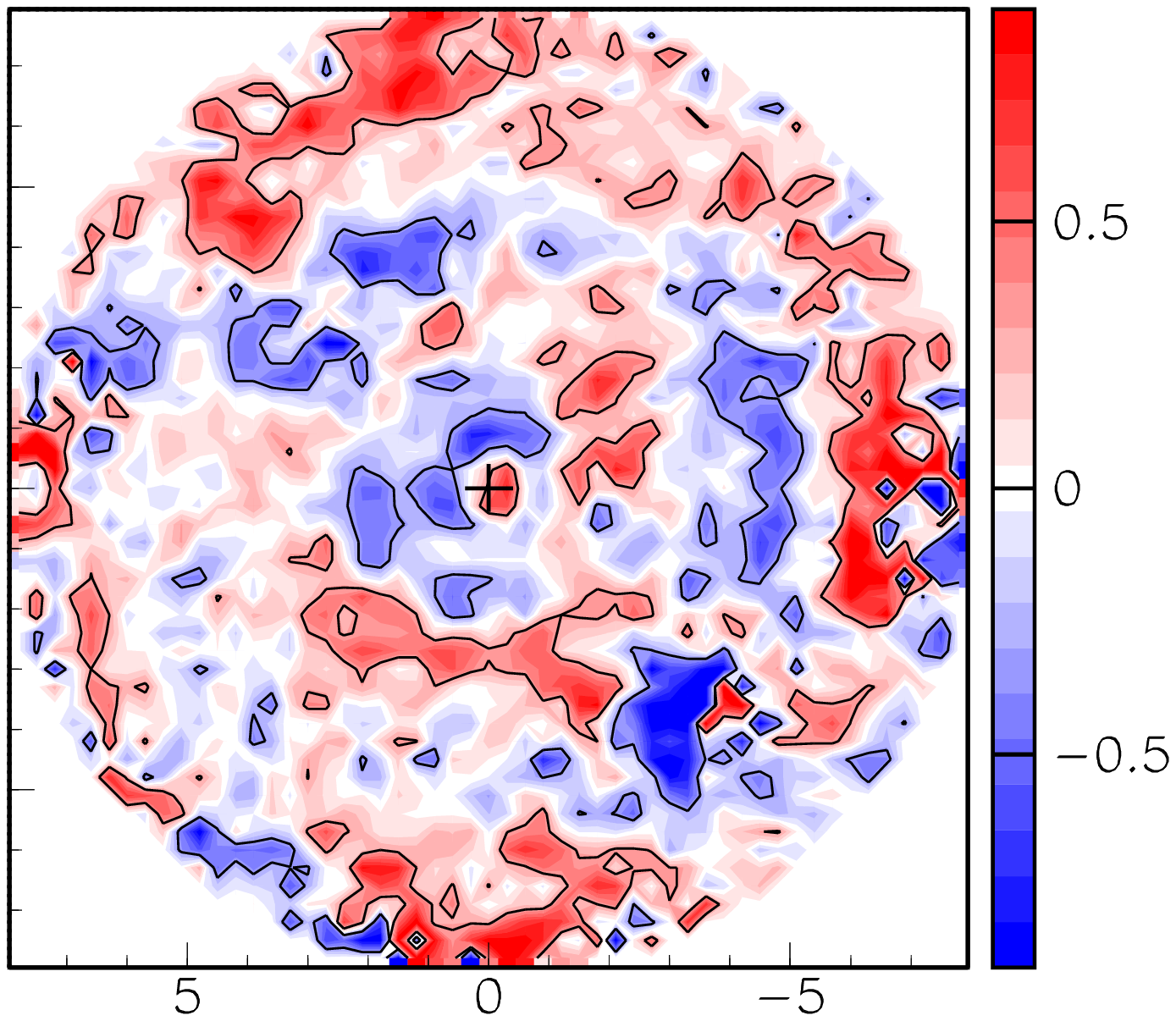}
  \caption{Left: distributions of $\delta{V}$ for \mbox{CO(1-0)} (black) and \mbox{CO(2-1)} (red) emissions. The curves are Gaussian fits (see text). Middle and right: sky maps of $\delta{V}$ (\kms) for \mbox{CO(1-0)} (middle) and \mbox{CO(2-1)} (right) emissions; contours are at $\pm 1\sigma$.}
  \label{fig7}
\end{figure*}

As the narrow component is associated with the equatorial outflow and is seen nearly face-on, much can be learned about its properties without having to de-project the data cube into space. The results displayed in Figure \ref{fig5} and listed in Table \ref{table2} have shown that the mean Doppler velocity is well described by a $\cos\!\psi$ term with amplitude $<\!\!V\!\!>_0=0.30$ (0.36) \kms\ for \mbox{CO(1-0)} (\mbox{CO(2-1)}); in order to reveal possible additional features, it is therefore pertinent to introduce, for each pixel, a quantity \mbox{$\delta{V(x,y)}=<\!\!V_z\!\!>-<\!\!V\!\!>_0 \cos\!\psi$}. The distribution and sky maps of $\delta{V}$ are displayed in Figure \ref{fig7} for \mbox{CO(1-0)} and \mbox{CO(2-1)} emissions separately. The mean and rms values of $\delta{V}$ are respectively 0.01 and 0.26 \kms\ for \mbox{CO(1-0)} and 0.05 and 0.32 \kms\ for \mbox{CO(2-1)}. The agreement between \mbox{CO(1-0)} and \mbox{CO(2-1)} observations gives confidence in the quality of the result. 

\begin{figure*}
  \centering
  \includegraphics[height=5cm,trim=.5cm .9cm 1.5cm 1.cm,clip]{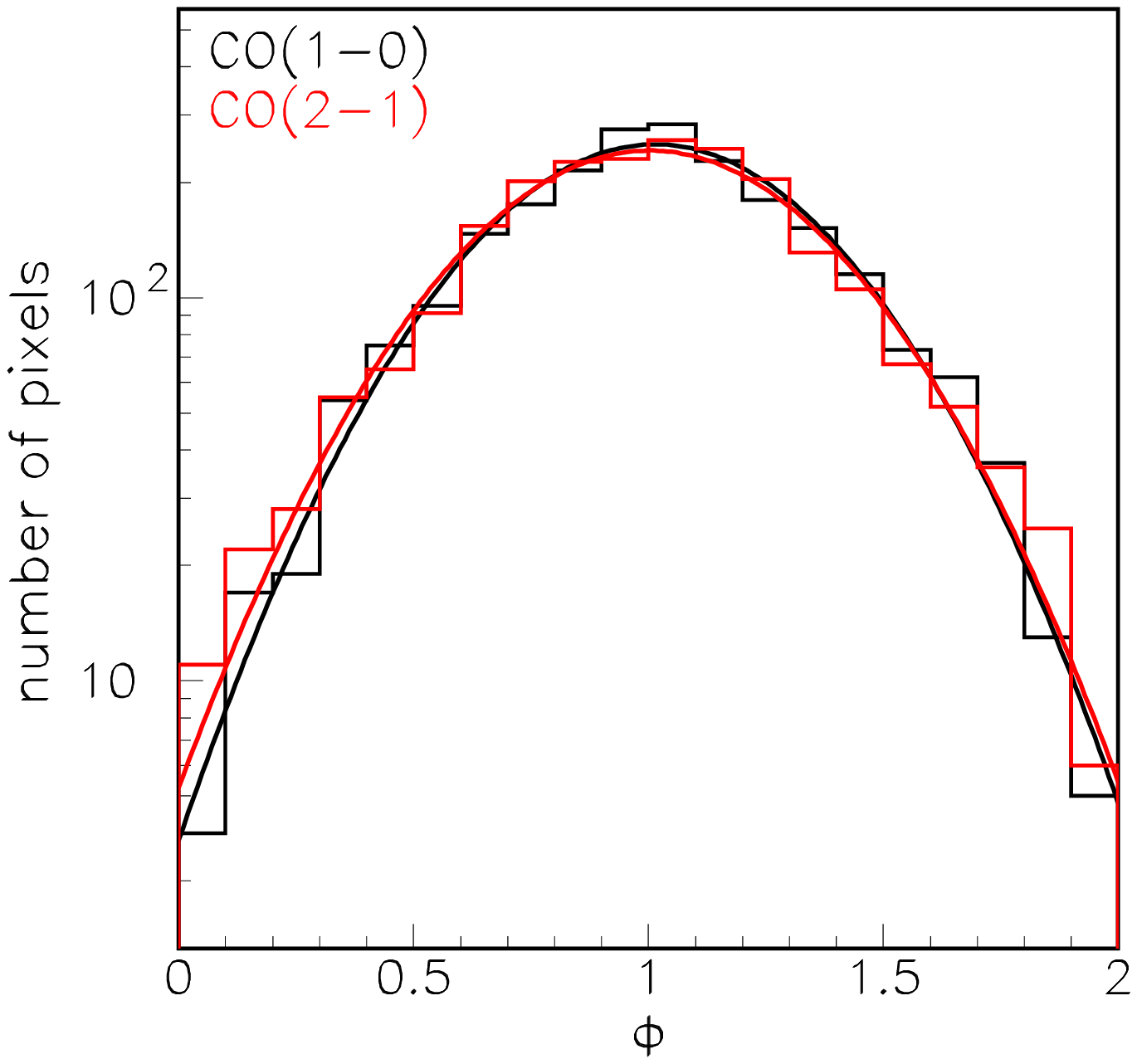}
  \includegraphics[height=5cm,trim=0.5cm 1.5cm 2cm 1.cm,clip]{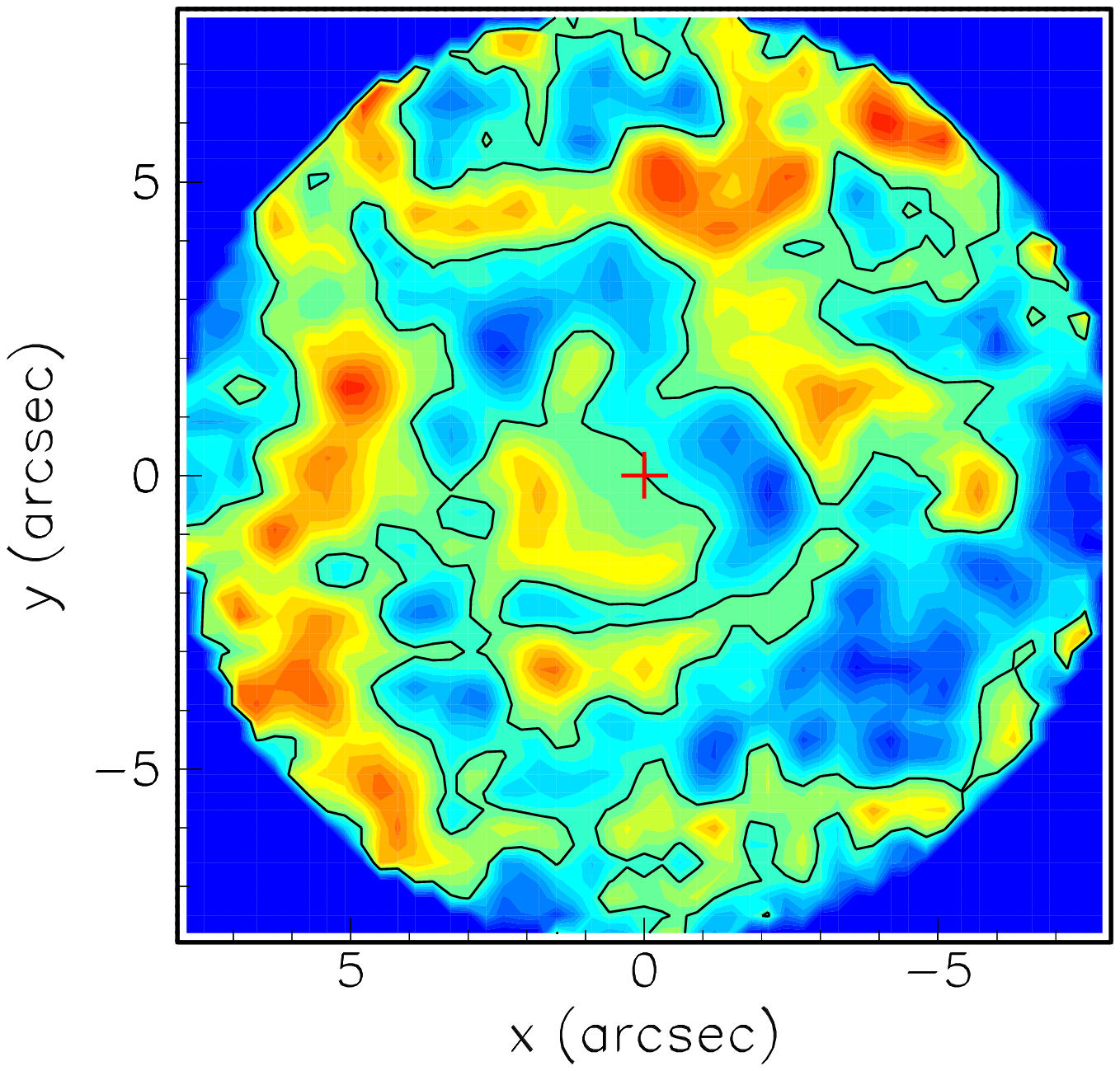}
  \includegraphics[height=5cm,trim=2cm 1.5cm .0cm 1.cm,clip]{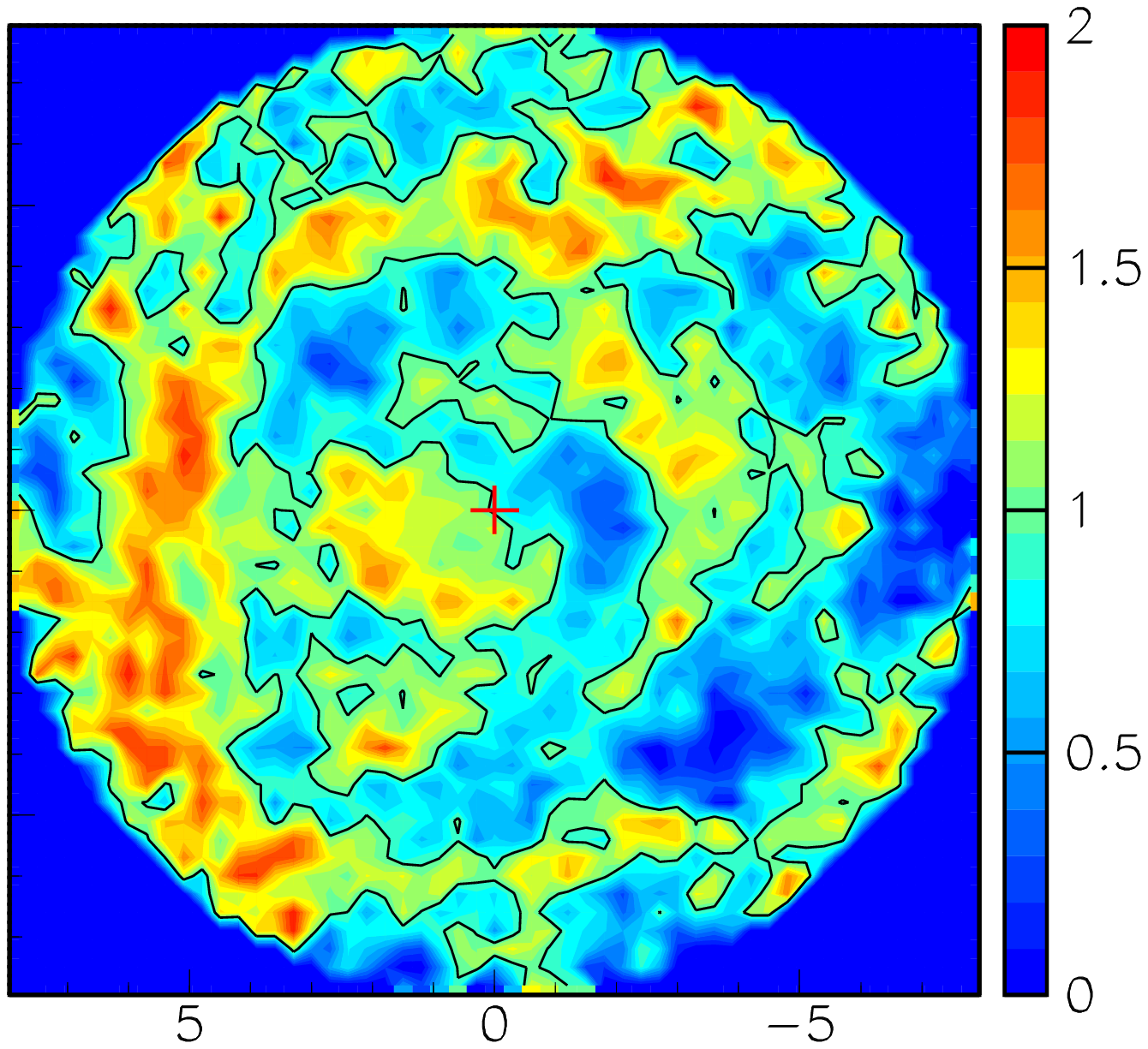}
    
  \caption{Narrow component. Left: distributions of $\mit{\Phi}$ for \mbox{CO(1-0)} (black) and \mbox{CO(2-1)} (red) emissions. Middle and right: sky maps of $\mit{\Phi}$ for \mbox{CO(1-0)} (middle) and \mbox{CO(2-1)} (right) emissions. Contours at $\mit{\Phi}=1$ are shown as black lines.}
  \label{fig8}
\end{figure*}

The sky maps display resolved lumps that are very similar between \mbox{CO(1-0)} and \mbox{CO(2-1)} emission, giving confidence in their reality: the kinematics of the narrow component is unambiguously shown to consist of lumps of slightly different Doppler velocities. Such lumpiness must, somehow, be associated with a related lumpiness of the morphology. In order to find out how, we define a function $\mit{\Phi}(x,y)= F(x,y)/\!\!<\!\!F\!\!>_R$ where $F(x,y)$ is the intensity measured in pixel $(x,y)$ and $<\!\!F\!\!>_R$ is its value averaged over $\psi$. In practice, we use parameterizations of $<\!\!F\!\!>_R$ in the form of a sum of two exponentials as given in the Appendix (Table \ref{table4}), which give good fits to the observations (Figure \ref{fig6}). The distribution of $\mit{\Phi}$ is displayed in Figure \ref{fig8} for \mbox{CO(1-0)} and \mbox{CO(2-1)} separately together with the sky maps. For both \mbox{CO(1-0)} and \mbox{CO(2-1)} it is well described by a Gaussian having mean 1 and $\sigma=0.36$. As in the case of $\delta{V}$ the maps of $\mit{\Phi}$ display features that are very similar in the \mbox{CO(1-0)} and \mbox{CO(2-1)} data, the effect of the smaller beam size being clearly apparent on the \mbox{CO(2-1)} map. This gives confidence in the method, showing in particular that the reference $<\!\!F\!\!>_R$ distributions are properly evaluated and introduce no significant bias. In both \mbox{CO(1-0)} and \mbox{CO(2-1)} emissions, an arc of enhanced intensity is seen to extend from $R\sim5$ to $\sim7$ arcsec in the north-eastern half of the sky and a deep depression is visible in the south-western quadrant.

\begin{figure*}
  \centering
  \includegraphics[height=5.cm,trim=.5cm 1.cm 3cm 0.cm,clip]{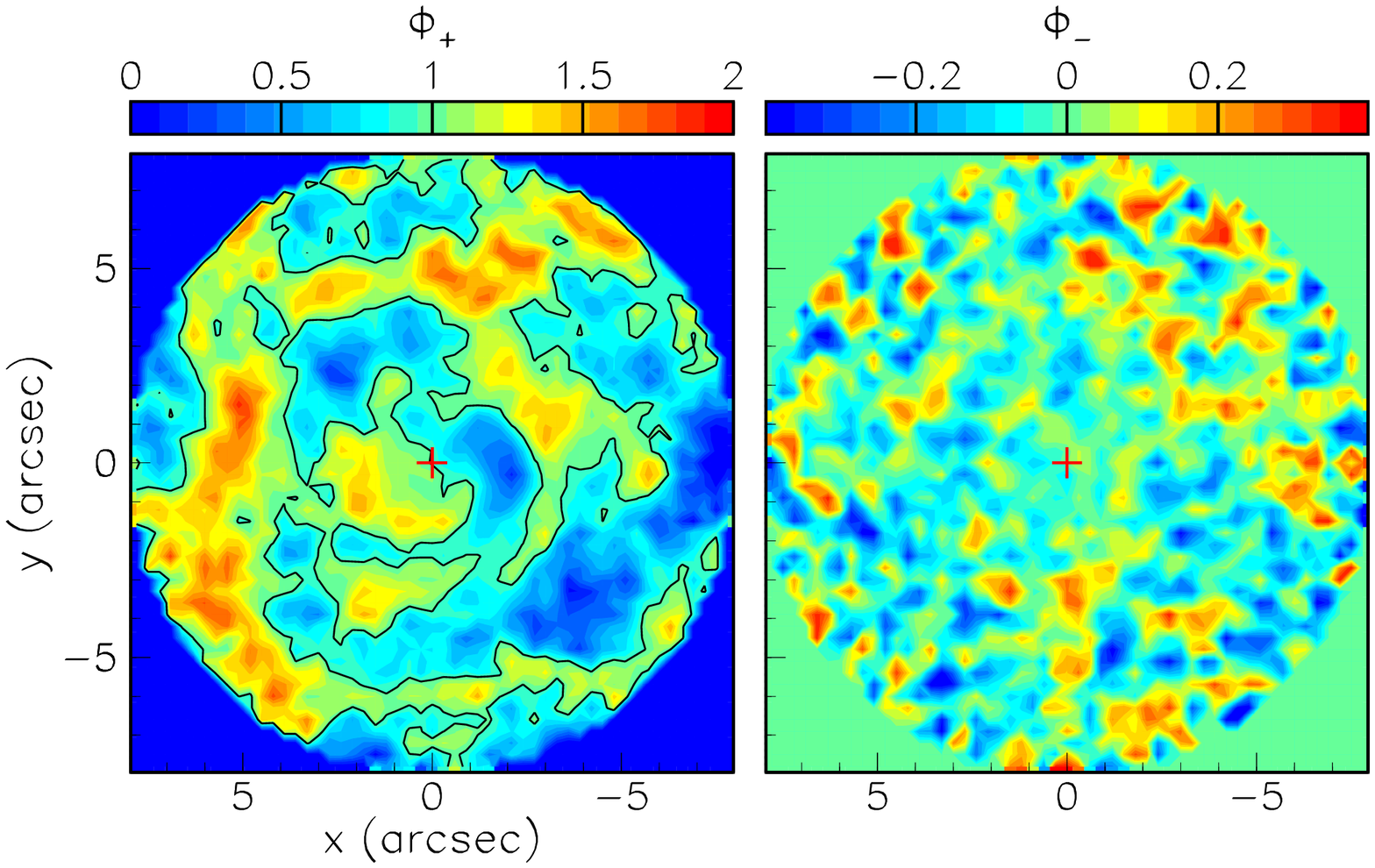}
  \includegraphics[height=5.cm,trim=2cm 1.cm 1.5cm 0.cm,clip]{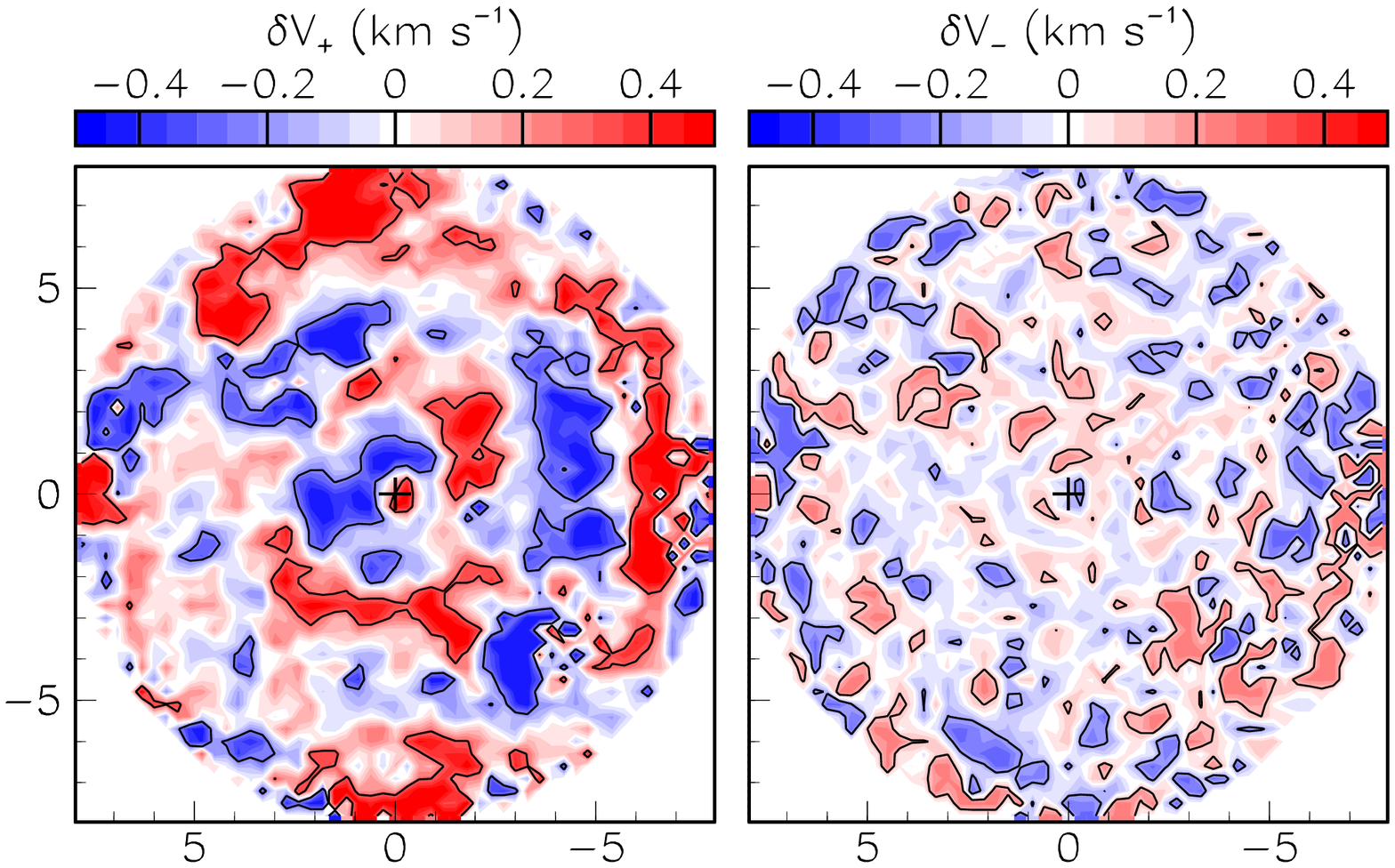}
\caption{Narrow component: sky maps of $\mit{\Phi}_+$ (left), $\mit{\Phi}_-$ (middle left), $\delta{V_+}$ (middle right) and $\delta{V_-}$ (right). Contours show level 1 for $\mit{\Phi}_+$ and mean $\pm \sigma$ (0.29 and $-$0.23 \kms) for $\delta{V_+}$ and (0.16 and $-$0.12 \kms) for $\delta{V_-}$.}
\label{fig9}
\end{figure*}

\begin{figure*}
  \centering
  \includegraphics[height=7.5cm,trim=0.cm 0cm 0cm 0.5cm,clip]{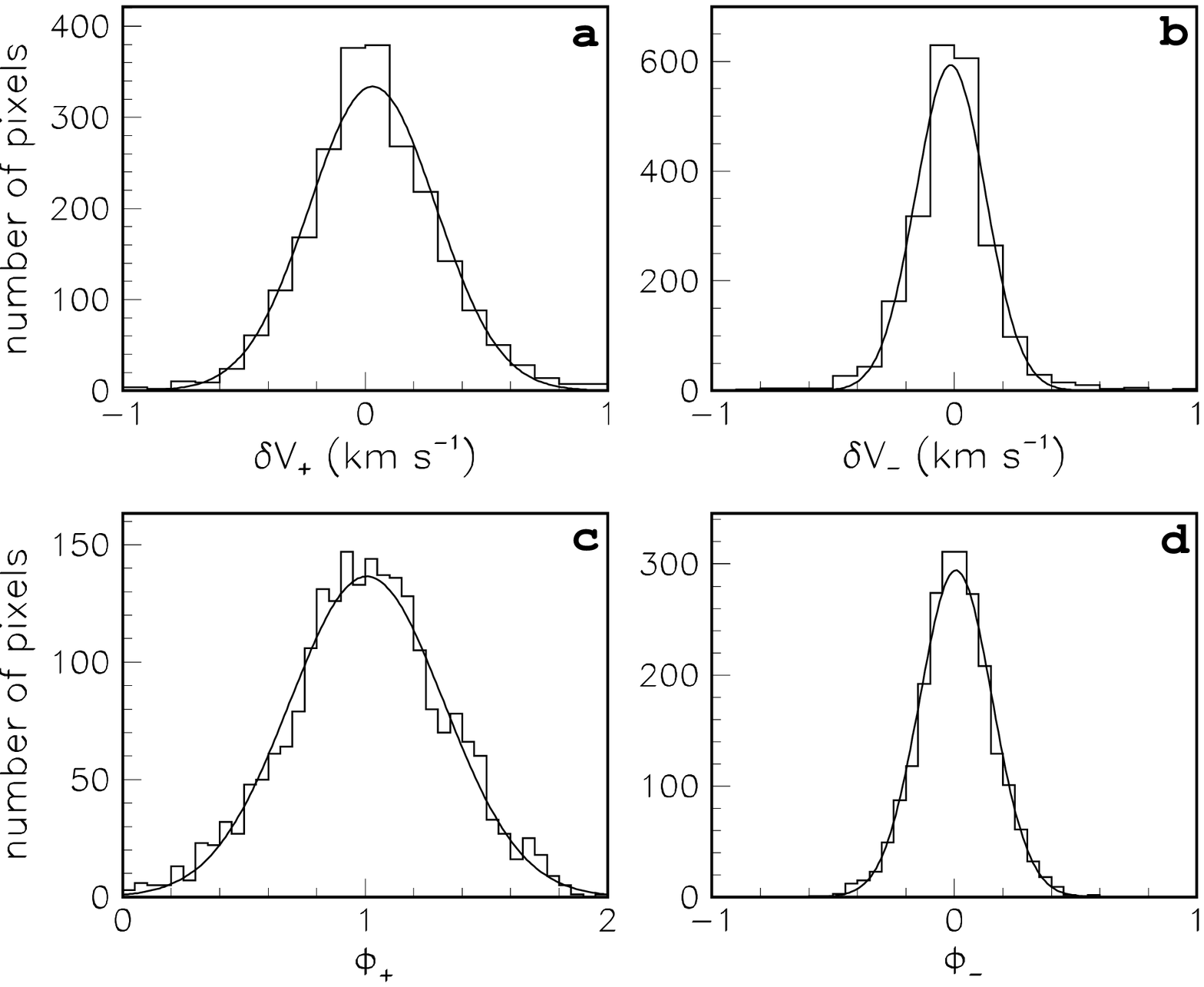}
  \includegraphics[height=7.5cm,trim=0.cm 0.5cm 1.5cm 0.cm,clip]{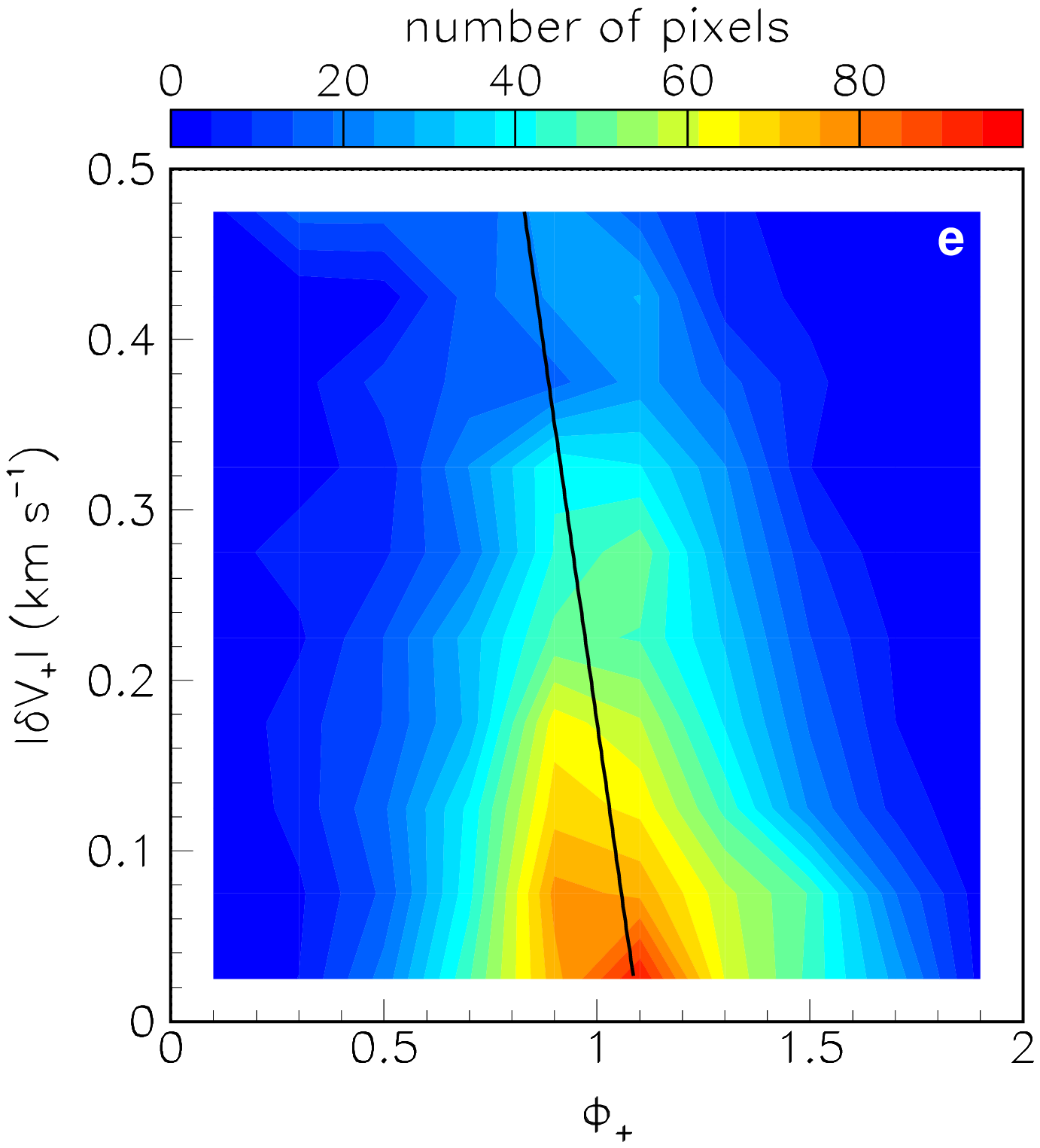}
  \caption{From a) to d): distributions of $\delta{V_+}$, $\delta{V_-}$, $\mit{\Phi}_+$ and $\mit{\Phi}_-$. The lines are Gaussian fits (see text). e) correlation between $|\delta{V_+}|$ and $\mit{\Phi}_+$. The line is for $\mit{\Phi}_+=1.10-0.57|\delta{V_+}|$.}
  \label{fig10}
\end{figure*}

To illustrate and quantify the similarity between \mbox{CO(1-0)} and \mbox{CO(2-1)} observations, we display in Figure \ref{fig9} maps of
\begin{equation}
  \begin{split}
    \delta{V_+}&=\frac{1}{2}[\delta{V}(\mbox{CO(1-0)})+\delta{V}(\mbox{CO(2-1)})],\nonumber\\
    \delta{V_-}&=\frac{1}{2}[\delta{V}(\mbox{CO(1-0)})-\delta{V}(\mbox{CO(2-1)})],\nonumber\\
    \mit{\Phi}_+&=\frac{1}{2}[\mit{\Phi}(\mbox{CO(1-0)})+\mit{\Phi}(\mbox{CO(2-1)})] \ \mbox{and}\nonumber\\
    \mit{\Phi}_-&=\frac{1}{2}[\mit{\Phi}(\mbox{CO(1-0)})-\mit{\Phi}(\mbox{CO(2-1)})].\nonumber
  \end{split}
\end{equation}

Both $\delta{V_-}$ and $\mit{\Phi}_-$ are seen to take small values. The results of Gaussian fits to the distributions of $\mit{\Phi}_\pm$ and $\delta{V_\pm}$ are listed in Table \ref{table5} and illustrated in Figure \ref{fig10}. In both cases the difference quantities, $\mit{\Phi}_-$ and $\delta{V_-}$, are about half as wide as the averaged quantities, $\mit{\Phi}_+$ and $\delta{V_+}$. A small but significant anti-correlation is found between $|\delta{V_+}|$ and $\mit{\Phi}_+$ with $\mit{\Phi}_+=1.10-0.57|\delta{V_+}|$ ($\delta{V_+}$ in \kms), as illustrated in the right panel of Figure \ref{fig10}.

\begin{figure}
\centering
\includegraphics[height=8.5cm,trim=.5cm 1.cm 1.5cm 1.cm,clip]{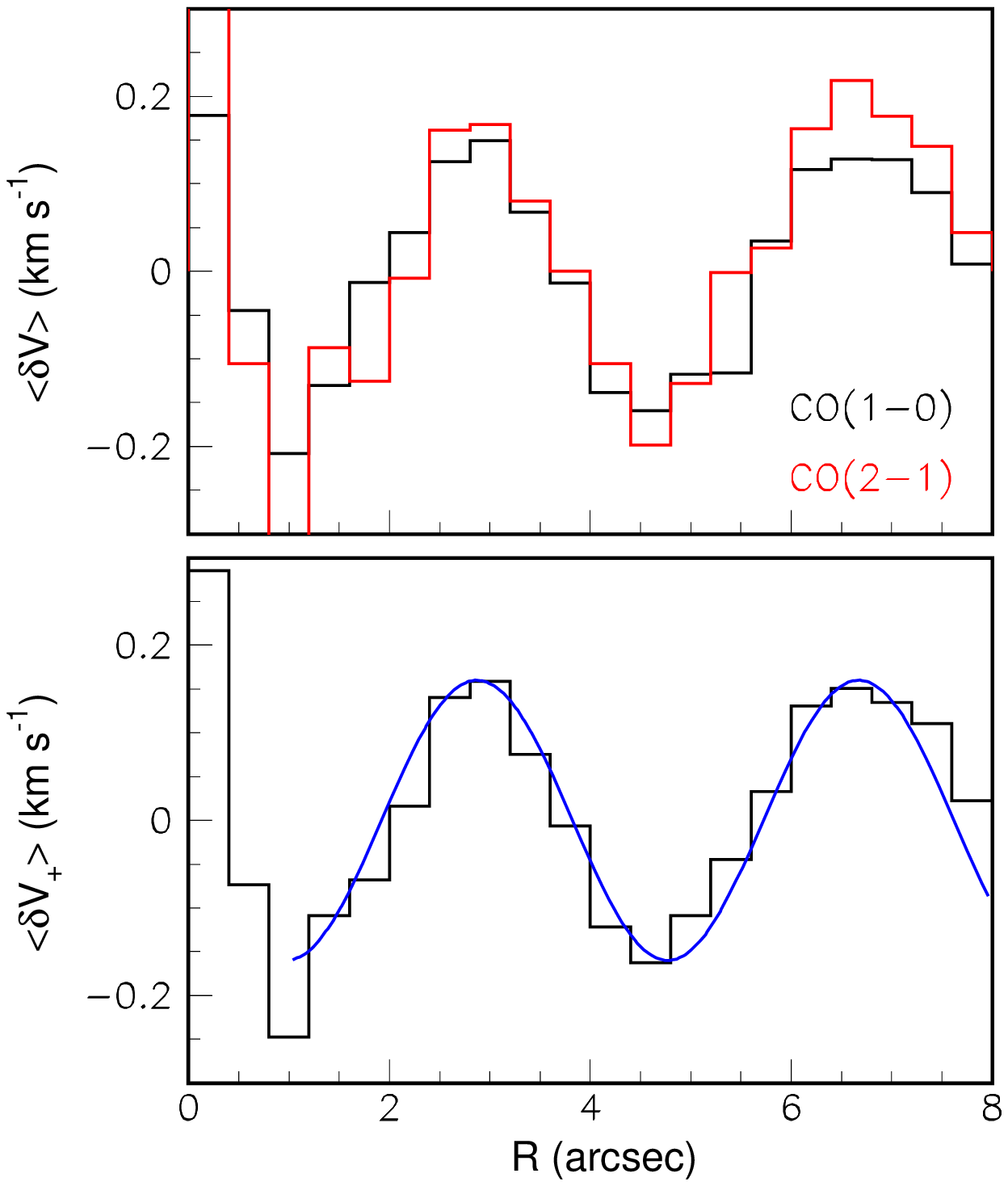}
\caption{Upper panel: dependence on $R$ of $<\!\!\delta{V}\!\!>$ for \mbox{CO(1-0)} (black) and \mbox{CO(2-1)} (red) emissions. Lower panel: dependence on $R$ of $<\!\!\delta{V_+}\!\!>$; the line is the result of a sine wave fit to the region $R>1$ arcsec of the form $<\!\!\delta{V_+}\!\!>=-0.16\sin(2\pi R/3.8-2^\circ)$.}
\label{fig11}
\end{figure}

The dependence on $R$ of $\delta{V}$ is displayed in the left panel of Figure \ref{fig11}, giving evidence for a strong modulation with maxima at $R\sim0$, 3 and 7 arcsec for both \mbox{CO(1-0)} and \mbox{CO(2-1)} emissions. The dependence on $R$ of $\delta{V_+}$ is shown in the middle panel with a sine wave fit having amplitude of $\pm0.16$ \kms, a period of 3.8 arcsec and a negligible phase-shift. Note that one might fear that such an oscillation would have been absorbed in the division by $<\!\!F\!\!>_R$ when dealing with the intensity; this is not the case, however, as $<\!\!F\!\!>_R$, being defined as shown in Figure \ref{fig6} and Table \ref{table4}, is free of such an oscillation.

\begin{table*}
  \centering
  \caption{Results of Gaussian fits to the $\mit{\Phi}_\pm$ and $\delta{V_\pm}$ distributions.}
  \begin{tabular}{|c|c|c|c|c|c|c|c|}
    \hline
    \multicolumn{2}{|c|}{$\mit{\Phi}_+$}&\multicolumn{2}{c|}{$\mit{\Phi}_-$}&\multicolumn{2}{c|}{$\delta{V_+}$ (\kms)}&\multicolumn{2}{c|}{$\delta{V_-}$ (\kms)}\\
    \hline
    Mean&$\sigma$&Mean&$\sigma$&Mean&$\sigma$&Mean&$\sigma$\\
    \hline
    1.01&0.32&0.01&0.15&0.03&0.26&$-$0.02&0.14\\
    \hline
  \end{tabular}
  \label{table5}
\end{table*}

\subsection{Spirals and arcs}\label{sec4.2}

\begin{figure*}
\centering
\includegraphics[height=8cm,trim=0cm 0cm 0cm 0cm,clip]{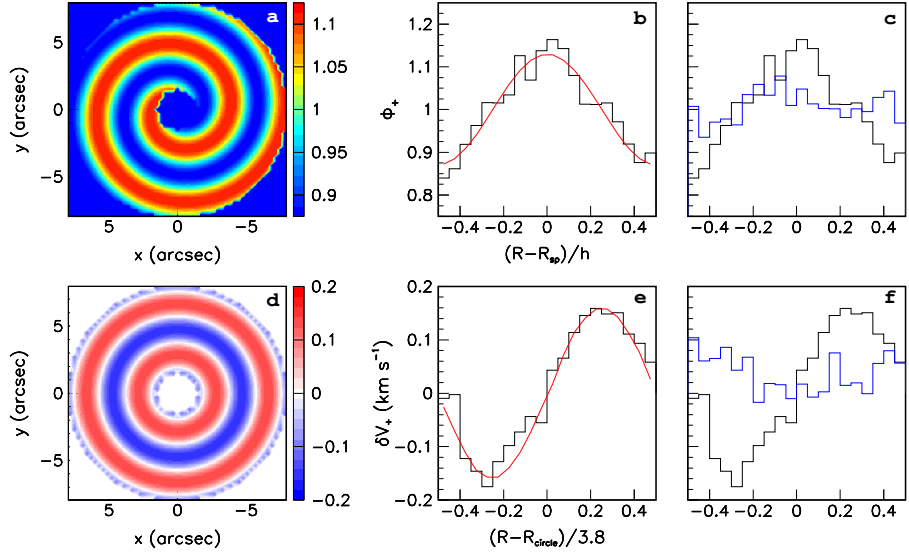}
\caption{a) sky map of the spiral model of $\mit{\Phi}_+$; b) measured (black) and model (red) dependence of $\mit{\Phi}_+$ on $(R-R_{sp})/h$; c) measured dependence of $\mit{\Phi}_+$ on $(R-R_{sp})/h$ (black) and on $(R-R_{circle})/3.8$ (blue); d) sky map of the circular model of $\delta{V_+}$; e) measured (black) and model (red) dependence of $\delta{V_+}$ on $(R-R_{circle})/3.8$; f) measured dependence of $\delta{V_+}$ on $(R-R_{circle})/3.8$ (black) and on $(R-R_{sp})/h$ (blue).}
\label{fig12}
\end{figure*}

The features displayed on the sky map of $\mit{\Phi}_+$ are reminiscent of the spiral studied by \citet{Homan2018} and the rings reported by \citet{Winters2007}. The question then arises of the relation between these and of their relation to the radial oscillations displayed by $\delta{V_+}$. In order to quantify the spiral assignment, we parameterize a possible spiral as $R_{sp}=R_0+\frac{1}{2}h\psi/\pi$ and use as a model $\mit{\Phi}_+(R,\psi)=1+\mit{\Phi}_0\cos[2\pi(R-R_{sp})/h]$. The best fit is obtained for $\mit{\Phi}_0=0.13$, $R_0=0.87$ arcsec and $h=4.0$ arcsec. Similarly, we define circles of radius $R_{circle}=1.9$ (modulo 3.8) arcsec meant to match the zeroes of $\delta{V_+}$ and model $\delta{V_+}$ as $\delta{V_+}=0.16\sin[2\pi(R-R_{circle})/3.8]$.  We restrict the analysis to projected distances from the star, $R$, in excess of 1.5 arcsec, where the spiral and circular assignments are more reliable. Figure \ref{fig12} displays the sky maps obtained from the resulting models (a spiral for $\mit{\Phi}_+$ and circles for $\delta{V_+}$) together with the dependence of $\mit{\Phi}_+$ and $\delta{V_+}$ on both $(R-R_{sp})/h$ and $(R-R_{circle})/3.8$ arcsec.

The result is clear: the spiral model fits $\mit{\Phi}_+$ but does not fit $\delta{V_+}$, the circular model fits $\delta{V_+}$ but does not fit $\mit{\Phi}_+$. However, the best-fit spiral is but a crude representation of the observations and differs from that observed by \citet{Homan2018}: it has a pitch of 4.0 arcsec instead of 2.8 arcsec (note that Homan et al. erroneously quote an inter-arm distance of only $\sim1$ arcsec in their paper); the reason is that it gives a better fit to the observations over the whole circle $R<8$ arcsec while the spiral described by Homan et al. covers only $R<5$ arcsec. This underlines the arbitrariness inherent to such spiral assignments when the observed features are insufficiently clear.

\begin{figure*}
\centering
\includegraphics[height=7cm,trim=0cm 0cm 0cm 0cm,clip]{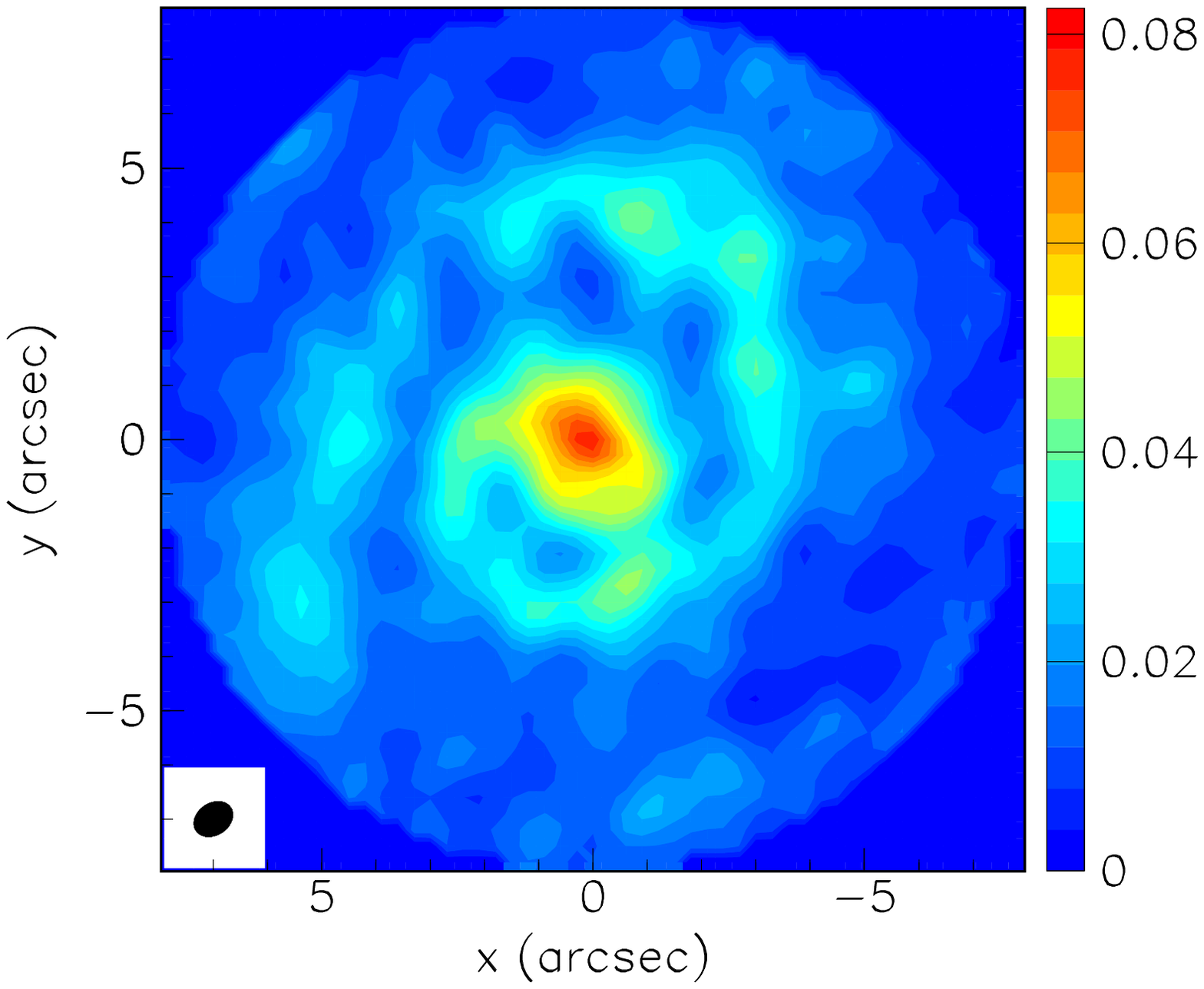}
\includegraphics[height=7cm,trim=0cm 0cm 0cm 0cm,clip]{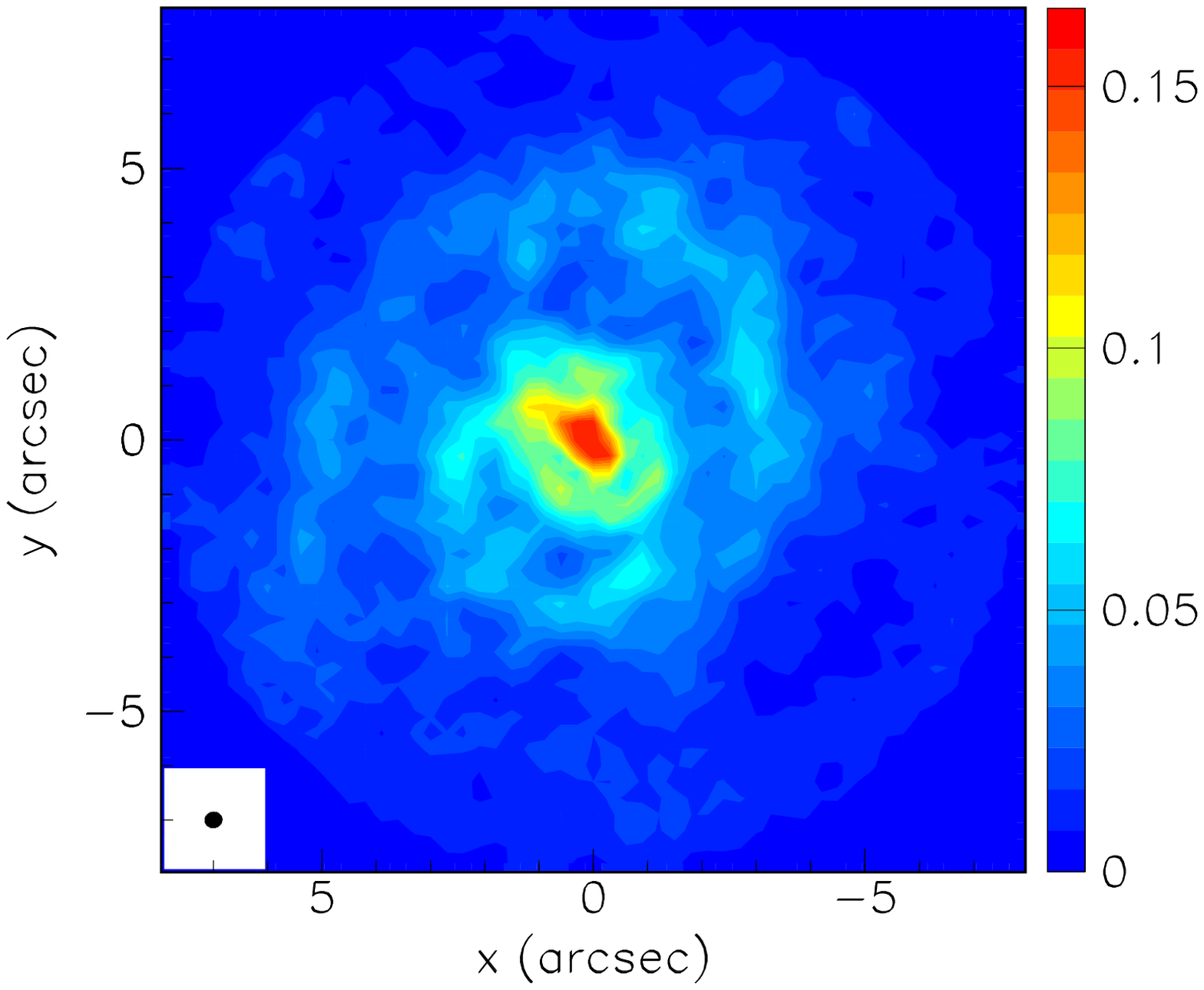}
\caption{Maps of the velocity integrated intensity (Jy\,beam$^{-1}$\,\kms) in the interval $|V_z|<0.3$ \kms. Narrow and broad components of \mbox{CO(1-0)} (left) and \mbox{CO(2-1)} (right) emissions are shown. The beams are shown in the lower left corners.}
\label{fig13}
\end{figure*}

\begin{figure*}
\centering
\includegraphics[height=8cm,trim=0cm 0cm 0cm 0cm,clip]{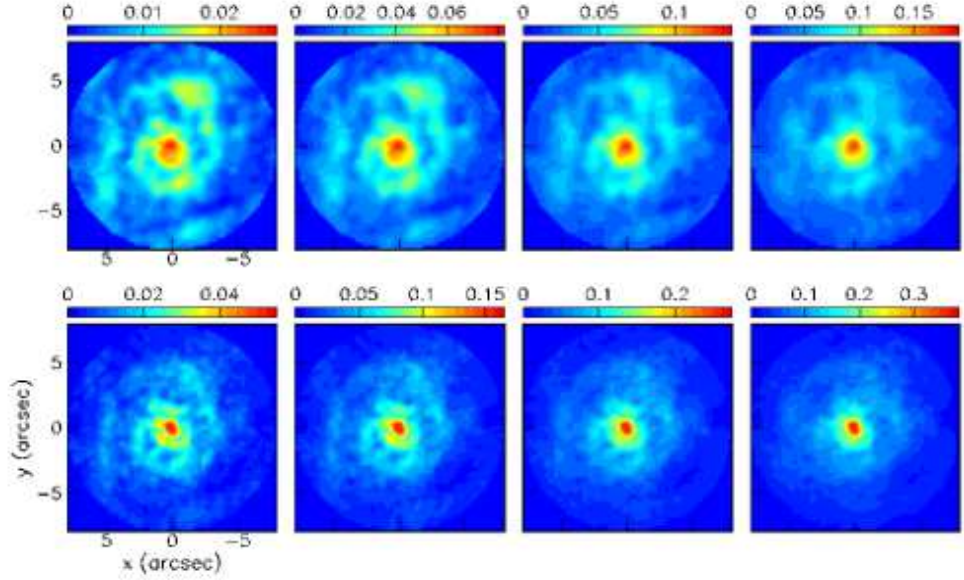}
\caption{Maps of the integrated intensity (Jy\,beam$^{-1}$\,\kms, both components together) corrected for inclination for \mbox{CO(1-0)} (upper panels) and \mbox{CO(2-1)} (lower panels). The Doppler velocity covers intervals $|V_z|<V_{cut}$ with $V_{cut}=0.1$, 0.3, 0.5 and 0.7 \kms\ from left to right.}
\label{fig14}
\end{figure*}

\begin{figure}
\centering
\includegraphics[height=8.5cm,trim=0cm 0cm 0cm 0cm,clip]{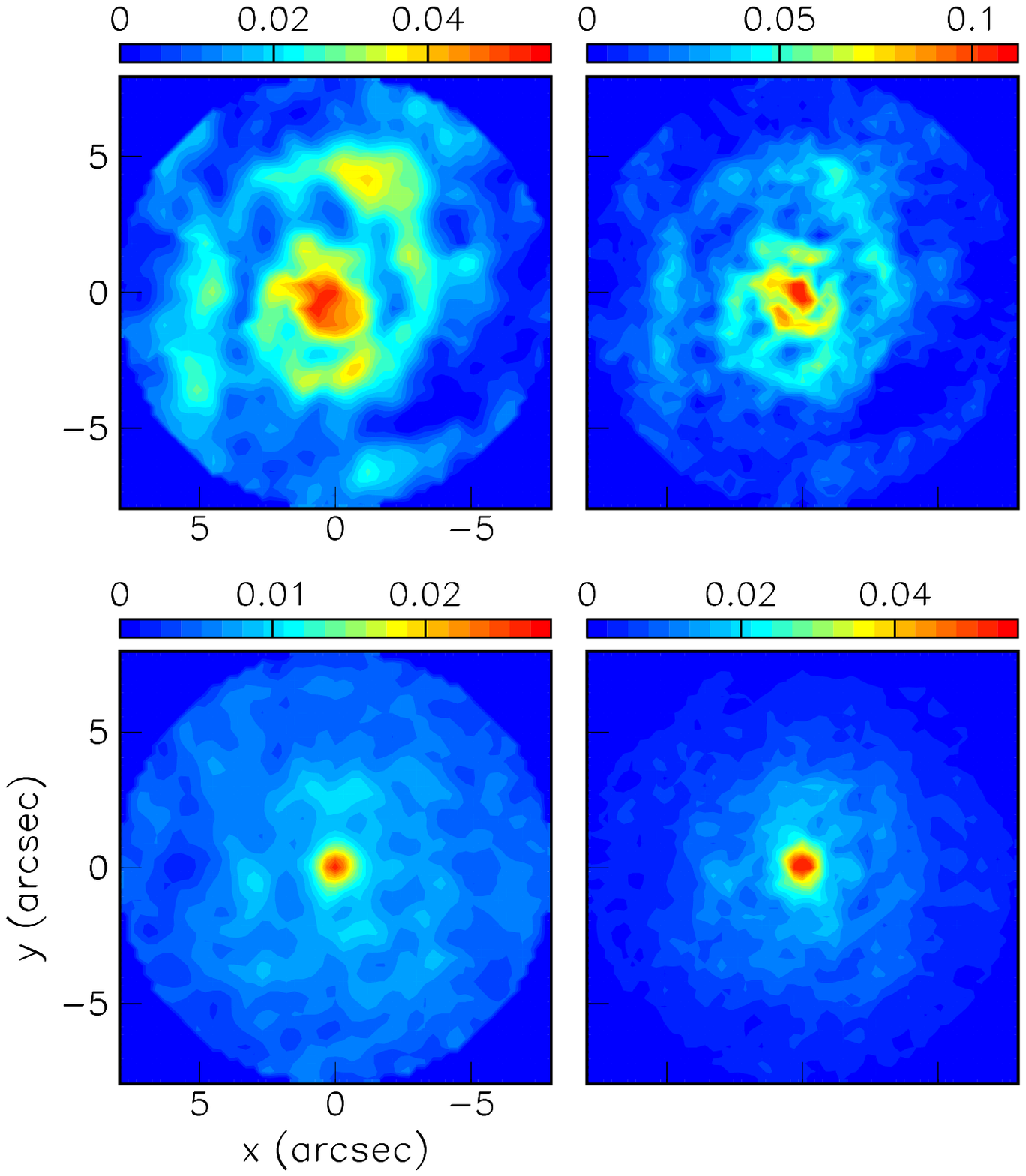}
\caption{Maps of the intensity integrated over $|V_z|<0.3$ \kms\ (Jy\,beam$^{-1}$\,\kms) of \mbox{CO(1-0)} (left) and \mbox{CO(2-1)} (right) for narrow (upper panels) and broad (lower panels) components. The narrow component maps are corrected for inclination.}
\label{fig15}
\end{figure}

As the claim for evidence for a spiral enhancement is the central argument developed by Homan et al., we illustrate this point further in Figure \ref{fig13} that displays sky maps of the intensity integrated over $|V_z|<0.3$ \kms, the central velocity bin used by these authors in their Figure 5. Both \mbox{CO(1-0)} and \mbox{CO(2-1)} maps are shown, the spiral appearance being clear in both cases. The small width of the velocity interval used in Figure \ref{fig13} is equal to the amplitude of the oscillation of the Doppler velocity as a function of position angle $\psi$; as this may be a cause of bias, we repeat the exercise by subtracting from the Doppler velocity a correction term of $0.3\cos\!\psi$ \kms\ before selecting $|V_z|<V_{cut}$, where $V_{cut}$ takes values of 0.1, 0.3, 0.5 and 0.7 \kms. The result is displayed in Figure \ref{fig14}: the effect of the correction is negligible. Finally, Figure \ref{fig15} displays maps of the intensity integrated over the Doppler velocity interval $|V_z|<0.3$ \kms\ for the broad and narrow components separately.

Bringing together the information contained in Figures \ref{fig13} to \ref{fig15}, a spiral-like enhancement is visible on the narrow component for both \mbox{CO(1-0)} and \mbox{CO(2-1)} emissions but the broad component displays no significant feature. Correcting for inclination has no significant effect and the narrow component enhancement is sharper for narrower velocity intervals. Radial enhancements are seen to link the spiral arms at irregular angular intervals of typically 50\dego\ to 60\dego; they may equally well be interpreted as successive enhancements of the intensity along the spiral arms. Different choices of colour scale produce different images revealing enhancements of apparently different shapes. As commented by \citet{Homan2018}, the spiral-like enhancement is far from an ideal Archimedes' spiral, at strong variance with observations made, for example, on carbon star LL Pegasi \citep[]{Morris2006}.

\begin{figure*}
\centering
\includegraphics[height=6.5cm,trim=0cm 0cm 0cm 0cm,clip]{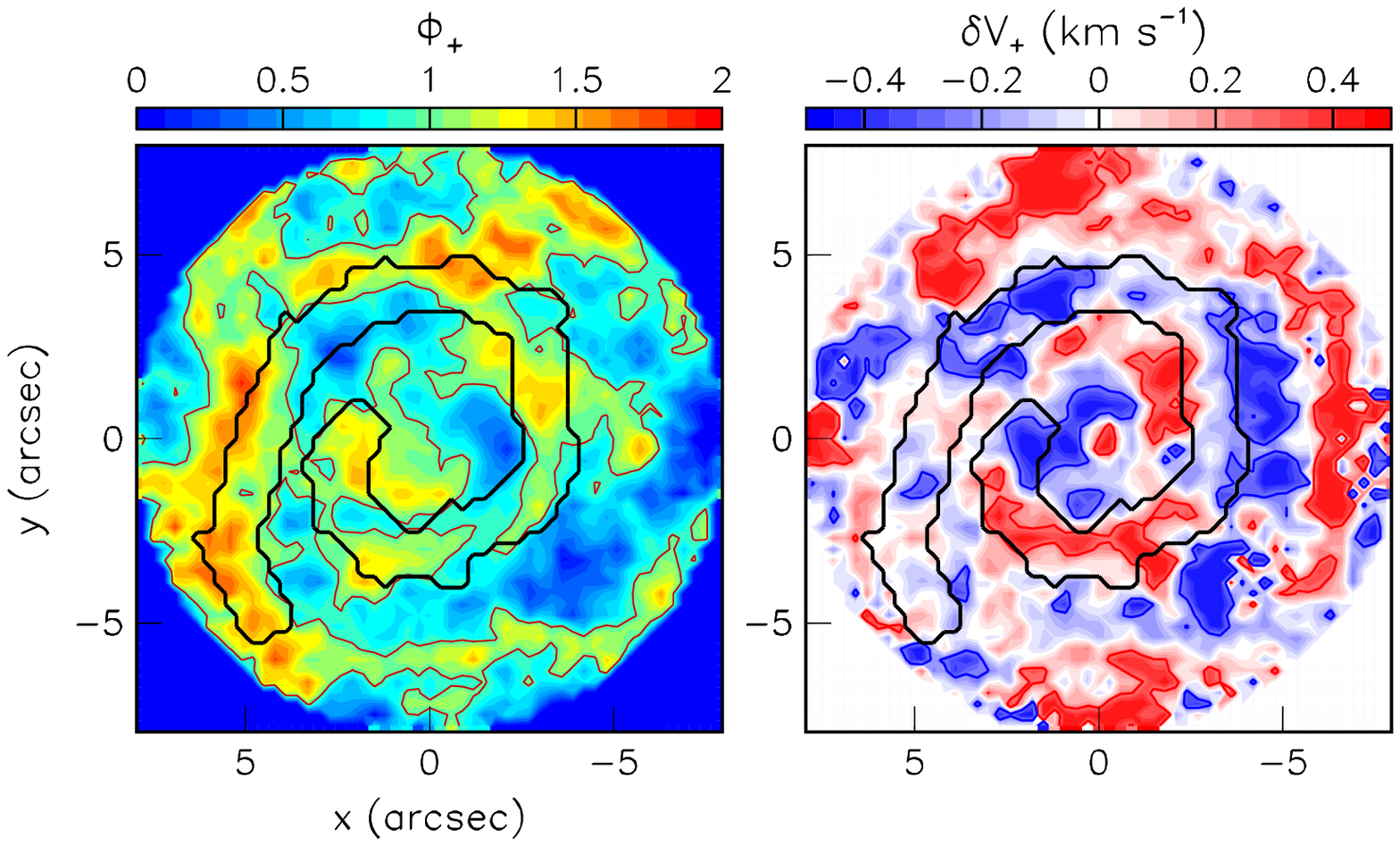}
\caption{Narrow component: sky maps of $\mit{\Phi}_+$ (left) and $\delta{V_+}$ (right). Coloured contours are as in Figure \ref{fig9}. The black contour shows pixels distant from $S$ by less than 0.7 arcsec.}
\label{fig16}
\end{figure*}

Finally, we use the \mbox{CO(1-0)} and \mbox{CO(2-1)} maps of Figure \ref{fig13} to draw a curve, $S$, that runs along maximal intensity, without attempting to find its equation, and compare it with the maps of $\mit{\Phi}_+$ and $\delta{V_+}$ displayed in Figure \ref{fig9}. Pixels having their centre at less than 0.7 arcsec from $S$ are delineated in Figure \ref{fig16}; they have average values of $\mit{\Phi}_+$ and $\delta{V_+}$
of respectively 1.14 and $-$0.02 \kms. As $\mit{\Phi}_+$ is the intensity divided by its average over position angle at the relevant value of $R$, it tends to be enhanced at slightly larger values of $R$ than $S$ is; but, apart from this difference, $S$ matches reasonably well the enhanced arc of $\mit{\Phi}_+$ emission. On the contrary, $S$ spans a broad range of values of $\delta{V_+}$, as could be expected from the circular pattern displayed by this quantity.

\subsection{ Line widths and flaring of the equatorial outflow}\label{sec4.3}

Having evaluated the average dependence of $<\!\!V_z\!\!>$ on $R$ and $\psi$ we correct the Doppler velocity spectra accordingly in each pixel, using a correction term of the form $\Delta{V_z}=a\cos\!\psi+b\sin(2\pi R/3.8)$. We find that $a$ and $b$ can be set to 0.30$\pm$0.05 and $-0.20\pm0.05$ \kms\ for both \mbox{\mbox{CO(1-0)}} and \mbox{CO(2-1)} emissions to obtain minimal line widths, in agreement with the results obtained above. Fitting Gaussians to the line profiles of \mbox{\mbox{CO(1-0)}} and \mbox{CO(2-1)} emissions gives $\sigma$ values of respectively 0.62 and 0.67 \kms\ before correction, decreasing to respectively 0.51 and 0.55 \kms\ after correction: the correction is found to significantly decrease the line width, by $\sim18$\%. If, instead of applying the correction we re-centre each Doppler velocity spectrum on its mean value in each pixel separately, we obtain $\sigma$ values of respectively 0.44 and 0.47 \kms, showing that after correction there is not much room left for an additional systematic contribution to the line width. Indeed, as a function of position angle $\psi$, the line widths fluctuate by less than 0.03 \kms\ with respect to their mean. Table \ref{table6} in the Appendix summarizes the results.

The corrected line widths receive contributions from several factors, instrumental resolution, thermal broadening, turbulence and flaring being expected to be the most important. Instrumental resolution ($\sigma$) contributes $\sim0.1$ \kms\ and thermal broadening a factor $\sqrt{2kT/M_{CO}}$ where $T$ is the temperature, $k$ Boltzman constant and $M_{CO}$ the mass of the CO molecule, meaning 0.16 \kms\ at $T=45$ K. Together, they contribute therefore $\sim0.2$ \kms, leaving some 0.4 \kms\ to be accounted for by turbulence and flaring, the uncertainty on this number not exceeding 0.1 \kms. We retain therefore the value of $0.4\pm0.1$ \kms\ for the ($\psi$-independent) contribution of flaring and turbulence. Defining flaring (FWHM) as $a_{flar}=2.35\times\sigma_{flar}=2.35\times \mbox{Rms}(\sin\!\alpha)$ and recalling that $V\sin\!\varphi\sim0.33$ \kms, the flaring contribution to the line width is, to leading order in $\varphi$, $(0.33/\sin\!\varphi) \,\sigma_{flar}$. Blaming on it the whole $0.4\pm0.1$ \kms\ means therefore $\sigma_{flar}= (1.2\pm0.3)\sin\!\varphi$. For $\varphi$=10\dego, $\sigma_{flar}=(0.20\pm0.05)$ giving an upper limit of $\sim0.6$ on $a_{flar}$ corresponding to a latitude of $\pm17$\dego. Note that, to leading order in $\varphi$, expansion alone contributes to the broadening due to flaring, the rotation velocity being always normal to the star axis. Note also that if effects such as turbulence contribute significantly to the line width, the flaring contribution and the limit obtained on the flaring angle will accordingly be simply smaller.

The line widths measured on and in between the spiral arms, using only the $\cos\!\psi$ term as correction and requiring $R$ to exceed 1.5 arcsec, differ by $\sim0.06$ \kms\ as do the line widths measured on and in between the concentric circles (see Table \ref{table6}): large $|\delta{V_+}|$ values (corresponding to $|R-R_{circle}|>0.95$), or large $\mit{\Phi}_+$ values (corresponding to $|R-R_{sp}|<h/4$) tend to produce narrower line profiles. This barely significant difference, if blamed on turbulence, would mean a contribution of $\sim0.24$ \kms. Line widths ($\sigma$) measured at less than 0.7 arcsec distance from the spiral-like curve $S$ are 0.40 \kms\ and 0.42 \kms\ for \mbox{CO(1-0)} and \mbox{CO(2-1)} emissions respectively, namely $\sim0.05$ \kms\ lower than average, again a barely significant difference.

\subsection{Summary}\label{sec4.4}

In summary, the picture of the narrow component that emerges from this analysis is of an equatorial outflow having a small inclination $\varphi$ with respect to the sky plane. On average, it expands radially with a uniform velocity of $\sim(0.33\pm0.03)/\sin\!\varphi$ \kms, namely 1.9$\pm$0.2 \kms\ for $\varphi$=10\dego. An upper limit of 0.38 has been obtained for the ratio of a possible rotation velocity to the radial expansion velocity. Lumps of small velocities normal to the sky plane have been detected in excess to the contribution of the expansion velocity; they are at the scale of 0.24 \kms\ (1$\sigma$) and display a clear modulation with $R$ of amplitude $\pm$0.16 \kms\ and period 3.8 arcsec. The intensity displays fluctuations at the level of $\pm$36\%, however not clearly correlated with the velocity fluctuations, the latter following an approximate concentric circular pattern and the former being dominated by an arc of enhanced intensity in the north-eastern half of the sky and a deep depression in the south-western quadrant, in qualitative agreement with the spiral-like feature described by \citet{Homan2018}. A small but significant anti-correlation is detected between the amplitude of the intensity fluctuations and the amplitude of the small velocity components detected in excess to the contribution of the expansion velocity. Line widths of $\sim$1.2 \kms\ (FWHM) have been measured, after subtraction of the effect of the inclination of the star equator with respect to the plane of the sky, consistent with a very small contribution of other effects such as flaring, instrumental resolution and thermal broadening. The dependence on $|\cos\!\psi|$ of the flaring contribution to the line width has provided an estimate of an upper limit on the flaring angle, $\pm$17\dego\ at half maximum for $\varphi$=10\dego. The study has failed to reveal new features that could strengthen the evidence for the significance of the spiral observed in the low $V_z$ channel maps. On the contrary it has revealed as significant a circular feature in the maps of the mean Doppler velocity, apparently uncorrelated with the spiral. As both are near the limit of sensivity of the observations, caution must be exercised when proposing an interpretation.

\section{The broad component}\label{sec5}

Contrary to the narrow component, the broad component, which is seen pole-on, needs to be de-projected in space to reveal the details of its morphology and kinematics. To do so, we normally use a radial expansion velocity of the form $V=V_0(1-\lambda\cos\!2\alpha)$, where $\alpha$ is the stellar latitude, to de-project the effective emissivity following the method developed in \citet{Nhung2018a,Nhung2018b}. We use a polar expansion velocity $V_{pole}=V_0(1+\lambda)=10.3$ (10.7) \kms\ for the blue(red)-shifted part of the spectrum (see Section \ref{sec5.3} below) and a prolateness parameter $\lambda=0.7$, meaning an equatorial velocity $V_{eq}=V_0(1-\lambda)=1.9$ \kms. The inclination of the star axis with respect to the line of sight is fixed as $\varphi$=10\dego.

The low value adopted for the equatorial radial velocity deserves some comments. It is obtained from the arguments developed in Section \ref{sec3.3} that measure the radial wind velocity at equator as (0.33$\pm$0.03)/$\sin\varphi$ \kms, and from the choice of 10\dego\ for the value of $\varphi$. The escape velocity from a solar mass star is 2.7 \kms\ at $r=250$ au (where broad and narrow components can be described separately, see below) and reaches 1.9 \kms\ at $r=490$ au, meaning 5.6 arcsec. To compensate gravity at $r=250$ au requires a rotation velocity 2.6 times larger than the maximum acceptable according to the arguments developed in Section \ref{sec3.4}. It is therefore reasonable to conclude that the equatorial wind velocity obtained in Section \ref{sec3} is below escape velocity over an important fraction of the radial range explored here. In the region where the wind is being accelerated, the competition between the mechanism of acceleration and the gravity of the star governs the gas dynamics. It is again reasonable to expect that the terminal velocity has been essentially reached at $r\sim250$ au. A realistic model of the wind velocity would need to take all these arguments in due account, in particular the effect of gravity causing a decrease of the radial velocity at the equator. However, a reliable treatment requires an understanding of the dynamics governing small distances to the star, where other tracers than CO molecules are necessary, which we leave for a later publication including analyses of SiO and SO$_2$ emissions (\citet{TuanAnh}, in preparation). In the present work, the form of the radial wind velocity used for de-projection is meant to simply help with the evaluation of the effective emissivity and ultimately the flux of matter, which it does well (see Section \ref{sec6.3}).

\subsection{Broad-narrow separation}\label{sec5.1}

\begin{figure}
\centering
\includegraphics[height=5.2cm,trim=.5cm 0.cm 0cm 1.cm,clip]{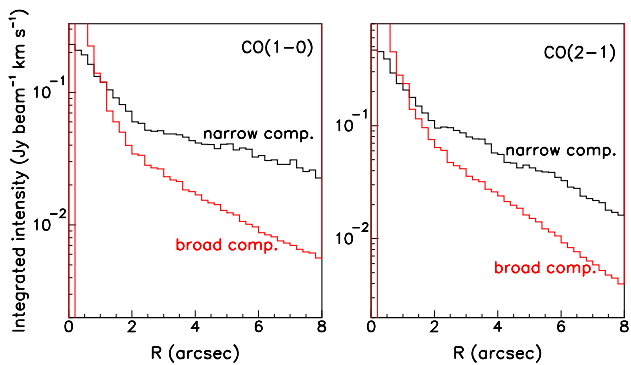}
\caption{Comparison between the $R$ dependence of the intensity of the narrow (black) and broad (red) components, the latter being integrated over a thickness of $\pm$0.5 arcsec with respect to the equator of the de-projected effective emissivity. The left panel is for \mbox{CO(1-0)} and the right panel for \mbox{CO(2-1)}. }
\label{fig17}
\end{figure}

We remark that our definition of the broad component includes part of the equatorial outflow, obtained by interpolation of the blue- and red-shifted wings of the Doppler velocity spectrum below the narrow central peak. As a result, there exists a region in space where both components merge and where the distinction between them becomes meaningless. This is illustrated in Figure \ref{fig17} that displays the dependence on $R$ of the de-projected emissivity of the broad component near the equator, averaged over $z'=\pm0.5$ arcsec (measured along the star axis), and compares it with the intensity $F_{narrow}$ measured for the narrow component. It shows dominance of the narrow component for values of $R$ exceeding typically 1 arcsec. We recall that for $\varphi$=10\dego\ $\sigma_{flar}\sim0.2$, implying that such a $z'$ interval (FWHM) is covered at a distance $R\sim2.1$ arcsec. In practice, we shall retain 2 arcsec as the distance from the star where the two components merge and where the separation between them becomes meaningless.

\subsection{General properties}\label{sec5.2}

\begin{figure*}
\centering
\includegraphics[width=0.9\textwidth,trim=.5cm 0.cm 1.5cm 1.cm,clip]{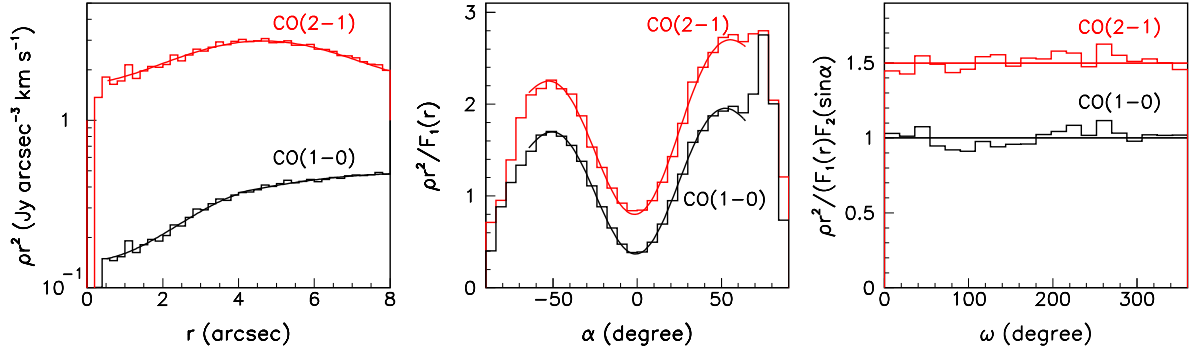}
\caption{ Broad component. Left: dependence of $\rho r^2$ on $r$; the curves show the $F_1(r)$ fits (see Table \ref{table7}) . Middle: dependence of $\rho r^2/F_1(r)$ on $\alpha$; the curves show the $F_2(\sin\!\alpha)$ fits (see Table \ref{table7}). Right: dependence of $\rho^*=\rho r^2/[F_1(r)F_2(\sin\!\alpha)]$ on $\omega$. In the rightmost panels, the \mbox{CO(2-1)} distribution has been shifted up by half a unit for convenience.}
\label{fig18}
\end{figure*}

The de-projected emissivity, $\rho$, multiplied by $r^2$, is displayed in Figure \ref{fig18} (left) for \mbox{CO(1-0)} and \mbox{CO(2-1)} emissions separately as a function of the distance to the star, $r$, together with polynomial fits, $F_1(r)$ (Table \ref{table7} in Appendix). The functions $\rho r^2/F_1(r)$ are displayed in Figure \ref{fig18} (middle) as a function of stellar latitude $\alpha$, together with polynomial fits $F_2(\sin\!\alpha)$ (Table \ref{table7}). They increase from the equator up to latitudes of the order of $\pm$45\dego\ where they reach a maximum before decreasing smoothly toward the poles. The reliability of this result is critically discussed in Section \ref{sec5.3} below.   
  
The normalised effective emissivity, $\rho^*(x,y)= \rho(x,y)r^2/[F_1(r)F_2(\sin\!\alpha)]$ is displayed in Figure \ref{fig18} (right) as a function of stellar longitude, $\omega$. It is nearly constant for both \mbox{CO(1-0)} and \mbox{CO(2-1)} emissions, providing evidence for axi-symmetry when averaged over other variables. The region $\omega>180^\circ$ is in very slight excess with respect to the region $\omega<180^\circ$. In each case, $\rho^*$ is averaged over the other two variables with the conditions $0.5<r<8$ arcsec and $|\sin\!\alpha|<0.9$.

\subsection{Dependence on latitude and polar depression}\label{sec5.3}

It is important to assess the reliability of the evidence for a polar depression of the effective emissivity presented in the preceding section. As was remarked by \citet{Nhung2018a} the shape of the Doppler velocity spectrum near its end points is closely related to the latitudinal dependence of the effective emissivity near the stellar poles. In particular, a locally isotropic and $r$-independent radial wind associated with a spherical emissivity proportional to $1/r^2$ produces a Doppler velocity spectrum proportional to $[1-(V_z/V)^2]^{-1/2}$ when integrated over a circle on the sky plane. This means sharp horns at the ends of the spectrum, which are absent from the present observations: the \mbox{CO(2-1)} spectrum displays no horn and the \mbox{CO(1-0)} spectrum displays horns much less sharp than produced by the above law. The small inclination of the star axis with respect to the line of sight can be ignored in this context. A natural interpretation is a depression of the effective emissivity near the stellar poles, which may be associated with a decrease of density and/or an effect of temperature and/or an effect of absorption. However, one might argue that this result is an artefact of the form of wind velocity used for de-projection: we address this issue below.

\begin{figure*}
\centering
\includegraphics[width=0.8\textwidth,trim=.5cm 0.cm 1.5cm 1.cm,clip]{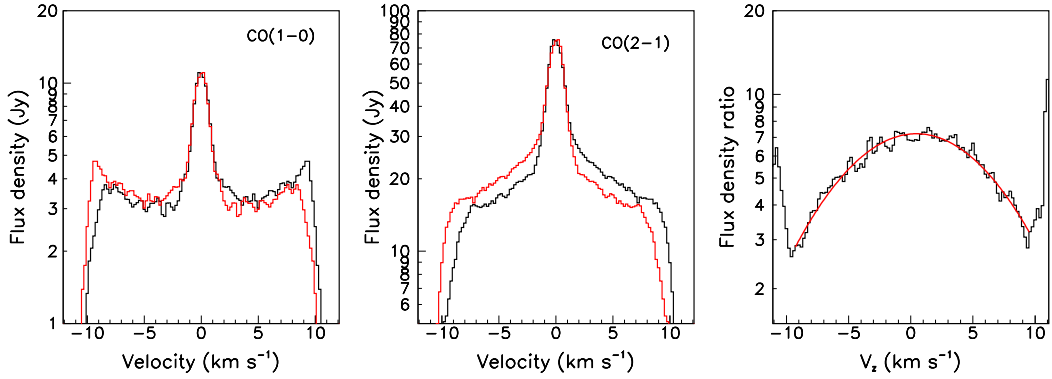}
\caption{Left and middle: comparison between the Doppler velocity spectra (black) and their symmetric with respect to the origin (red, illustrating the blue-red asymmetry) for \mbox{CO(1-0)} (left) and \mbox{CO(2-1)} (middle) emissions. Right: ratio between the \mbox{CO(2-1)} and \mbox{CO(1-0)} Doppler velocity spectra; the curve is a Gaussian fit with mean 0.5 \kms\ and $\sigma$ 7.3 \kms. Narrow and broad components are shown together.}
\label{fig19}
\end{figure*}

In addition to a marked difference between the \mbox{CO(1-0)} and \mbox{CO(2-1)} Doppler velocity distributions (Figure \ref{fig2} right), a notable feature of the broad component is the asymmetry between the red-shifted and blue-shifted brightness, which had been previously noted by \citet{Nhung2015a}. It is illustrated in Figure \ref{fig19} for \mbox{CO(1-0)} and \mbox{CO(2-1)} emissions separately, together with the ratio of the respective spectra. The latter displays a remarkably smooth shape if one excludes end effects where noise is important. A Gaussian fit in the interval $-9<V_z<9$ \kms\ gives a mean at 0.5 \kms\ and a $\sigma$ of 7.3 \kms. We evaluate the position of the blue end-point at $\sim-10.2$ \kms\ and that of the red end-point at $\sim10.6$ \kms. The precise location of the end points of the Doppler velocity spectrum fixes the polar wind velocity just above their values, but not higher. Using too large a polar velocity, larger than the end point velocity by a fraction $\varepsilon$, causes the de-projected effective emissivity to cancel at latitudes exceeding $\sin^{-1}(1-\varepsilon)\sim\pi/2-\sqrt{2\varepsilon}$: $\epsilon=5$\% means a cut-off latitude of 72\dego. 

\begin{figure*}
\centering
\includegraphics[height=5.cm,trim=1cm 0.cm 0cm 1.cm,clip]{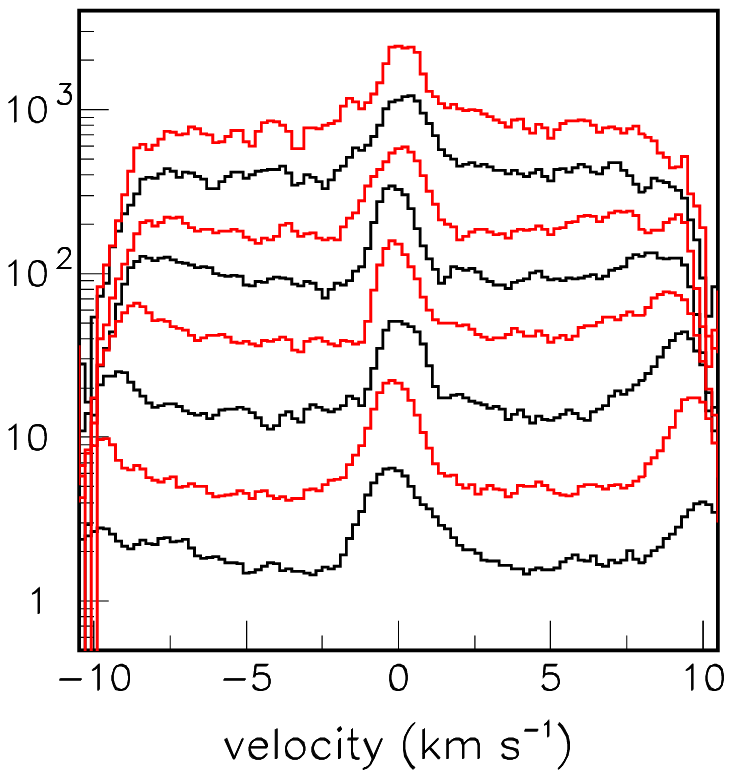}
\includegraphics[height=5.cm,trim=1cm 0.cm 0cm 1.cm,clip]{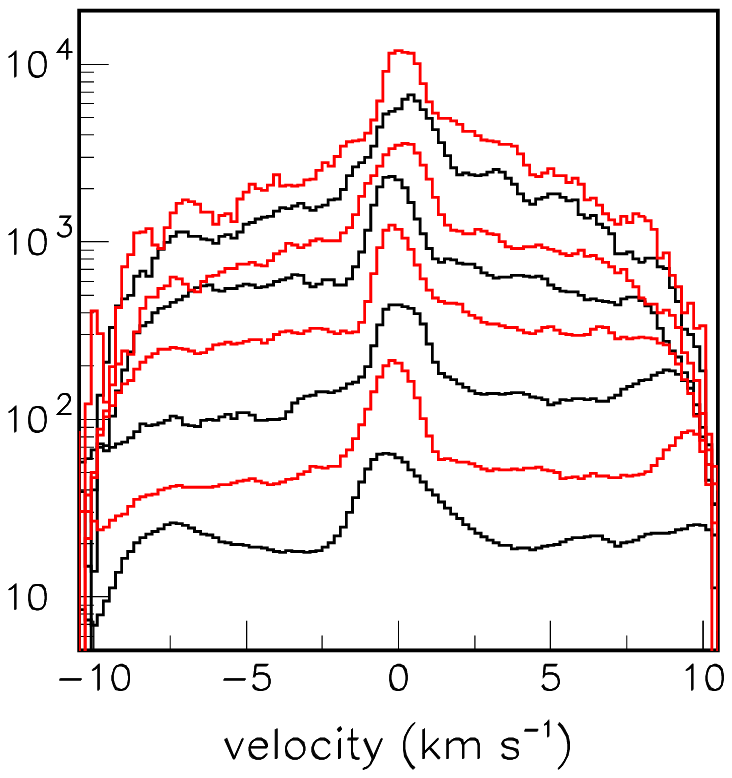}
\includegraphics[height=5.cm,trim=0.5cm 0.cm 0cm 1.cm,clip]{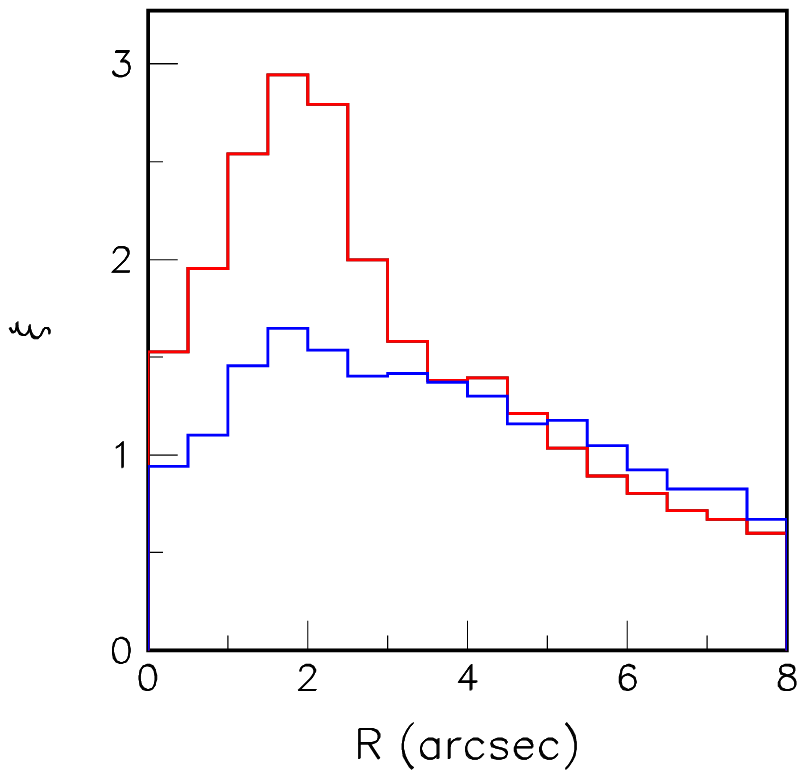}
\caption{Left and middle: Doppler velocity distributions in successive intervals of $R$, each 1 arcsec wide, from $0<R<1$ arcsec (lower curves) to $7<R<8$ arcsec (upper curves). The distributions have been arbitrarily shifted upward for clarity. Narrow and broad components are shown together. \mbox{CO(1-0)} and \mbox{CO(2-1)} spectra are displayed in the left and middle panels respectively. Right: dependence on $R$ of the parameter $\xi$ defined in the text for \mbox{CO(1-0)} emission, measuring how sharp and enhanced are the horns at the ends of the velocity spectrum. Blue is for the blue-shifted horn, red for the red-shifted horn.}
\label{fig20}
\end{figure*}

Another notable property of the Doppler velocity spectra concerns their evolution when moving away from the star. While being approximately independent of position angle, they narrow down when $R$ increases, as illustrated in Figure \ref{fig20}. This is not surprising, larger values of $R$ probing regions of the outflow farther away from its axis. More surprising, however, is the appearance or enhancement of horns at the ends of the spectra in the $R$ intervals between 1 and 3 arcsec. A spherical emissivity distribution would produce instead horns that decrease in amplitude and sharpness when $R$ increases: this, again, suggests that the emissivity is depressed in the polar regions. To further illustrate the evolution of the spectrum horns with $R$ we define a parameter $\xi$, meant to describe how sharp and enhanced they are, as the ratio between the content of the spectrum in the 2 \kms\ wide interval below the end point and in the interval covering between 5 and 3 \kms\ below the end point. The dependence of $\xi$ on $R$ is displayed in the right panel of Figure \ref{fig20} for \mbox{CO(1-0)} emission. The evidence for peaking just below $R\sim2$ arcsec is very strong. The effect is also present but weaker for the \mbox{CO(2-1)} spectrum. The ratio between the \mbox{CO(2-1)} and \mbox{CO(1-0)} spectra remains well behaved in separate $R$ intervals and evolves from a nearly flat distribution near the star to a distribution reaching a maximum at low $|V_z|$ for large values of $R$.

\begin{figure*}
\centering
\includegraphics[height=5.5cm,trim=.5cm 0.cm 1.5cm 1.cm,clip]{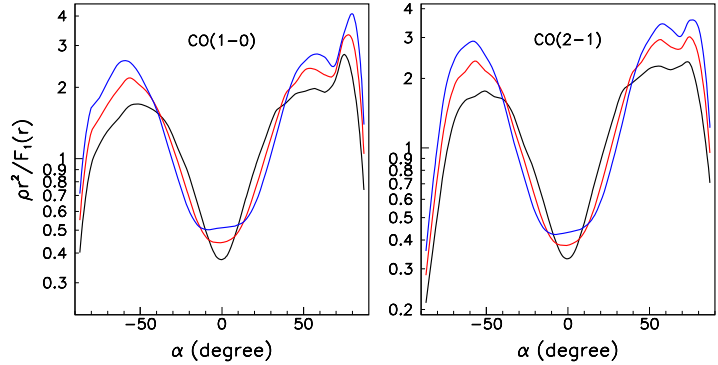}
\caption{ Dependence on stellar latitude $\alpha$ of the de-projected emissivity for \mbox{CO(1-0)} (left) and \mbox{CO(2-1)} (right) using winds of the form $V_{eq}+(V_{pole}-V_{eq})|\sin\!\alpha|^n$ with $n=2$ (black), 3 (red) and 4 (blue) respectively.}
\label{fig21}
\end{figure*}

As a further illustration of the evidence for polar depression of the emissivity, we show in Figure \ref{fig21} the dependence on stellar latitude $\alpha$ of the broad component de-projected emissivity for wind velocity distributions of different width around the poles. The more collimated is the wind velocity, the closer to the poles is the maximum of the emissivity, but the polar depression is always present.

\subsection{Inhomogeneities}\label{sec5.4}

\begin{figure*}
\centering
\includegraphics[height=5.5cm,trim=.5cm 0.cm 1.5cm 1.cm,clip]{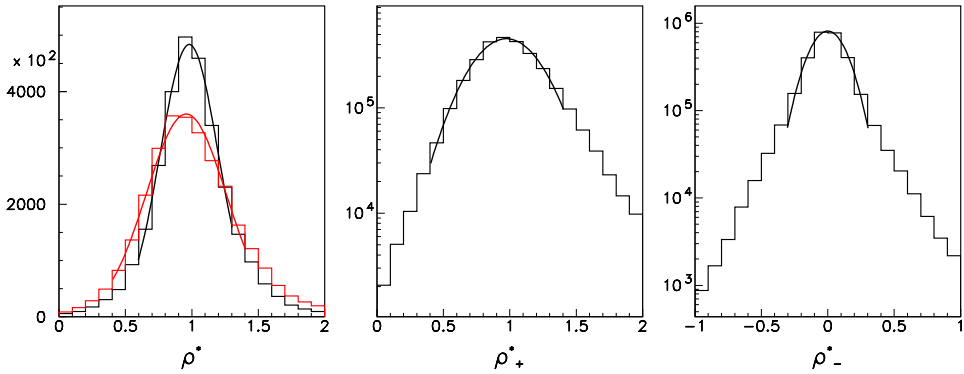}
\caption{Left: distribution of $\rho^*=\rho\,r^2/[F_1(r)F_2(\sin\alpha)]$; black is for \mbox{CO(1-0)} and red for \mbox{CO(2-1)}.  Middle: distribution of $\rho^*_+$; the curve is a Gaussian fit with (mean, $\sigma$)=(0.97, 0.24). Right: distribution of $\rho^*_-$; the curve is a Gaussian fit with (mean, $\sigma$)=(0, 0.14).}
\label{fig22}
\end{figure*}

The distributions of $\rho^*$ are shown in the left panel of Figure \ref{fig22}, together with Gaussian fits having (mean, rms) values of (0.98, 0.21) for \mbox{CO(1-0)} and (0.95, 0.30) for \mbox{\mbox{CO(2-1)}}. Averaging between \mbox{CO(1-0)} and \mbox{CO(2-1)}, the lumpiness is even smaller than for the narrow component: $\sigma=0.26$ instead of 0.36. In order to quantify this statement, we also show in Figure \ref{fig22} the distributions of the quantities $\rho^*_+=\frac{1}{2}(\rho^*_{\rm{CO(2-1)}}+\rho^*_{\rm{CO(1-0)}})$ and $\rho^*_-=\frac{1}{2}(\rho^*_{\rm{CO(2-1)}}-\rho^*_{\rm{CO(1-0)}})$. Gaussian fits give (mean, $\sigma$)=(0.97, 0.24) for $\rho^*_+$ and (0, 0.14) for $\rho^*_-$: the dispersion of $\rho^*_+$ owes as much to the difference between \mbox{CO(1-0)} and CO(2-1) data as to its intrinsic dispersion, $\sim0.19=\sqrt{0.24^2-0.14^2}$. At variance with the narrow component, the broad component displays more global uniformity and less similarity between \mbox{CO(1-0)} and \mbox{CO(2-1)} data, preventing a reliable identification of significant inhomogeneity of the $\rho^*_+$ distribution.

\subsection{Summary}\label{sec5.5}

In summary, the broad component has been shown to be physically distinct from the narrow component at distances from the star in excess of 2 arcsec or so. At distances smaller than 1 arcsec, both components are undistinguishable and must be studied jointly. At variance with the narrow component, the broad component displays significant differences between \mbox{CO(1-0)} and \mbox{CO(2-1)} emissions, and smaller intensity inhomogeneity. Moreover, it shows a small but significant asymmetry between its blue-shifted and red-shifted parts. The effective emissivity is axi-symmetric. The wind velocity, assumed to be small near the equator in order to join smoothly with the equatorial outflow, must increase toward the poles in order to produce Doppler velocities reaching $\sim11$ \kms. The absence of strong horns at the extremities of the measured Doppler velocity spectra generally implies significant depressions of the effective emissivity near the stellar poles, causing it to reach maximum at intermediate latitudes of the order of $\sim45$\dego\ to 60\dego. 

\section{Joint description}\label{sec6}

\subsection{Aim and method}\label{sec6.1}

The aim of the present section is to draw a unified picture of the morphology and kinematics of the circumstellar envelope, as realistic and reliable as possible. The preceding sections have provided detailed descriptions of its main components, a narrow component associated with an equatorial outflow in slow expansion and a broad component associated with a bipolar, biconical outflow in more rapid expansion. However, the effect of temperature and absorption has been ignored; accounting for it requires a radiative transfer calculation and a joint description of the morpho-kinematics of the central region, $r<\sim2$ arcsec, where the narrow and broad components merge.

The two-component description has been very effective in clarifying the properties of the morpho-kinematics of the circumstellar envelope. At distances from the star in excess of $\sim2$ arcsec, it has been shown to correspond to two different regions of space, each having its own identity. However, this does not imply that both components are separated by a gap; on the contrary, we shall see that they join smoothly. Moreover, it is important to be conscious of the arbitrariness inherent to the de-projection of radio observations, which has been discussed in detail by \citet{Diep2016} and \citet{Nhung2018b}. In particular, once a specific form has been assumed for the field of wind velocities, effective emissivities that reproduce exactly the observed data cubes can be easily de-projected: what matters then is to identify the main features that can be reliably ascertained and to discuss critically their reality.

In the remainder of the present section, we first use the two-component description to discuss the effects of temperature, evaluate its distribution in space and calculate the flux of matter and the mass-loss rate after de-projection of the broad component; we then use a joint description to evaluate the effect of absorption, and, taking it in due account, we calculate the distribution of temperature, density and flux of matter in stellar coordinates; finally, we discuss the results obtained. 

\subsection{Temperature}\label{sec6.2}
        
\begin{figure*}
\centering
\includegraphics[height=5.cm,trim=1cm 1.cm 1.5cm 1.cm,clip]{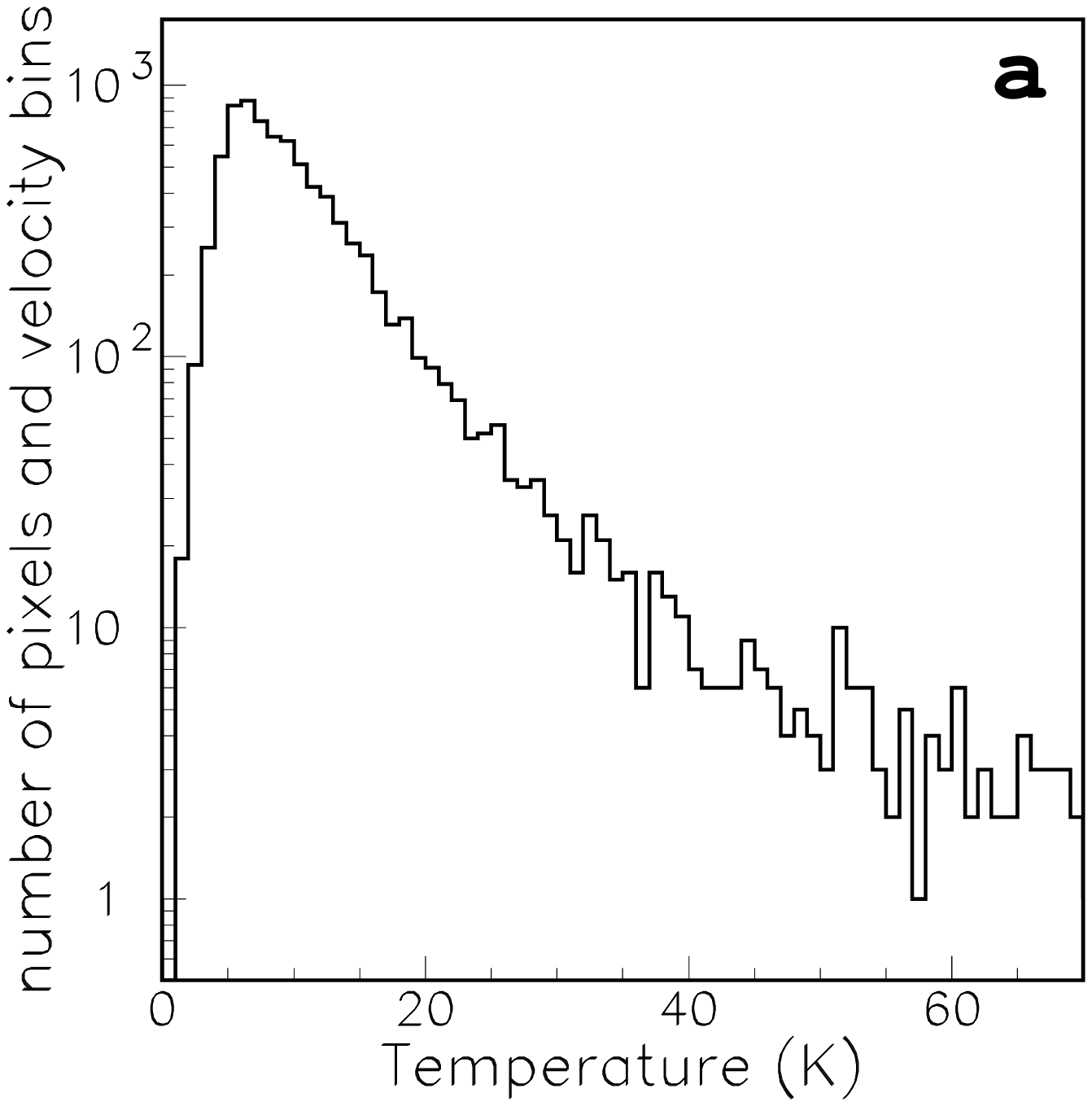}
\includegraphics[height=5.cm,trim=1cm 1.cm 1.5cm 1.cm,clip]{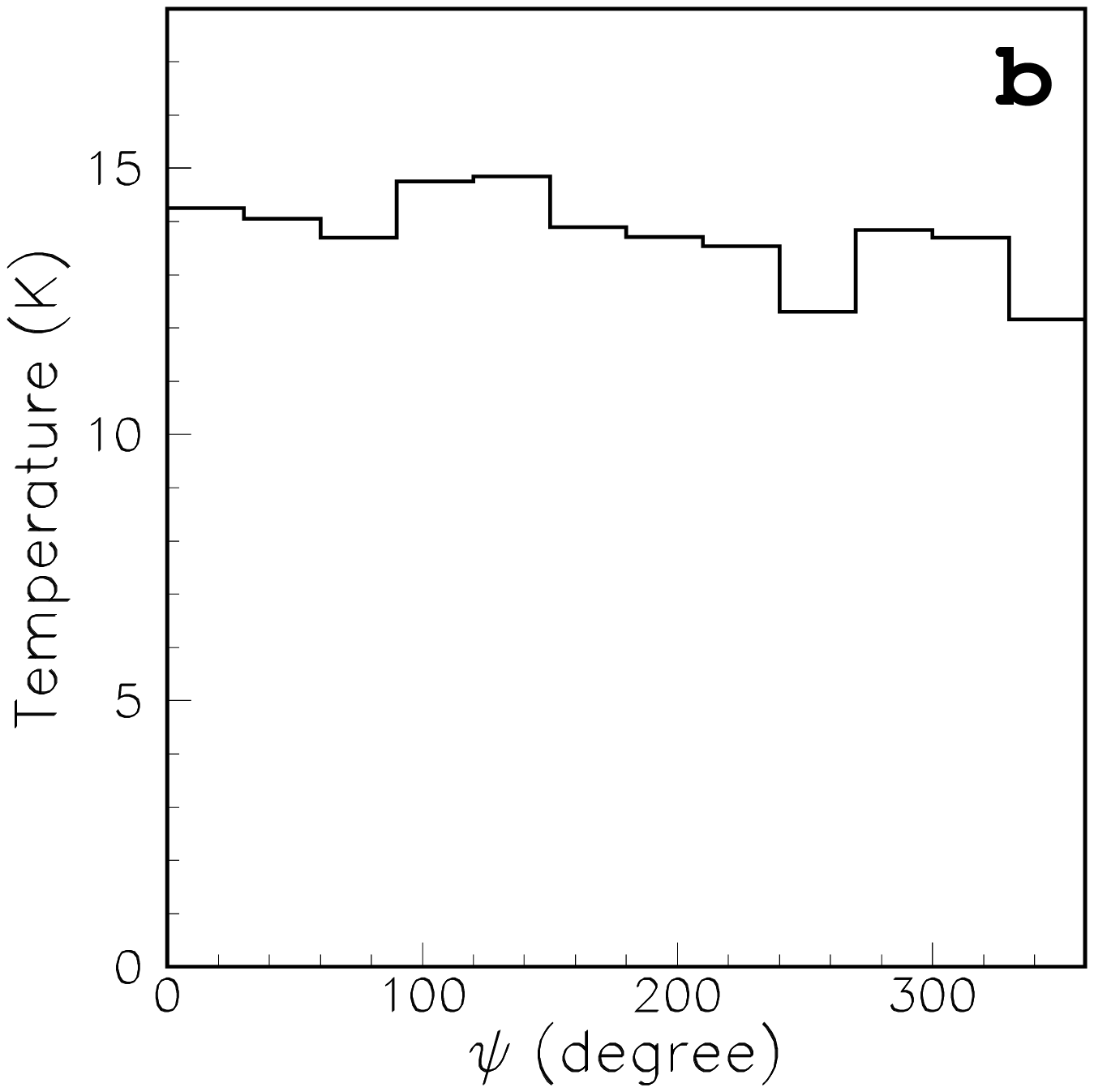}
\includegraphics[height=5.cm,trim=1cm 1.cm 1.5cm 1.cm,clip]{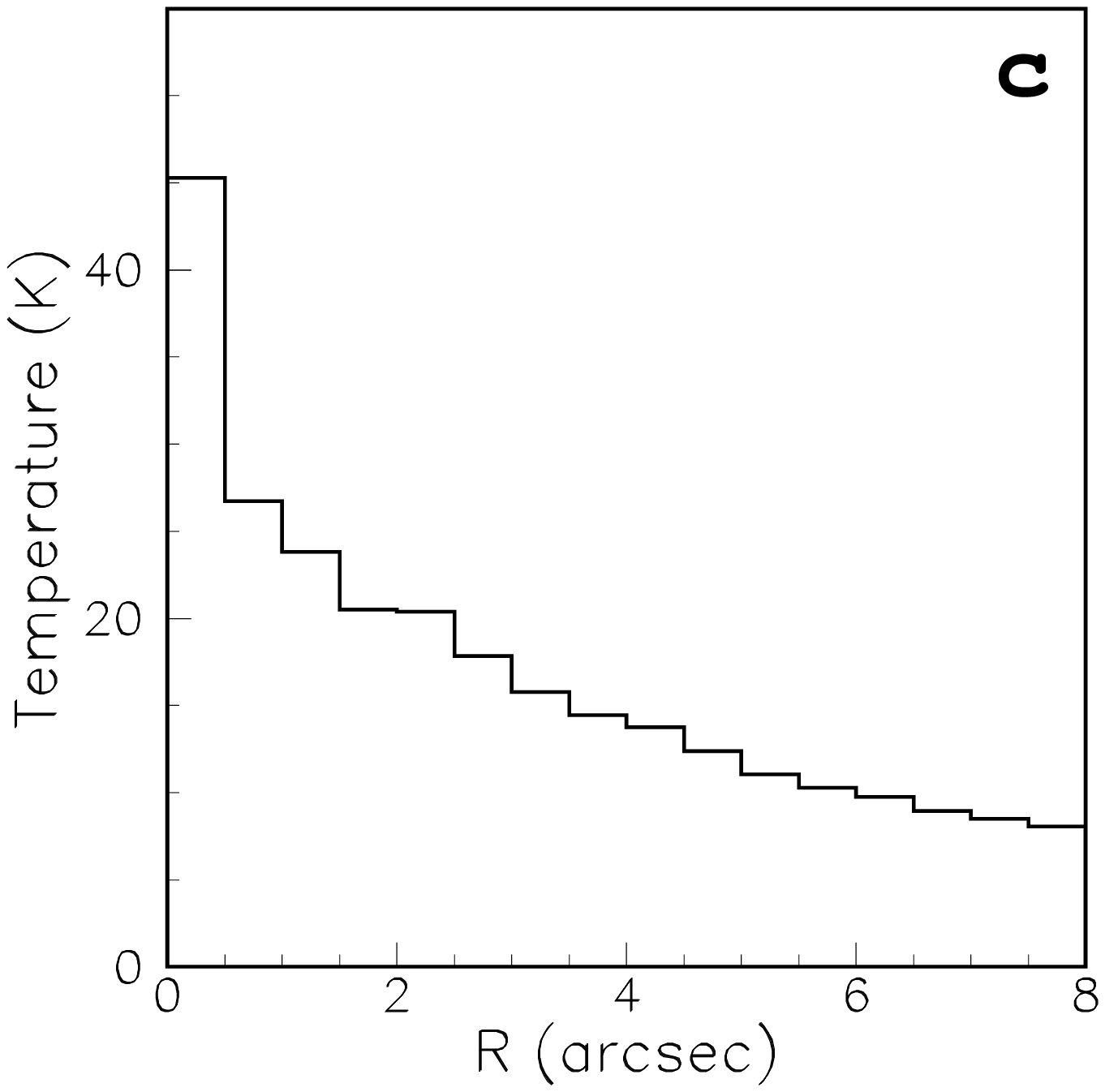}\\
\includegraphics[height=5.cm,trim=1cm 1.cm 0cm 1.cm,clip]{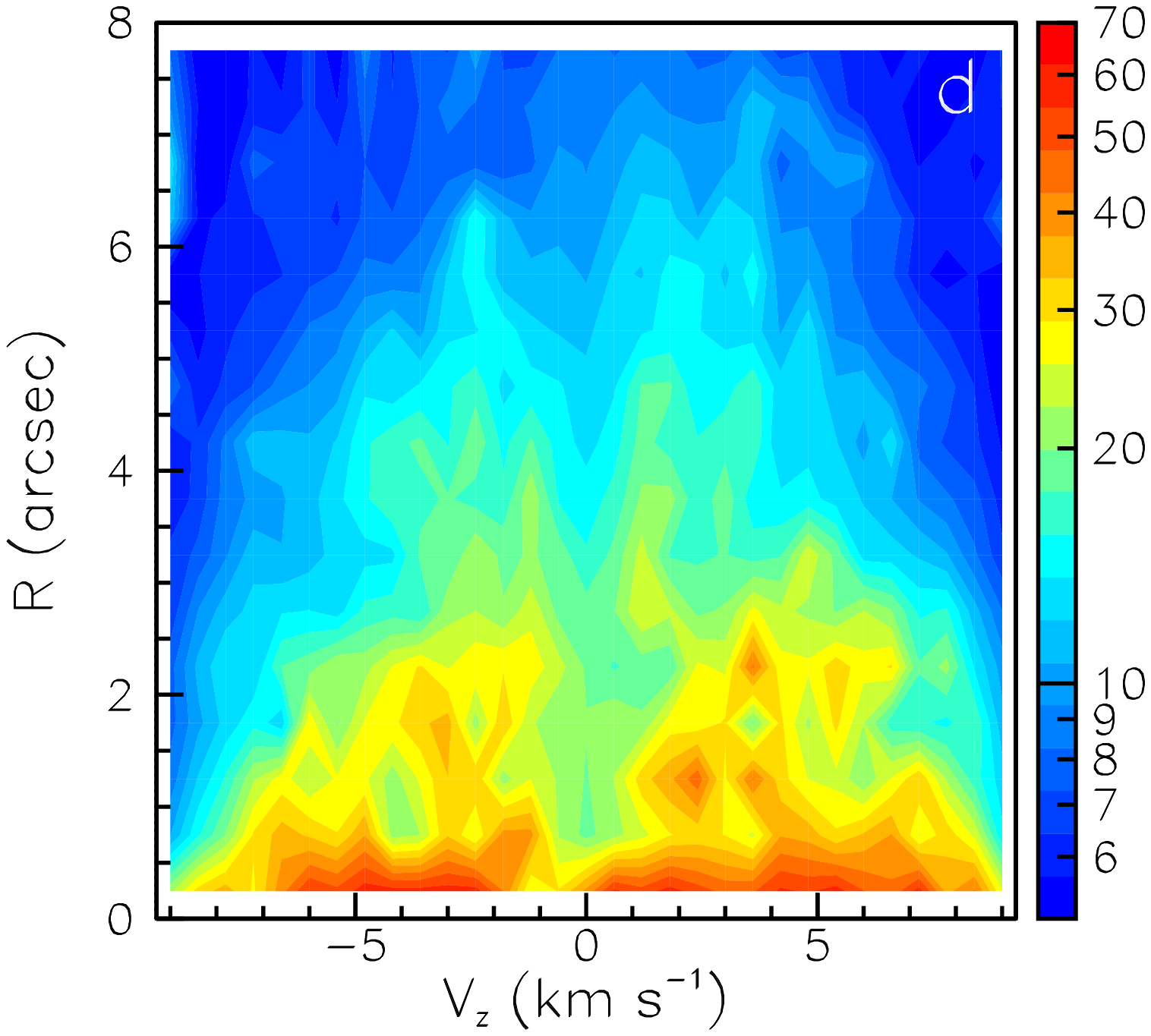}
\includegraphics[height=5.cm,trim=.5cm 1.cm 1.5cm 1.cm,clip]{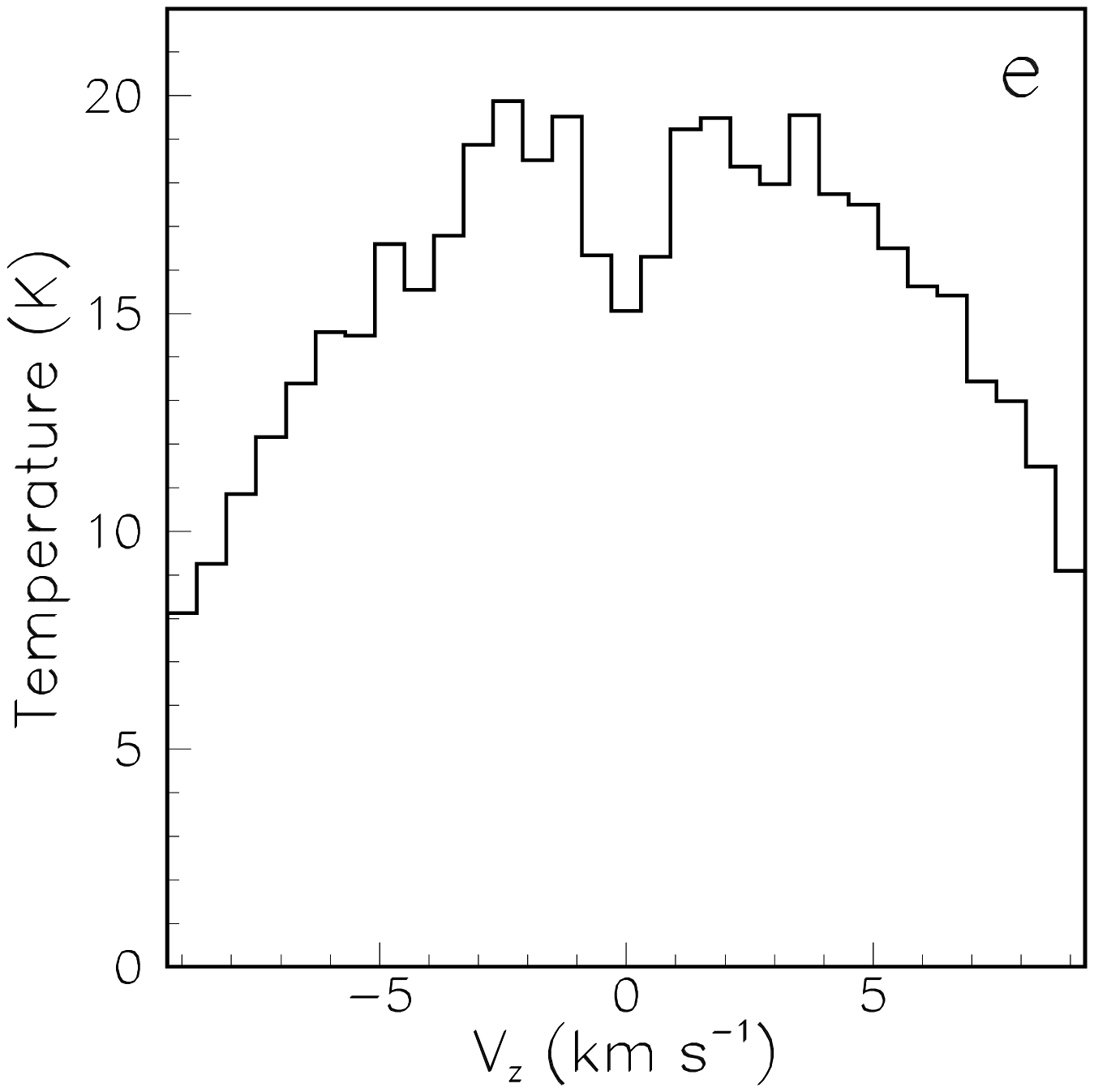}
\caption{a) temperature distribution over the whole data cube; b) dependence of the temperature, averaged over $R$ and $V_z$, on position angle $\psi$ ($1<R<8$ arcsec); c) dependence of the temperature, averaged over $\psi$ and $V_z$, on $R$; d) dependence of the temperature (K), averaged over $\psi$, on $R$ and $V_z$; e) dependence of the temperature, averaged over $\psi$ and $R$, on $V_z$.}
\label{fig23}
\end{figure*}

Ignoring absorption, the \mbox{CO(2-1)} and \mbox{CO(1-0)} brightnesses are expected to be in a ratio $Q_{2-1}/Q_{1-0}=16 \exp(-11.1/T [K])$ where 16 is the ratio between the respective $Q_0(2J+1)$ factors and 11.1 K the difference between the respective $E_J$ temperatures, as discussed in Section \ref{sec3.6}. Namely, ignoring absorption, the brightness ratio $R_T=f_{\rm{CO(2-1)}}/f_{\rm{CO(1-0)}}$ provides a measure of the temperature: $T[{\rm K}] =-11.1/\ln(R_T/16)$ in each data cube element separately. The dependence of $T$, calculated in pixels of 0.9$\times$0.9 arcsec$^2$ and Doppler velocity bins of 0.6 \kms, on $R$, $\psi$ and $V_z$ is displayed in Figure \ref{fig23}. It is approximately independent of $\psi$ and takes values mostly below 20 K. It decreases with $R$ from $\sim30$ K at $R=1$ arcsec to $\sim10$ K at $R=8$ arcsec. It is higher on the broad component than on the narrow component, reaching maxima at intermediate values of $|V_z|$. In what follows, we ignore the 9 central pixels having $R<0.5$ arcsec and require as usual $R$ not to exceed 8 arcsec.

\begin{figure*}
\centering
\includegraphics[width=0.75\textwidth,trim=0.5cm .5cm 1cm 1.cm,clip]{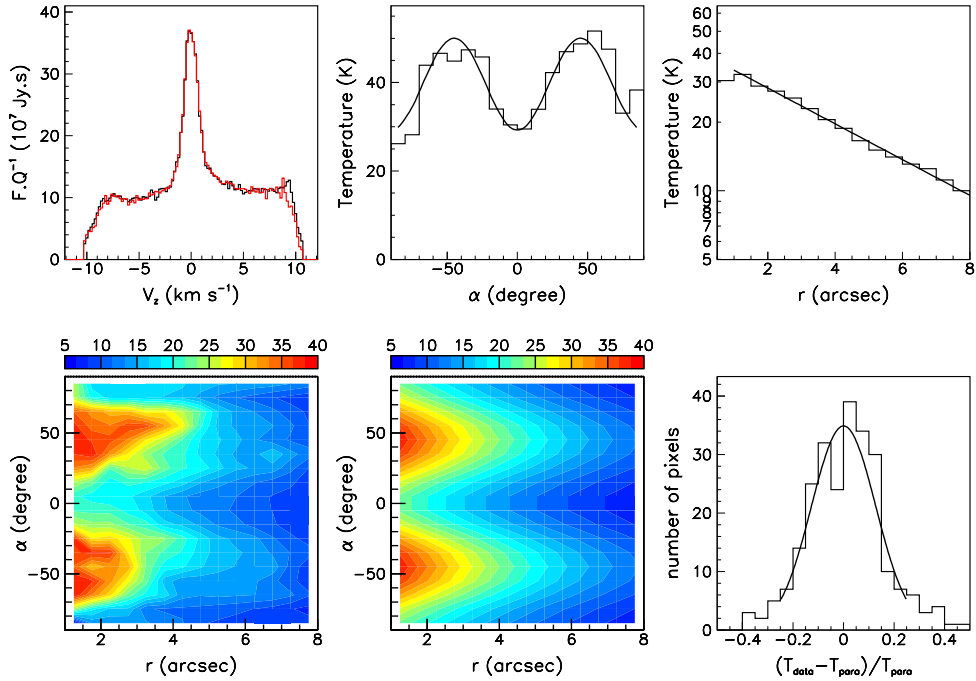}
\caption{ Upper left: temperature corrected Doppler velocity spectra of \mbox{CO(1-0)} (black) and \mbox{CO(2-1)} (red) emissions. Upper middle and right: temperature dependence on respectively $\alpha$ and $r$; the curves are a fit of the form $40(1-0.25\cos\!4\alpha)\exp(-r/5.52)$. Lower left and middle: dependence of $T$ on $r$ and $\alpha$; left shows the data and middle shows the parameterisation; lower right: distribution of the residuals $(T_{data}-T_{para})/T_{para}$ with a Gaussian fit of mean 0 and $\sigma=13$\%.}
\label{fig24}
\end{figure*}

As a check of the method, we correct the \mbox{CO(1-0)} and \mbox{CO(2-1)} data cubes for temperature: we divide each of them by the associated $Q$ factor calculated using the temperature measured as described in the preceding paragraph. The result is displayed in the upper left panel of Figure \ref{fig24}. A small difference between the \mbox{CO(1-0)} and \mbox{CO(2-1)} corrected spectra can be seen at the red-shifted end; it is due to uncertainties associated with the temperature evaluation near the end points.

The temperature-corrected data cubes are then de-projected in space using an axi-symmetric, $r$-independent radial velocity of the form $V=V_{eq}+(V_{pole}-V_{eq})\sin^2\!\alpha$ with $V_{eq}$ and $V_{pole}$ as defined in Section \ref{sec5.1}: $V_{eq}=1.9$ \kms\ and $V_{pole}= 10.3 (10.7)$ \kms\ for the blue(red)-shifted part of the spectrum. De-projection measures now the density rather than the effective emissivity. It is used to average the temperature as a function of $r$ and $\alpha$ instead of $R$ and $V_z$. The result is illustrated in Figure \ref{fig24}, showing that a good fit is obtained using a form $T=40(1-0.25\cos\!4\alpha)\exp(-r/5.52)$ with $T$ in Kelvin and $r$ in arcsec (an exponential gives a better fit than a power law). The relative uncertainty is $\sim13$\% as obtained from the $\sigma$ of a Gaussian fit to the residuals. 

\subsection{Flux of matter and mass-loss rate}\label{sec6.3}

The data cube, averaged between \mbox{CO(1-0)} and \mbox{CO(2-1)} emissions after correction for temperature as described in the preceding section, is now split into its narrow and broad components following the method described in Section \ref{sec3.2}. We use the results obtained in Sections \ref{sec4} and \ref{sec5} to construct as simple as possible a model of their morphology and kinematics in the region where they are distinct, $R>2$ arcsec. For both components, the wind velocity is taken of the form used in the preceding section (independence from stellar longitude $\omega$ is assumed throughout) and the inclination angle $\varphi$ is fixed at 10\dego.

\begin{figure*}
\centering
\includegraphics[width=0.95\textwidth,trim=0cm 1.cm 0cm 1.cm,clip]{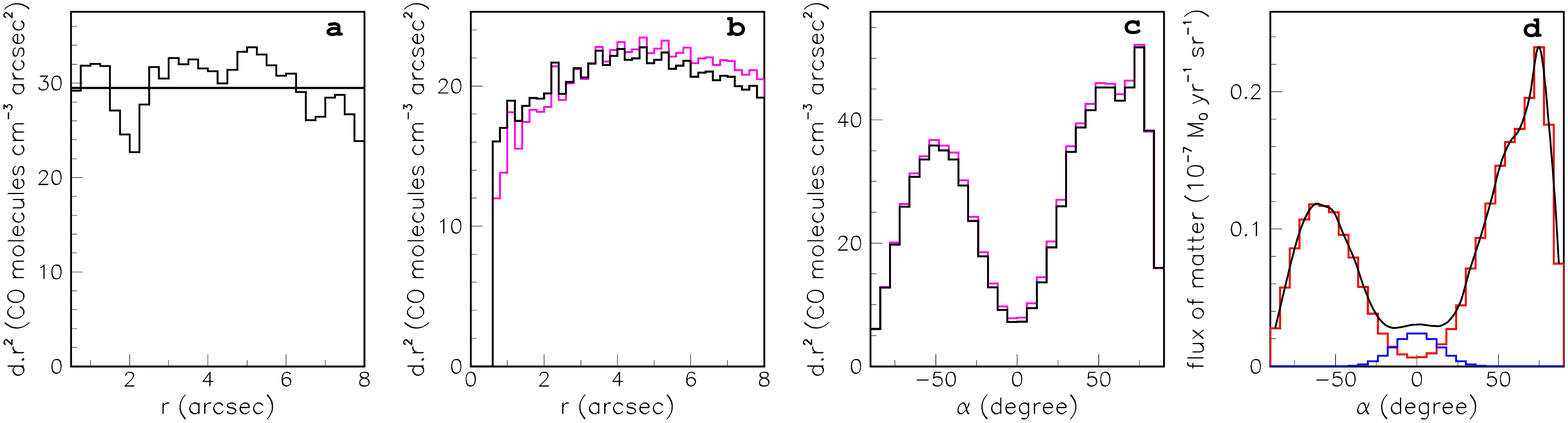}
\caption{a) dependence on $r$ of the density of the narrow component multiplied by $r^2$; b) and c): dependence on $r$ (b) and $\alpha$ (c) of the density of the broad component multiplied by $r^2$. The black curve is for standard de-projection using $V_{eq}=1.9$ \kms, the magenta curve uses instead $V_{eq}=V_{esc}$, the escape velocity of a solar mass star; d) dependence on stellar latitude $\alpha$ of the flux of matter for the broad (red) and narrow (blue) components as well as for their sum (black).}
\label{fig25}
\end{figure*}

From the temperature corrected intensity of the narrow component, assuming its density to be a Gaussian in $\sin\!\alpha$ of $\sigma=\sigma_{flar}=0.20$ (see Section \ref{sec4.3}), we obtain its dependence on $R$ or, to a good approximation, on $r$. Its product by $r^2$ is displayed in the left panel of Figure \ref{fig25}. It has an average value of 29 CO molecules\,cm$^{-3}$arcsec$^2$ and shows that the equatorial outflow has been developing at approximately constant mass-loss rate during the past 2 to 3 thousand years. However, it displays significant deviations with respect to such an approximation, at the typical level of $\pm10$\%, with maxima at $\sim3.5$ and 5.2 arcsec and minima at $\sim2$ and 8 arcsec.

The dependence on $r$ and $\alpha$ of the broad component de-projected density multiplied by $r^2$ is displayed in the middle panels of Figure \ref{fig25}. It is approximately constant for $r>2$ arcsec, at a level of 22 CO molecules\,cm$^{-3}$arcsec$^2$, showing that the bipolar outflow, as in the case of the equatorial outflow, has been developing at approximately constant mass-loss rate during the past 2 to 3 thousand years. Its dependence on $\alpha$ displays  maxima at intermediate latitudes.

The dependence on stellar latitude $\alpha$ of the flux of matter $\rm d\mit{\dot{M}}/\rm d\mit{\Omega}$ per solid angle $\rm d\mit{\Omega}=2\pi\cos\!\alpha \,\rm d\alpha$ can be obtained from the product of the density by $r^2$ and by $2\pi V\cos\!\alpha$ where $V$ is the wind radial velocity. The result is displayed in the right panel of Figure \ref{fig25} for the broad and narrow components separately and for their sum. Assuming a solar-like abundance ratio, $\rm{CO/H}\sim 2.5\,10^{-4}$, the total mass-loss rate is measured this way to be $0.9\,10^{-7}$ M$_{\odot}$yr$^{-1}$, with the contribution of the narrow component being only 10\% of the total. This result is independent of what happens at short distances from the star and needs only to be corrected for absorption.

In Section \ref{sec5}, we commented on the low value of the equatorial wind velocity and on its use for de-projection and for the evaluation of the effective emissivity and flux of matter; we were then claiming that concerns related to the equatorial wind velocity being below escape velocity over a significant fraction of the explored radial range were irrelevant in the present context. To illustrate this claim we display in panels b) and c) of Figure \ref{fig25} the results of a de-projection using as $V_{eq}$ the escape velocity evaluated for a solar mass star, $V_{esc}$ [\kms]$=42/\sqrt{r [\rm au]}$. The difference with the $V_{eq}$ result is indeed very small and does not significantly affect any of the conclusions made in the present Section.

\subsection{Absorption}\label{sec6.4}

In all previous considerations, absorption was ignored. It was implicitly assumed that its effect would be small, would not vary too much across the data cube and would not differ too much between \mbox{CO(1-0)} and \mbox{CO(2-1)} emissions. We are now in a position to evaluate it for each line and in each pixel independently. In each pixel, the relation between $V_z$ and $z$ is defined by choosing the standard wind velocity distribution ($V_{eq}=1.9$ \kms\ and $V_{pole}= 10.3$ (10.7) \kms\ for the blue(red)-shifted part of the spectrum); the temperature $T$ and the density $d$ are then well defined as functions of $z$, or $\sin\!\alpha=z'/r$ or $V_z$. In order to avoid too low signal-to-noise values, we use large data cube elements: 0.9 arcsec wide in $x$ and $y$ and 0.6 \kms\ wide in Doppler velocity. We moreover assume that the wind reaches terminal velocity at $r=0.5$ arcsec and increases in proportion to $r$ below this value; in practice, this is largely irrelevant to the arguments of the present article, which do not depend significantly on the morpho-kinematics in the close environment of the star.

\begin{figure*}
\centering
\includegraphics[height=4.5cm,trim=.5cm 1.cm 0cm 2.cm,clip]{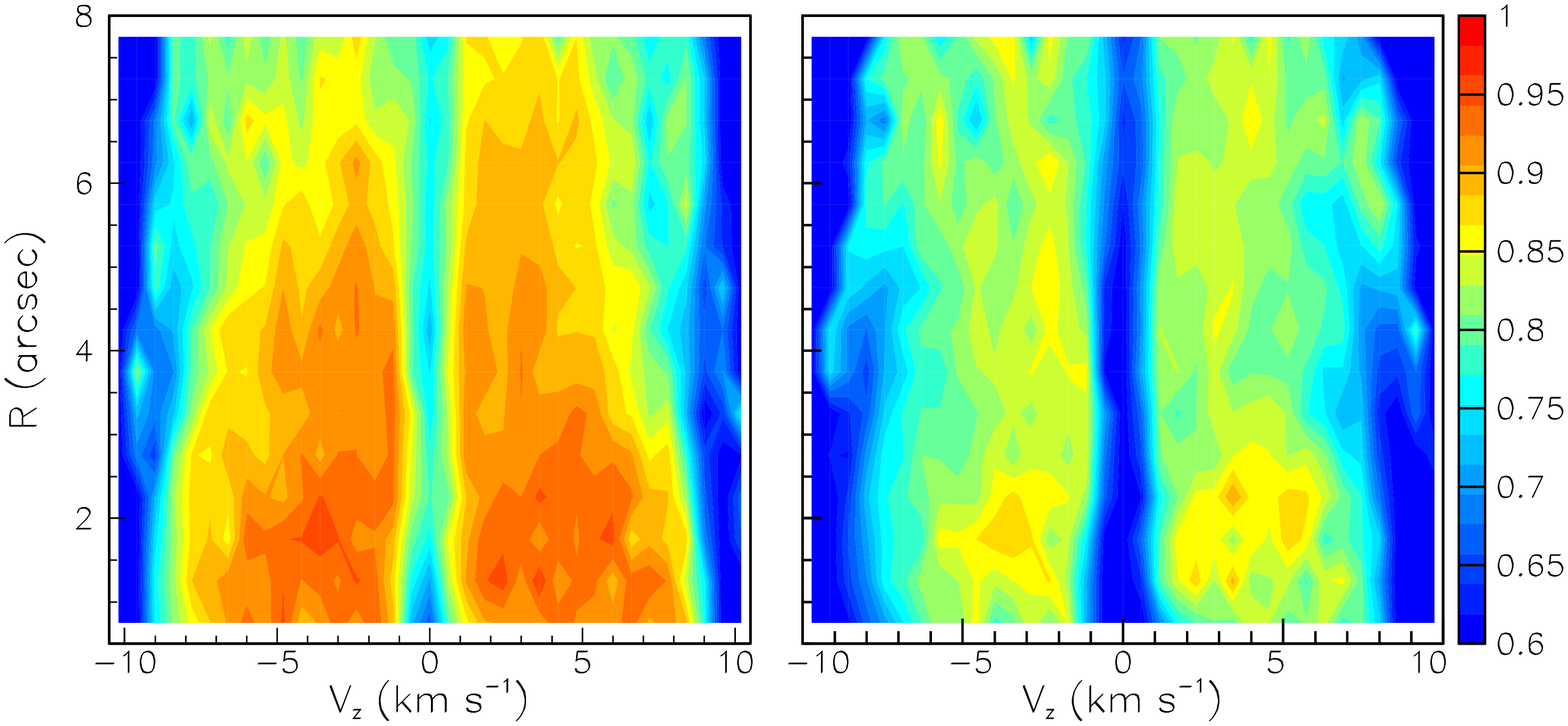}
\includegraphics[height=4.5cm,trim=.5cm 1.cm 0cm 2.cm,clip]{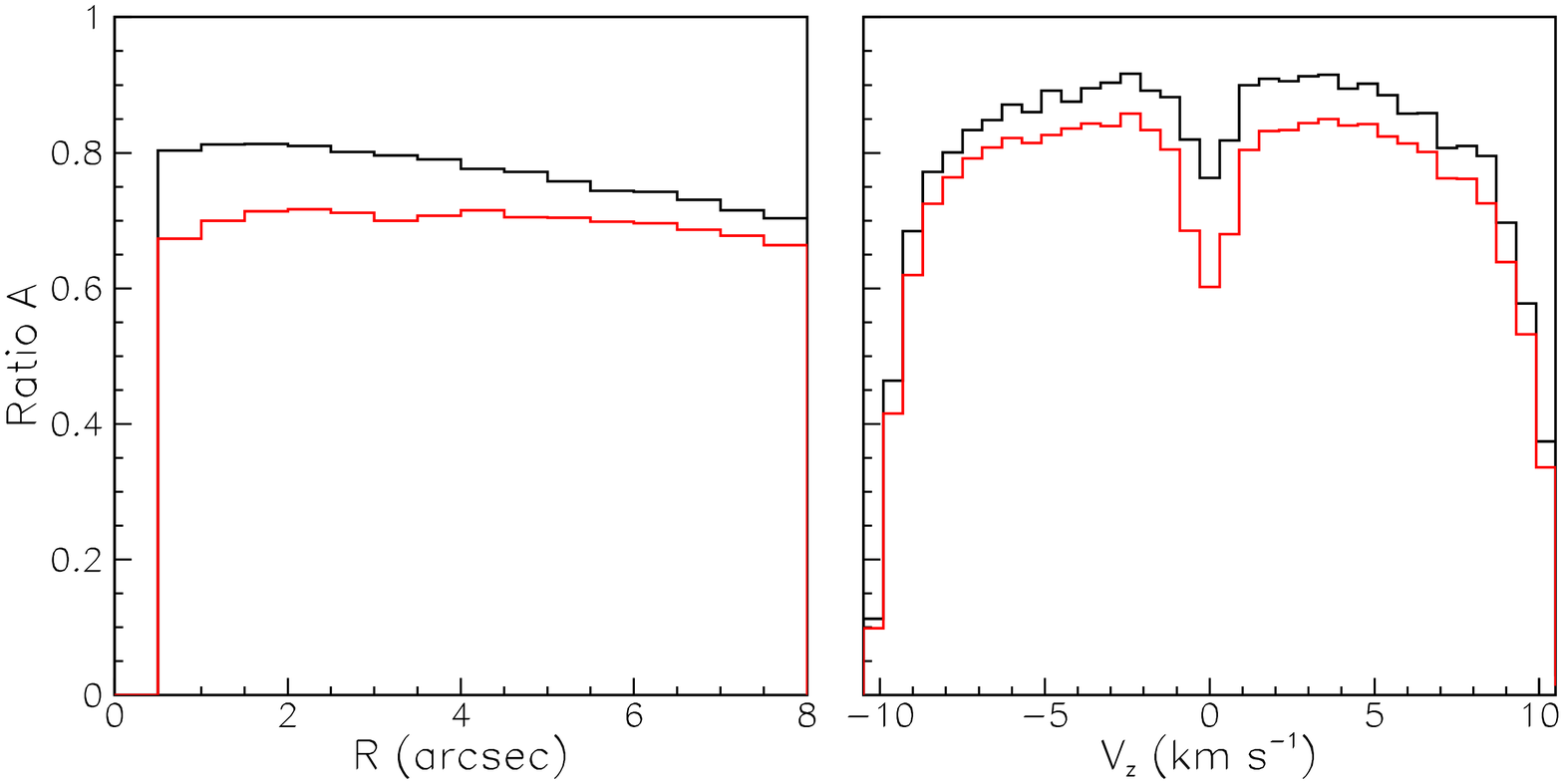}
\caption{Ratio $A$ between the flux densities averaged over position angle with and without absorption. Upper panels: distributions in the $R$ vs $V_z$ plane for \mbox{CO(1-0)} (left) and \mbox{CO(2-1)} (right) emissions. Lower panels (\mbox{CO(1-0)} black and \mbox{CO(2-1)} red): projections on the $R$ axis (left) and on the $V_z$ axis (right).}
\label{fig26}
\end{figure*}

We start from the results of Section \ref{sec6.2}, including de-projected effective emissivity for each line and an estimate of the temperature in the limit of zero absorption (Figures \ref{fig23} and \ref{fig24}). We then use the de-projected effective emissivity, $\rho$, to calculate the brightness, this time accounting for absorption (ignoring absorption gives the same values as we started from). This is done in each pixel and for each line separately by scanning along the line of sight in small $z$ steps of width $\Delta{z}=0.004$ arcsec, starting from the red-shifted end ($z=30$ arcsec). At each step, we calculate $V_z$ and add to the corresponding bin of the brightness distribution the contribution of the new $z$ step, $\rho\Delta z$, while subtracting the attenuation of the contributions of the earlier $z$ steps by a factor $f_{abs}\rho\Delta z$ where
$f_{abs}=F_J[\exp(\lambda_J)-1]$; here, $F_J$ is a constant inversely proportional to the third power of the line frequency and $\lambda_J$ depends on temperature as $\lambda_J=2E_J/[T(J+1)]$. We obtain this way a brightness $f'=Af$, smaller than the measured brightness $f$ by a factor $A$ evaluated in each data cube element separately. Its distribution in the $R$ vs $V_z$ plane and its projection on the $R$ and $V_z$ axes are displayed in Figure \ref{fig26} for each line separately. Absorption is measured at typical levels of 10 to 30\%. It is larger for the narrow than for the broad component because in the former case $V_z$ varies little when $z$ scans along the line of sight and line emission keeps self-absorbing. The difference between \mbox{CO(1-0)} and \mbox{CO(2-1)} absorptions is small, it rarely exceeds 10\%.

\subsection{Results}\label{sec6.5}

\begin{figure*}
\centering
\includegraphics[height=5.7cm,trim=.5cm 1.cm 0cm 1.cm,clip]{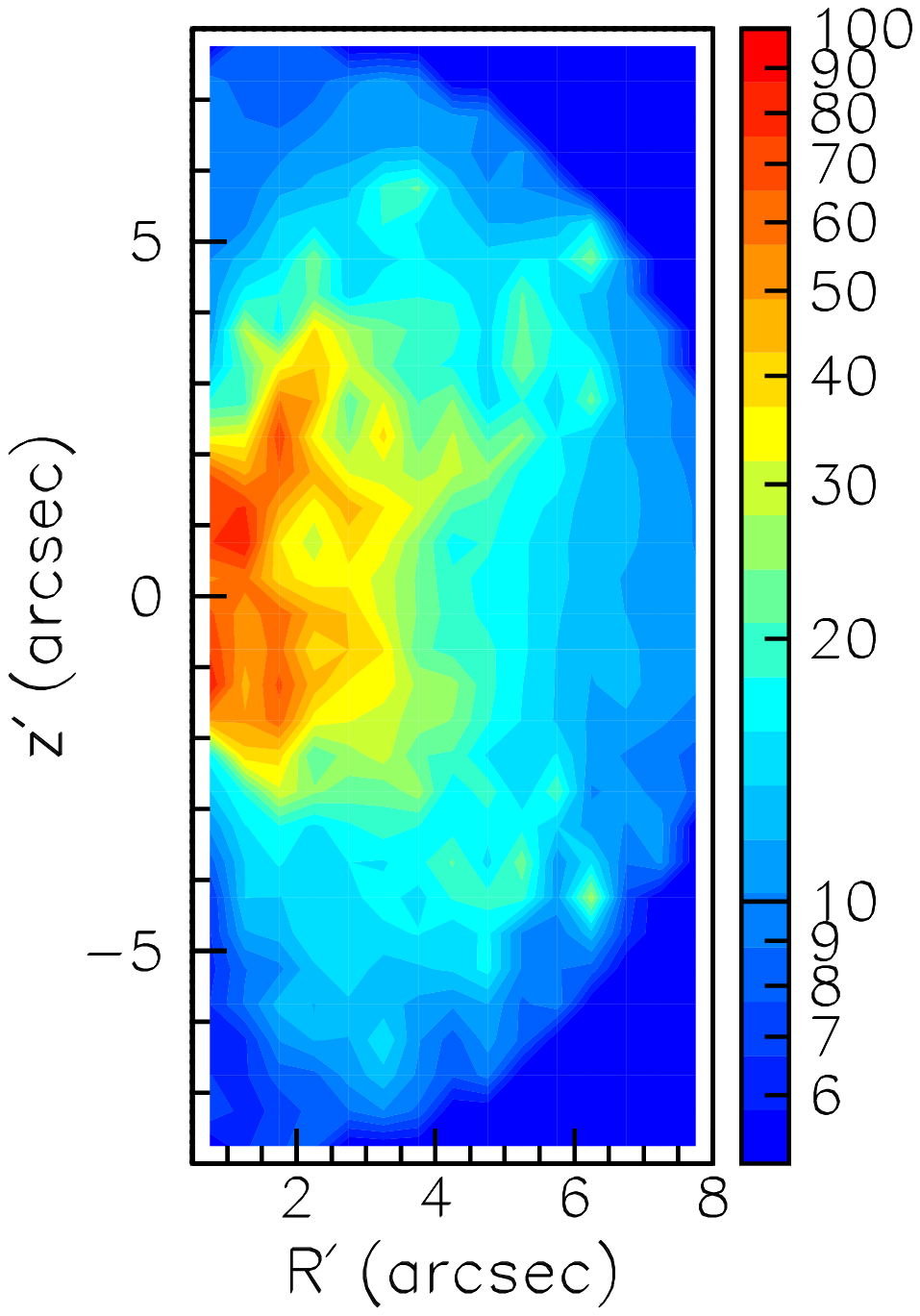}
\includegraphics[height=5.7cm,trim=.5cm 1.cm 1.5cm 1.cm,clip]{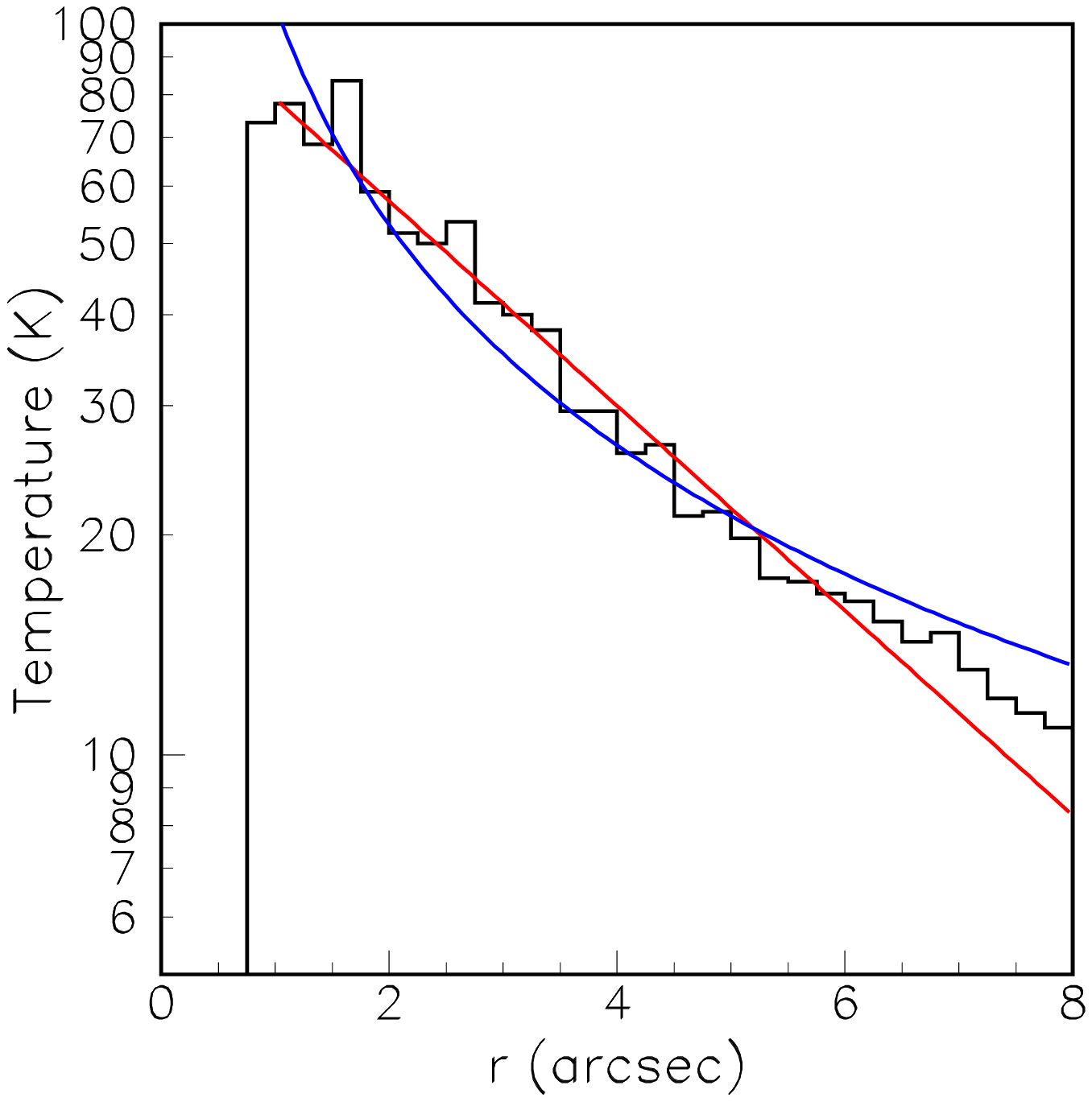}
\includegraphics[height=5.7cm,trim=.5cm 1.cm 1.5cm 1.cm,clip]{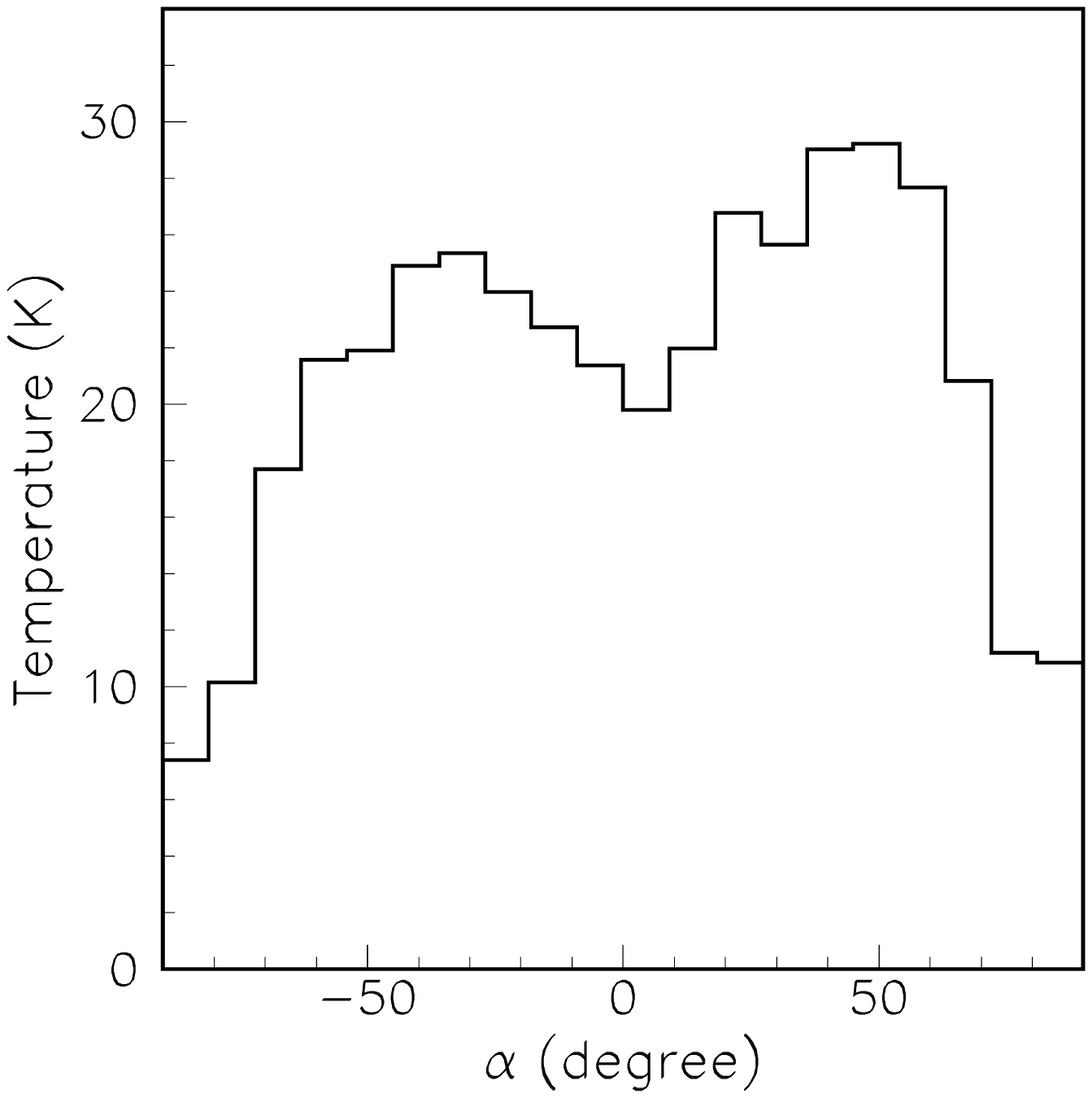}
\caption{Distributions of the temperature $T^*$ (K) in the star meridian plane (left, $z'$ vs $R'$), as a function of $r$ (arcsec, middle) and of stellar latitude $\alpha$ (degrees, right). The lines in the middle panel are for the best exponential fit (red) and the best power fit (blue).}
\label{fig27}
\end{figure*}

\begin{figure*}
\centering
\includegraphics[height=5.cm,trim=.5cm 1.cm 0cm 1.5cm,clip]{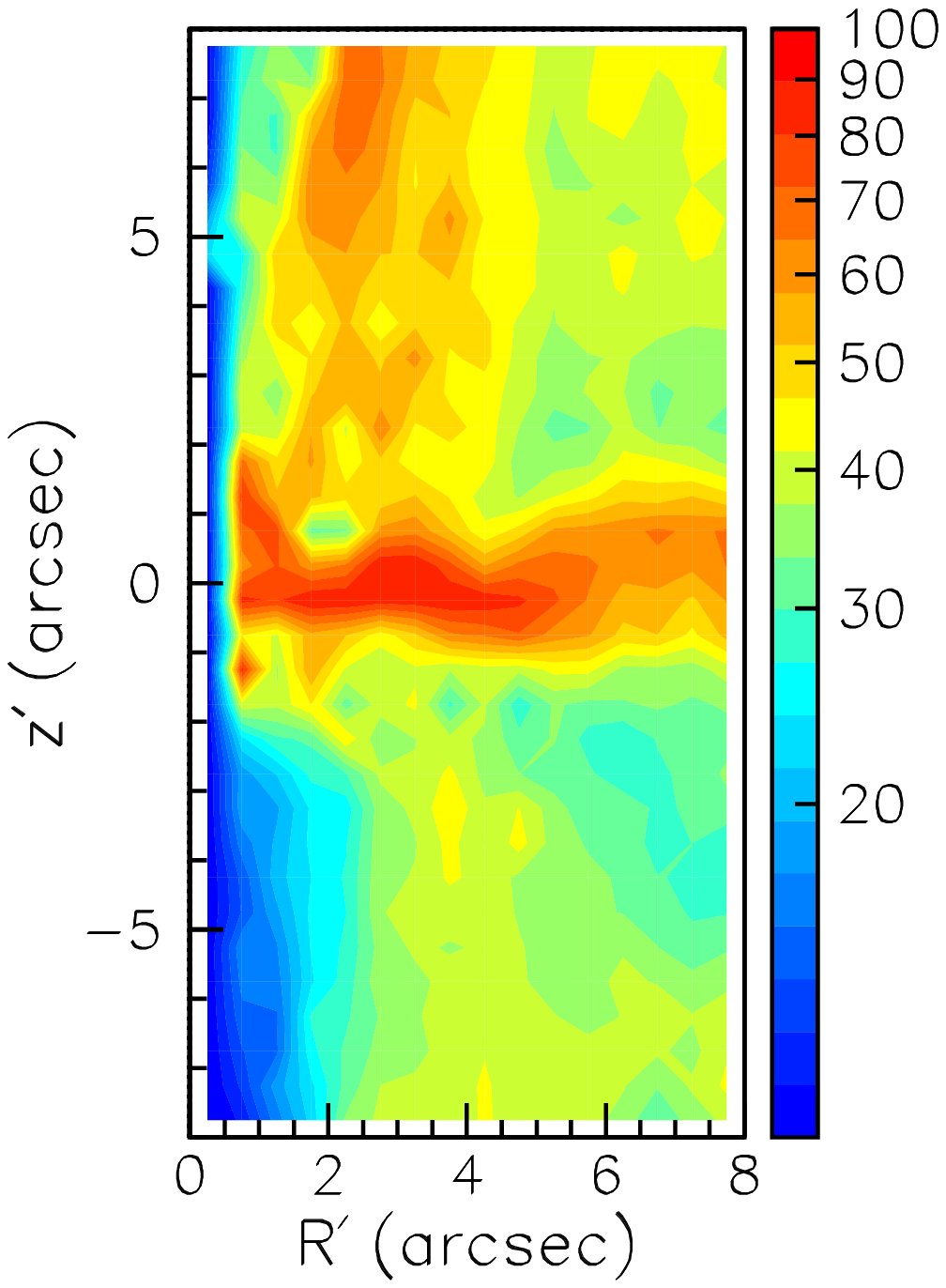}
\includegraphics[height=5.cm,trim=.5cm 1.cm 1.5cm 1.5cm,clip]{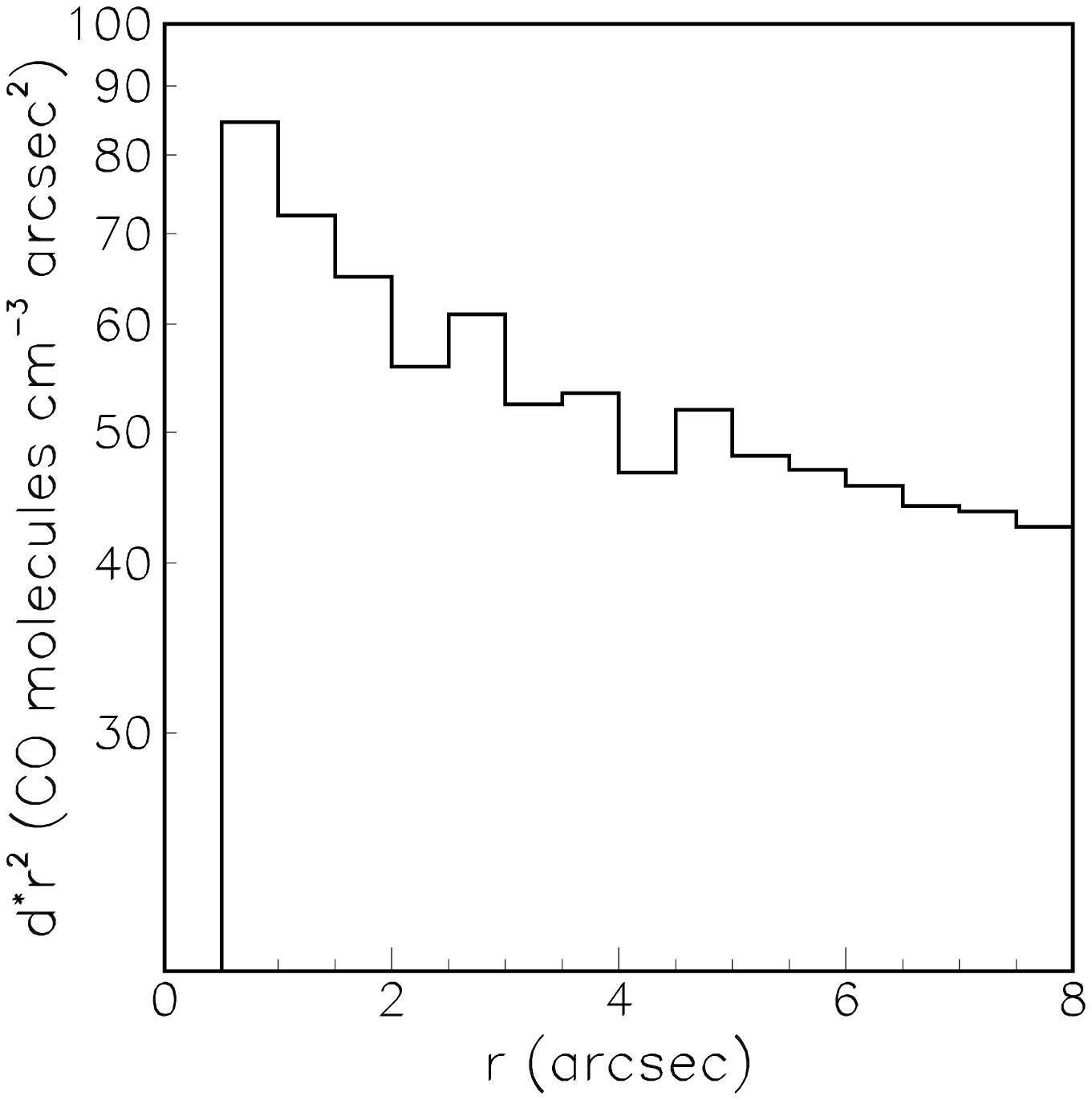}
\includegraphics[height=5.cm,trim=.5cm 1.cm 1.5cm 1.5cm,clip]{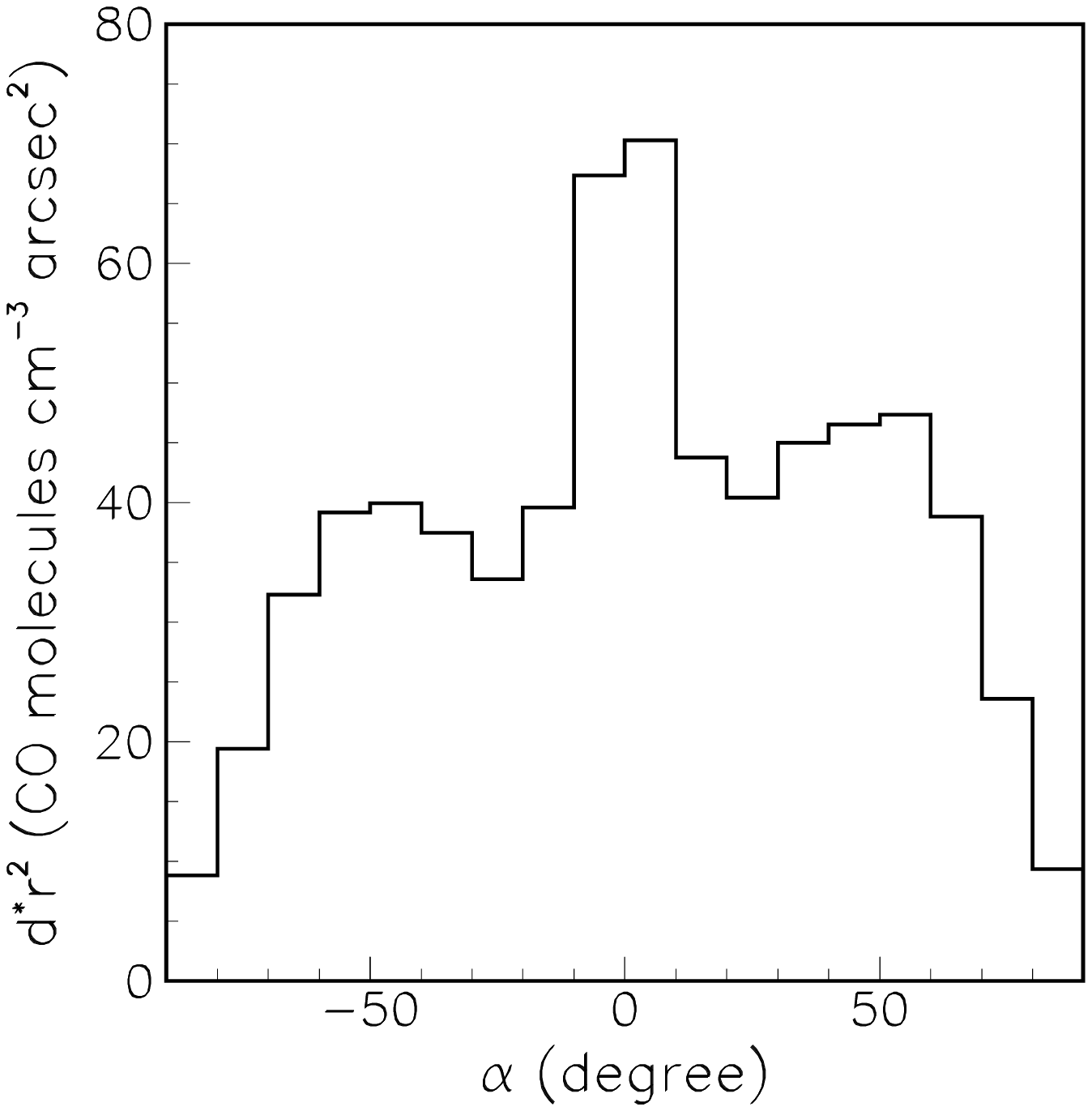} \\
\includegraphics[height=5.cm,trim=.5cm 1.cm 0cm 1.5cm,clip]{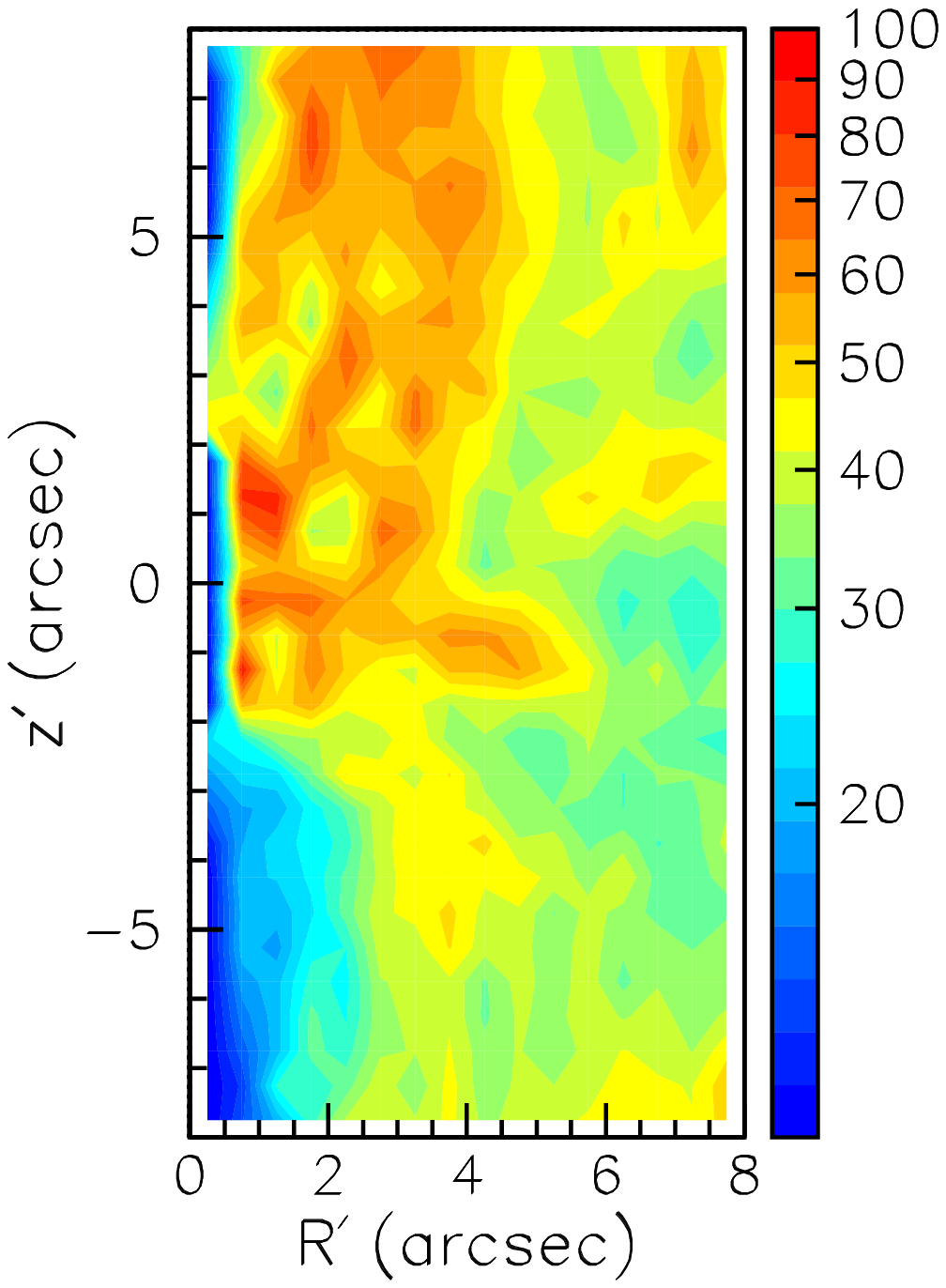}
\includegraphics[height=5.cm,trim=.5cm 1.cm 1.5cm 1.5cm,clip]{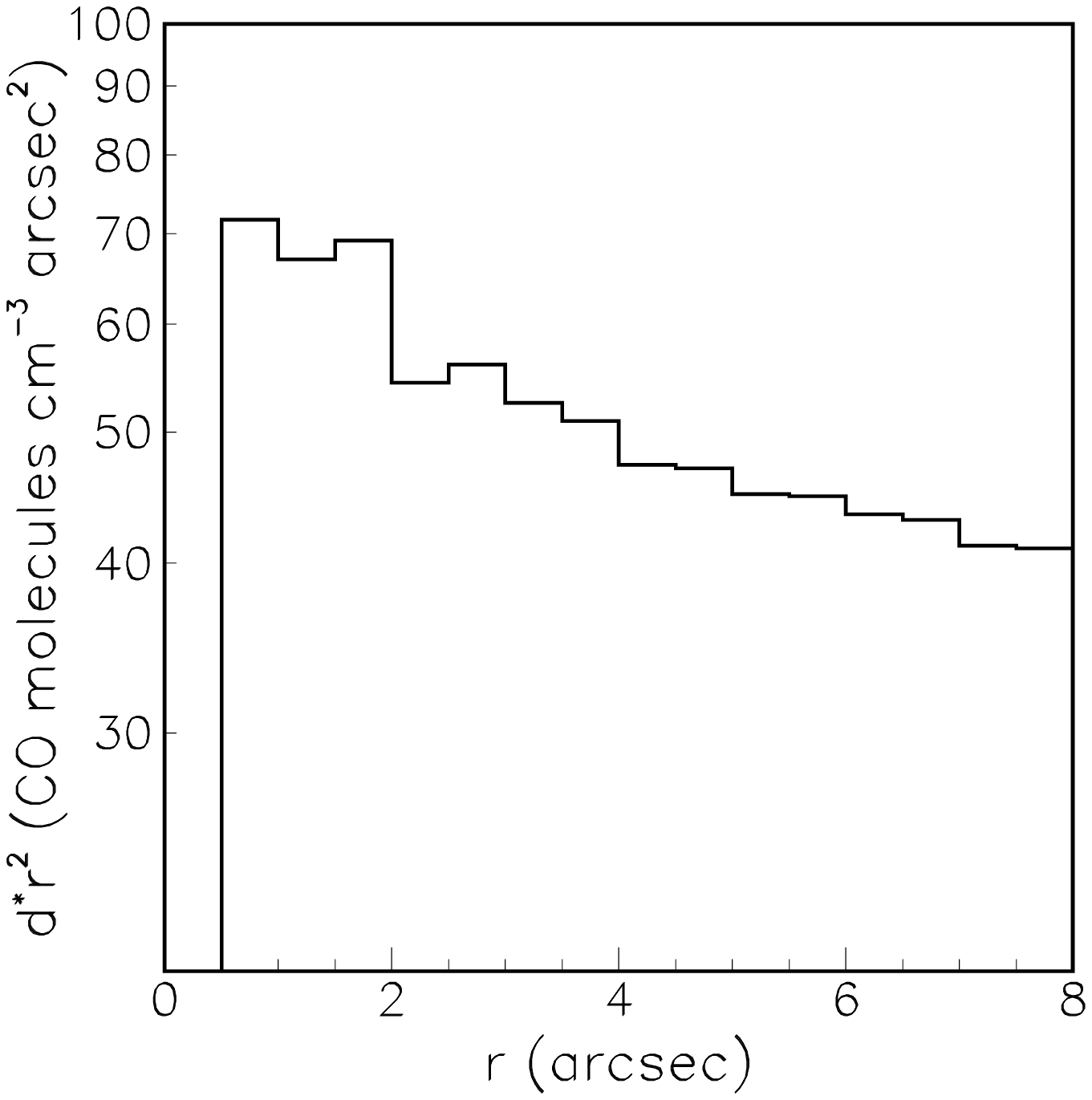}
\includegraphics[height=5.cm,trim=.5cm 1.cm 1.5cm 1.5cm,clip]{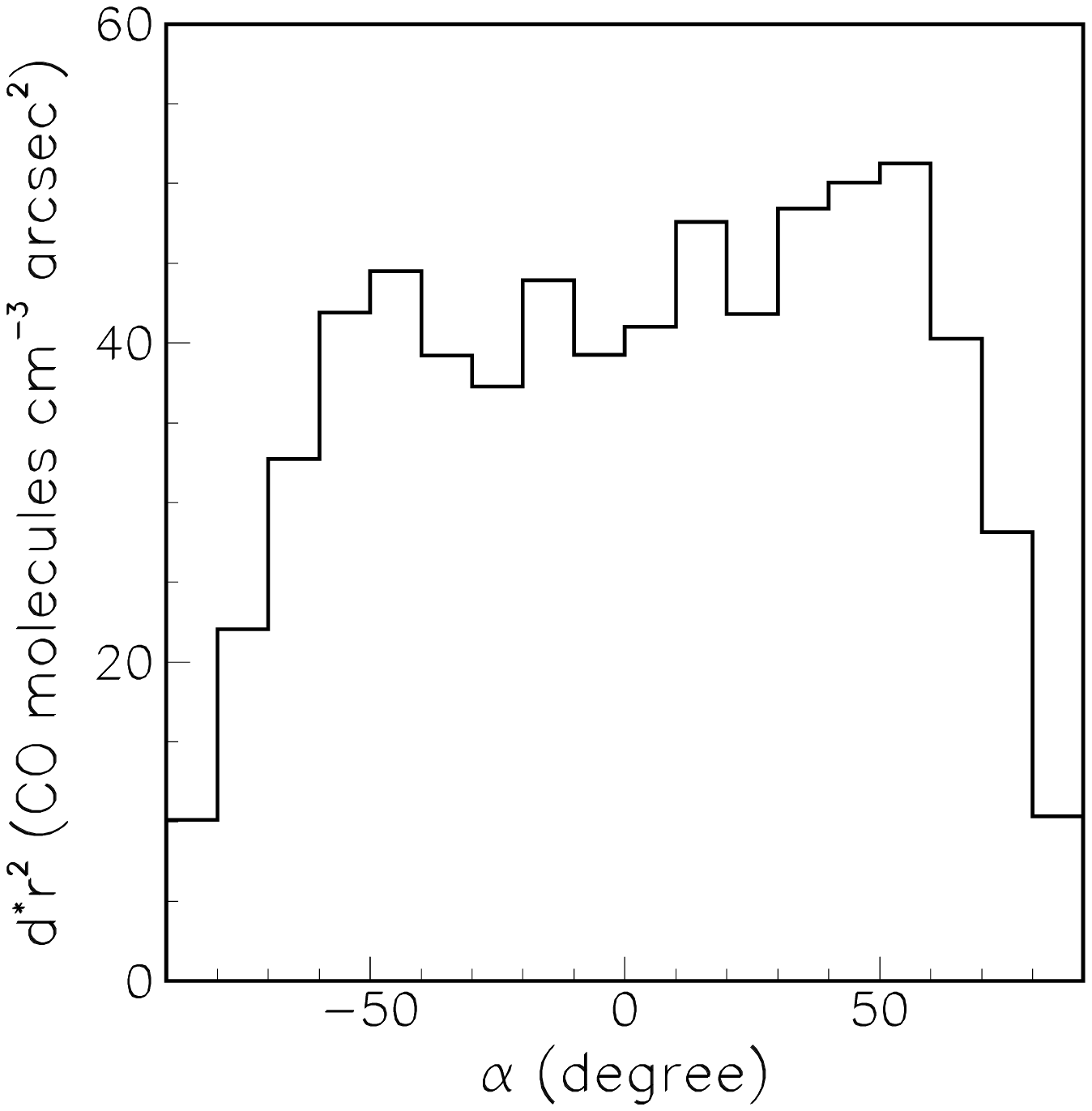}\\
\includegraphics[height=5.cm,trim=.5cm 1.cm 0cm 1.5cm,clip]{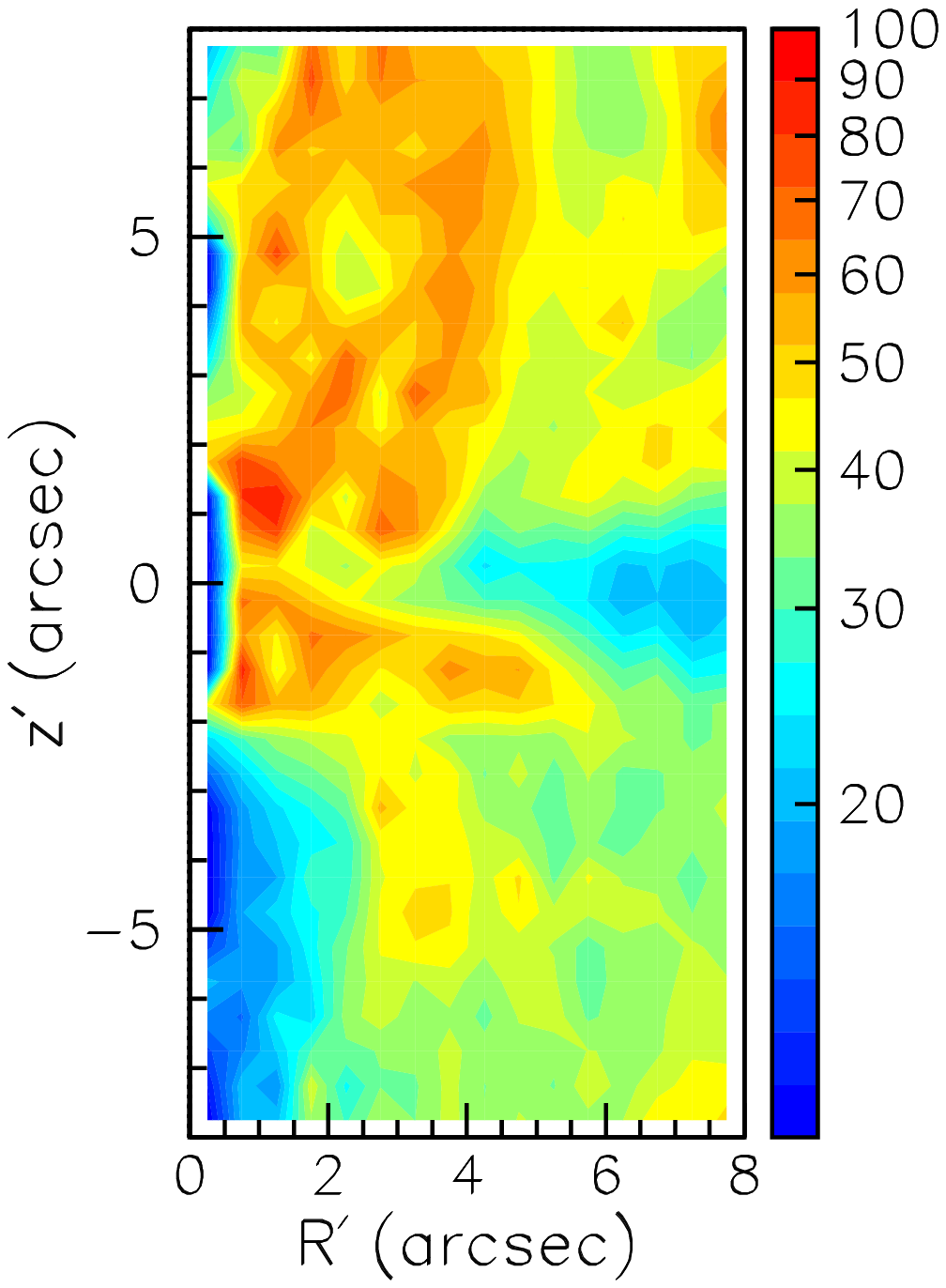}
\includegraphics[height=5.cm,trim=.5cm 1.cm 1.5cm 1.5cm,clip]{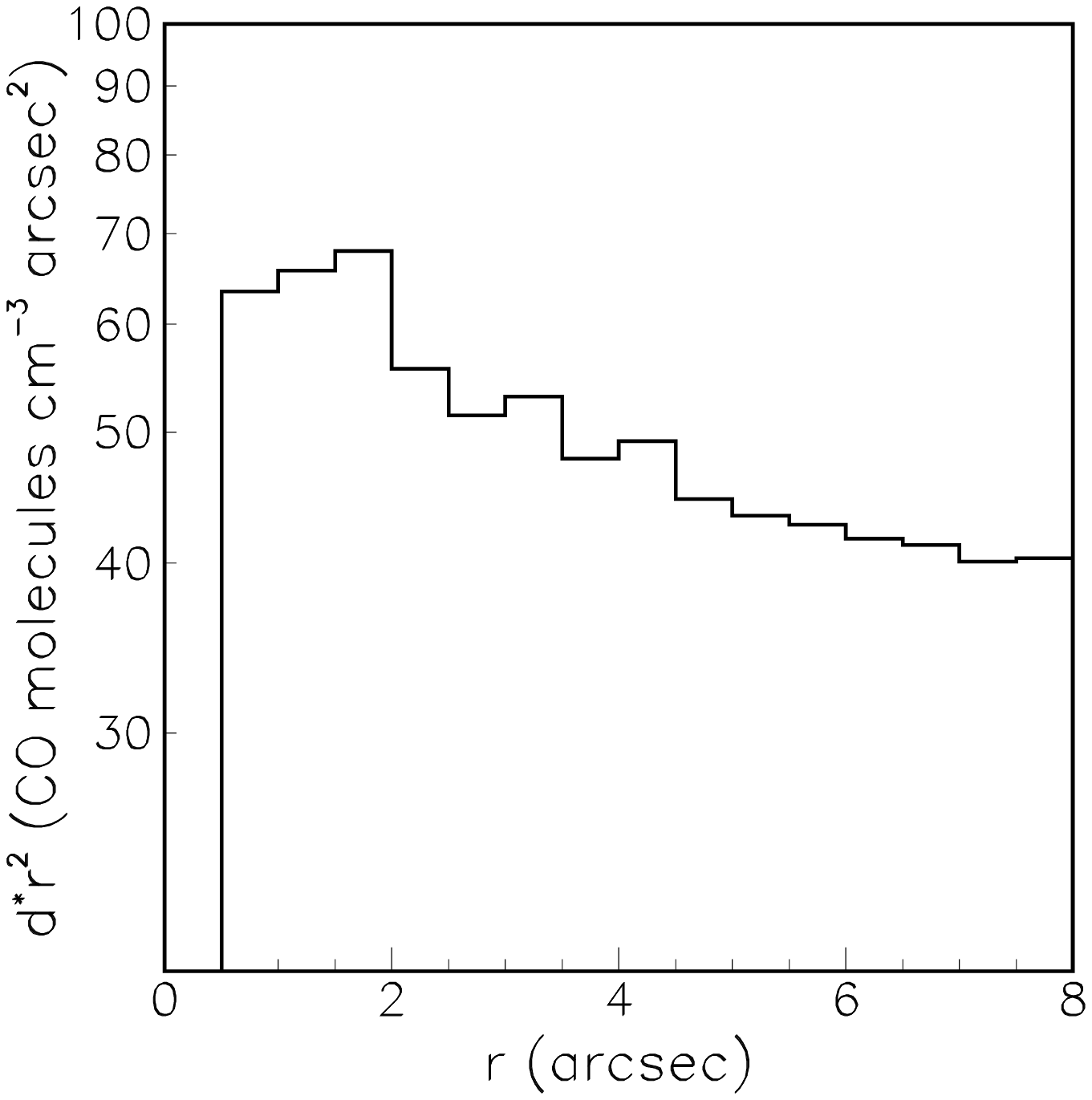}
\includegraphics[height=5.cm,trim=.5cm 1.cm 1.5cm 1.5cm,clip]{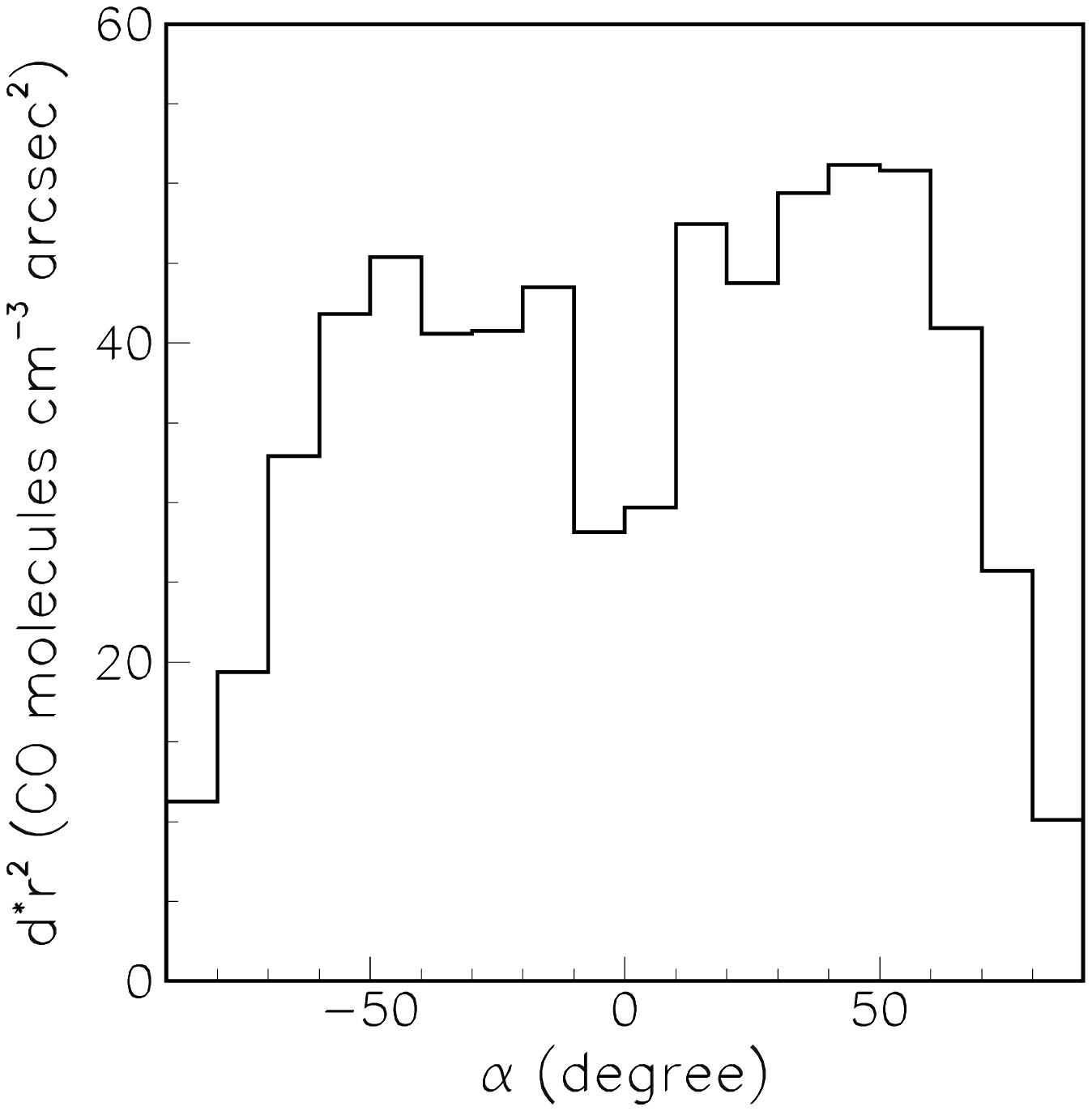}\\
\caption{Density $d^*$ (CO molecules cm$^{-3}$) multiplied by $r^2$. The inclination angle $\varphi$ is set at 5\dego\ (upper panels), 10\dego\ (middle panels) or 15\dego\ (lower panels). Left panels display the map of $d^*r^2$ in the meridian plane. Middle panels display the dependence of $d^*r^2$ on $r$.  Right panels display the dependence of $d^*r^2$ on stellar latitude $\alpha$.}
\label{fig28}
\end{figure*}

Having evaluated the effect of absorption and found it relatively small and well-behaved, we obtain new evaluations of the temperature and density using absorption-corrected flux densities, $f^*=f/A$. To avoid introducing undesired fluctuations, we use as $A$ the product of its projected distributions on the $R$ and $V_z$ axes, as displayed in Figure \ref{fig26}, properly normalized. This is equal to the value of $A$ evaluated in each data cube separately to within 6\% (rms). The temperatures $T^*$ evaluated from the ratio $R^*_T$ between the \mbox{CO(2-1)} and \mbox{CO(1-0)} values of $f^*$ differ from the former values $T$ to the extent that the absorptions of \mbox{CO(1-0)} and \mbox{CO(2-1)} emissions are different. The resulting distributions of temperature and density, $T^*$ and $d^*$, averaged over stellar longitude $\omega$, are displayed in the star meridian plane, $z'$ vs $R'$, in Figures \ref{fig27} and \ref{fig28}; also shown are their projections on the $r$ and $\alpha$ axes. By construction, they describe equally well \mbox{CO(1-0)} and \mbox{CO(2-1)} observations. Absorption is small enough to make further iteration unnecessary: using the $T^*$ and $d^*$ distributions displayed in Figures \ref{fig27} and \ref{fig28} to perform a radiative transfer calculation, we reproduce the measured data cubes to excellent precision.

The temperature (Figure \ref{fig27}), averaged over stellar latitude, decreases with $r$ nearly exponentially, by a factor of $\sim2$ over $\sim2$ arcsec, more precisely as $T^*\sim109\exp(-r/3.1)$ with $T^*$ in Kelvins and $r$ in arcsec; the power law best fit gives an exponent equal to unity, namely $T^*\sim106/r$; averaged over $r$, the temperature displays maxima at intermediate stellar latitudes, being slightly warmer in the red-shifted than in the blue-shifted hemisphere. 

The density (Figure \ref{fig28}) has been evaluated for three values of the inclination angle $\varphi$ of the star axis with respect to the line of sight, changing $V_{eq}$ accordingly: $V_{eq}=(0.33\pm0.03)/\sin\!\varphi$ \kms. It displays an excess of $\sim30$\% at distances from the star smaller than $\sim 2$ arcsec and then decreases slightly faster than $1/r^2$. Such a decrease is expected as the result of dissociation of the CO molecules by interstellar UV radiation. However, the present data cannot measure its value precisely. As was illustrated in Figure 25 b, it depends on the form used to describe the equatorial wind velocity, which is arbitrary as long as it averages to some 2 \kms\ over the explored radial range. We note that estimates of the decrease have been given by \citet{Mamon1988} and \citet{Groenewegen2017}; while the former predicts a decrease of $\sim$9\% at $r=5$ arcsec and $\sim$33\% at $r=10$ arcsec, the latter predicts instead a negligible effect: 1\% at 5 arcsec and 8\% at 10 arcsec. The dependence on $\alpha$ displays maxima at intermediate latitudes but the value at the equator depends on $\varphi$: the smaller $\varphi$ the larger the enhancement of the equatorial density, trivially scaling with $V_{eq}$ (as $\rho/f=\mbox{d}V_z/\mbox{d}z$ scales with $V_{eq}$ at the equator).

\begin{figure}
\centering
\includegraphics[height=5.5cm,trim=.5cm 1.cm 1.5cm 1.cm,clip]{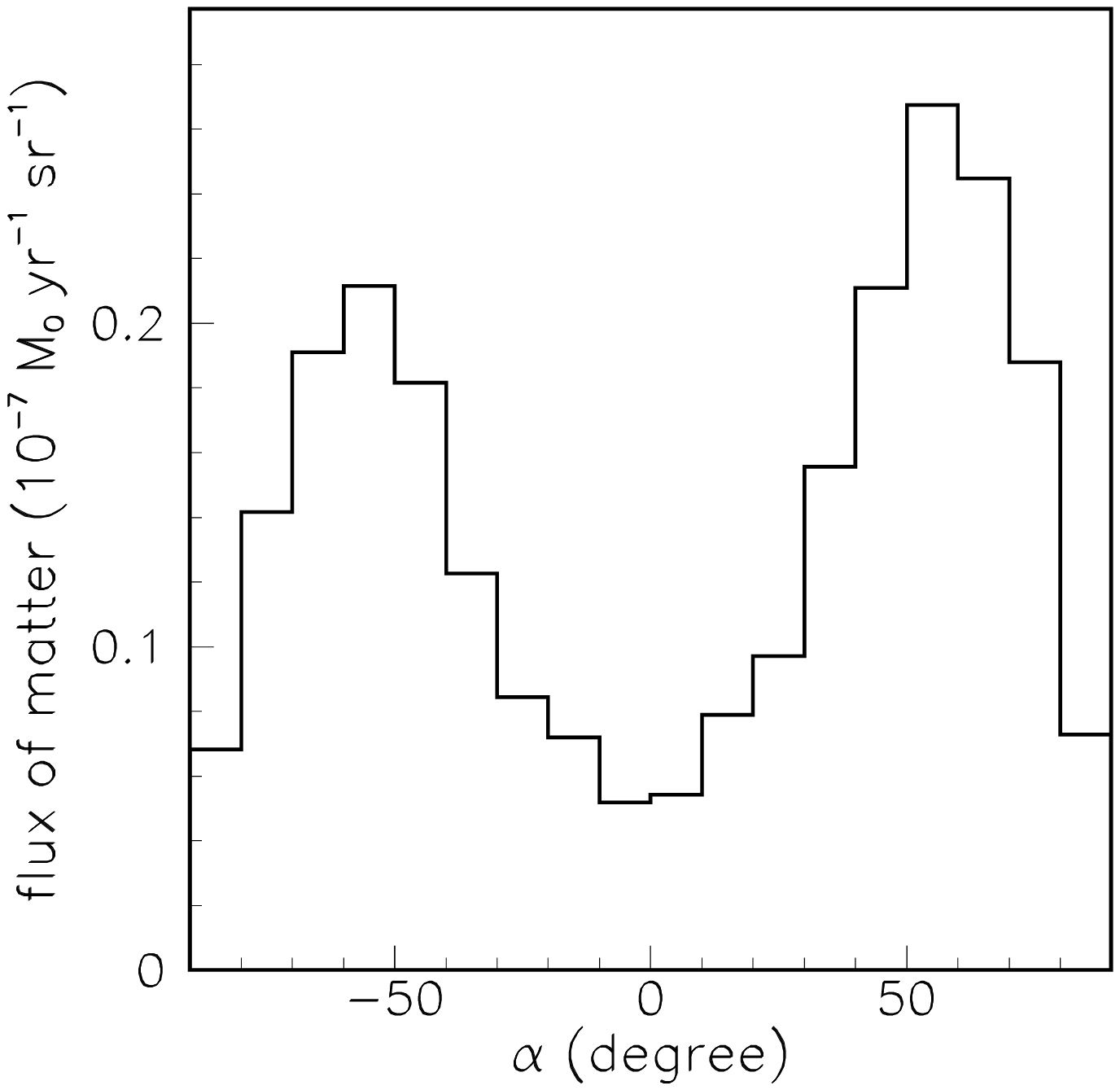}
\caption{Dependence on stellar latitude of the flux of matter corrected for absorption. The inclination angle $\varphi$ is set at 10\dego.}
\label{fig29}
\end{figure}

The mass-loss rate, which was evaluated as $0.9\,10^{-7}$ M$_\odot$\,yr$^{-1}$ in Section \ref{sec6.3} ignoring absorption, becomes ($1.6\pm0.4$) $10^{-7}$ M$_\odot$\,yr$^{-1}$ when using the absorption corrected values of the temperature and density, $T^*$ and $d^*$. The uncertainty is estimated by including effects of absorption and inclination angle. The dependence of the flux of matter on stellar latitude $\alpha$ is displayed in Figure \ref{fig29}.

\begin{figure*}
\centering
\includegraphics[height=5.3cm,trim=.5cm 1.cm 1.5cm 1.cm,clip]{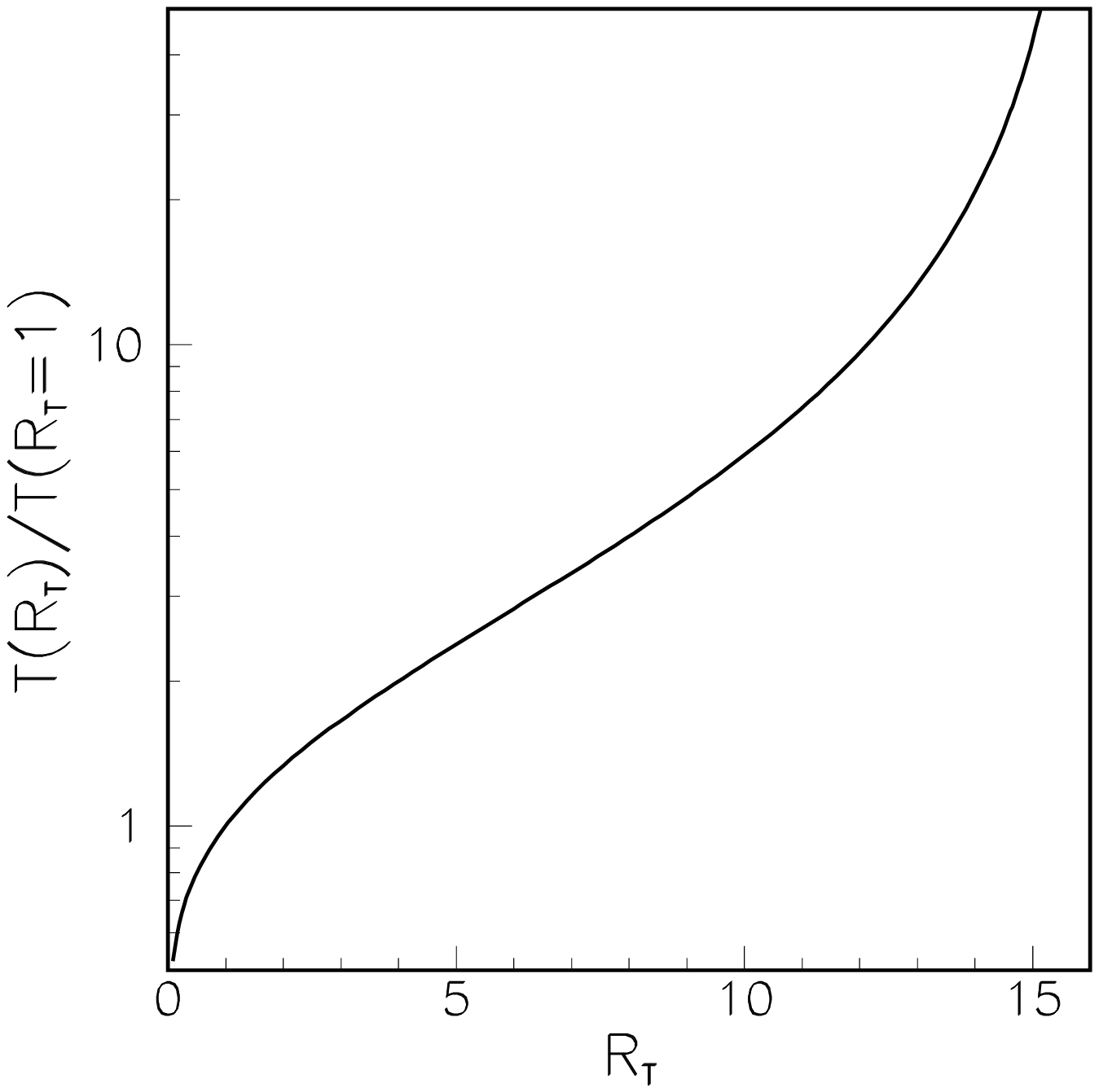}
\includegraphics[height=5.3cm,trim=.5cm 1.cm 1.5cm 1.cm,clip]{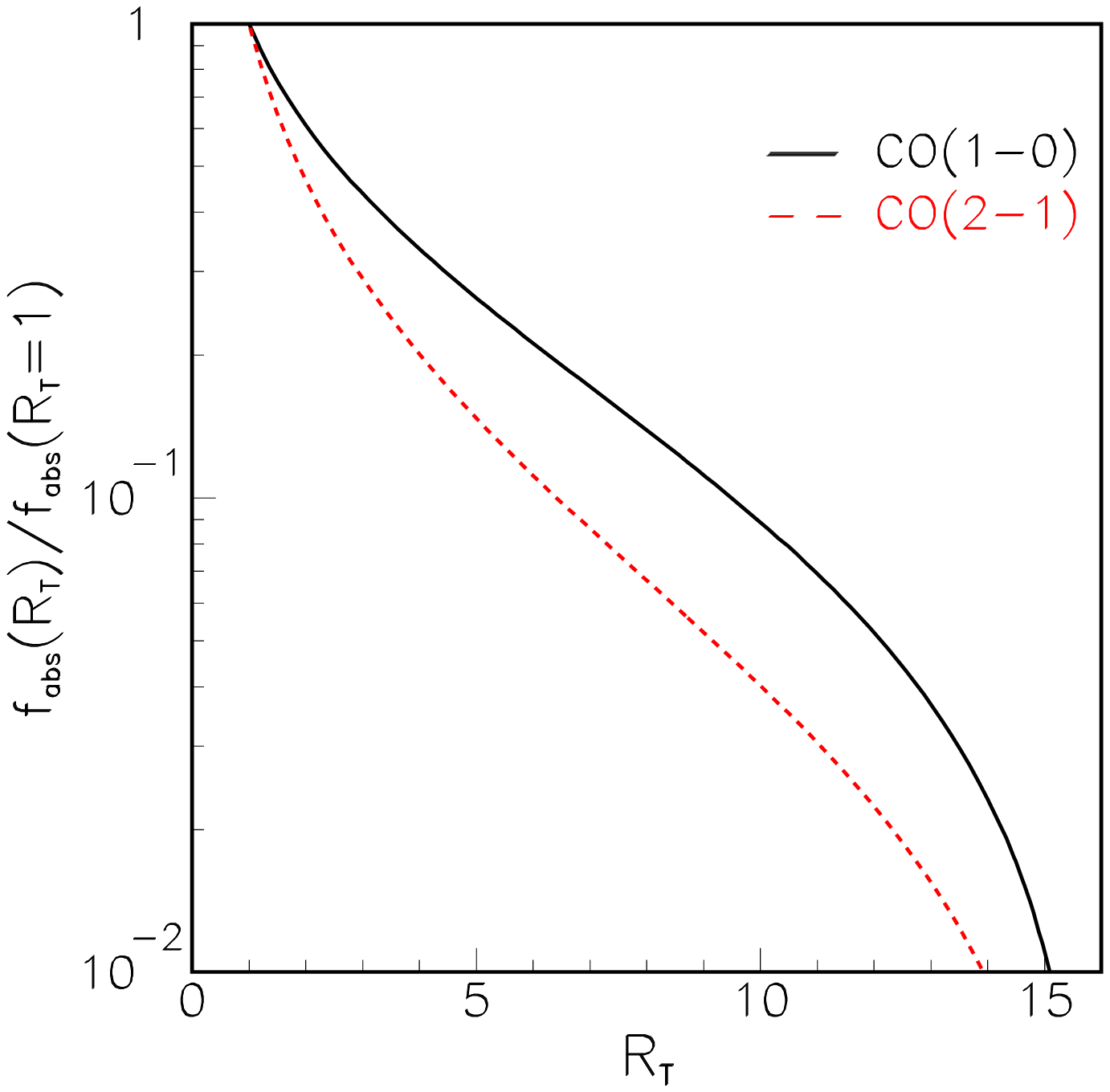}
\includegraphics[height=5.3cm,trim=.5cm 1.cm 1.5cm 1.cm,clip]{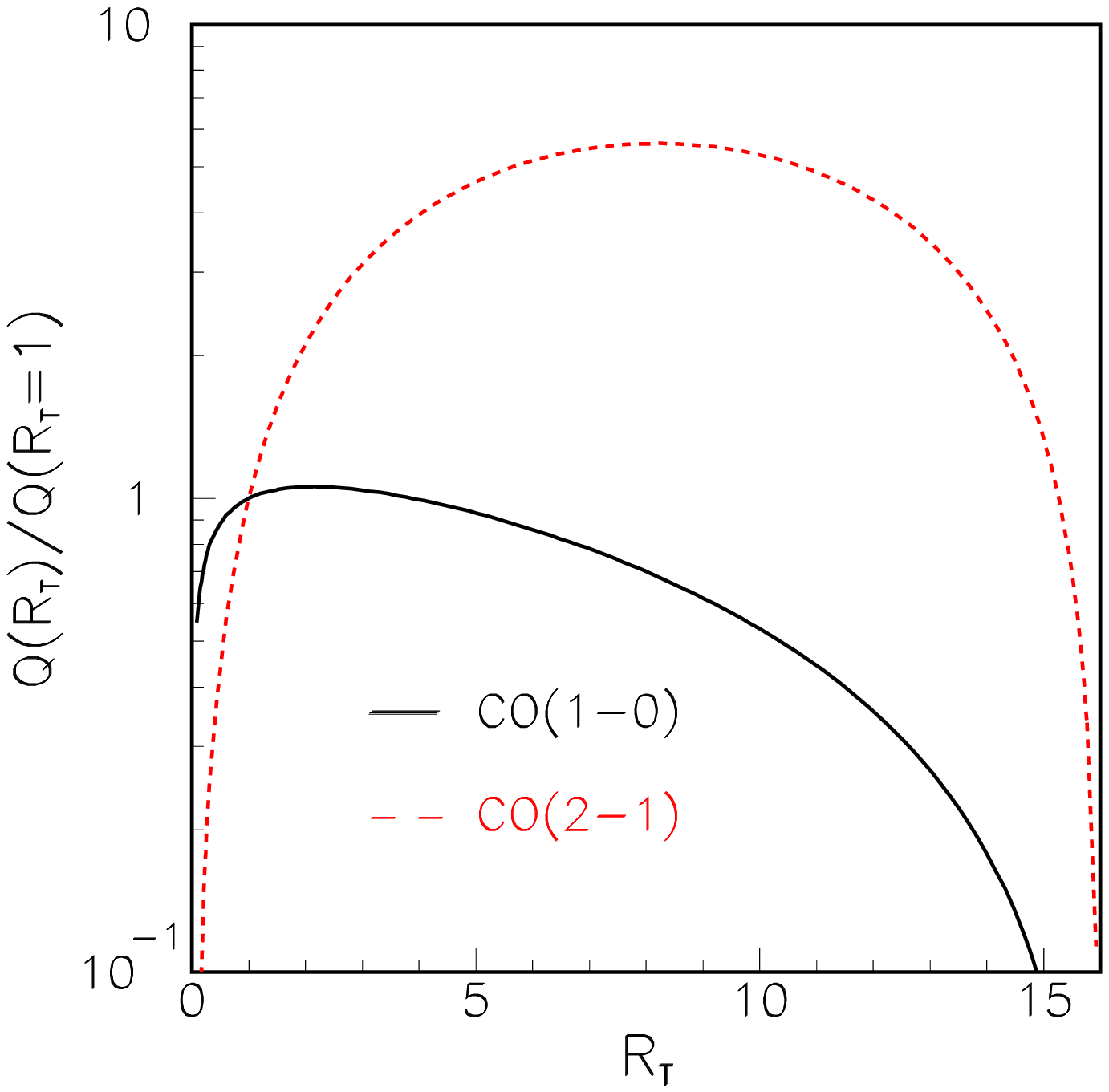}
\caption{Dependence on $R_T$ of the temperature (left), the absorption coefficients $f_{abs}$ (middle) and the $Q$ factors (right). Black is for \mbox{CO(1-0)} and red for \mbox{CO(2-1)}. In all three panels the axis of ordinate spans two decades.}
\label{fig30}
\end{figure*}

Several uncertainties are attached to these results. In addition to those inherent to the arbitrariness of the form adopted for the wind velocity, the evaluation of the temperature becomes less reliable at higher values. To illustrate this point, we display in Figure \ref{fig30} the dependence on $R_T$ of temperature dependent quantities of relevance to the radiative transfer calculation: the temperature $T$, the absorption coefficients, $f_{abs(1-0)}$ and $f_{abs(2-1)}$ and the factors $Q_{(1-0)}$ and $Q_{(2-1)}$ that control the temperature dependence of the emissivity. For convenience, we normalize all quantities to their values at $R_T=1$; the steep dependence on $R_T$ near the ends of the allowed range (0 to 16) implies large uncertainties on the evaluation of the temperature: for $T=30$ K, $R_T=11$ but $T$ increases to 40 K when $R_T$ increases by only 10\%: large temperatures are therefore affected by very large uncertainties. This has a strong effect on the evaluation of the emission probability but relatively little effect on the evaluation of the absorption because the $f_{abs}$ coefficients become very small at high temperatures, the exponential factor $\exp(E_J/T)$ acting in opposite directions in the expressions for $Q$ and $f_{abs}$.

The choice of a radial wind velocity independent from $r$, while sensible, is arbitrary. Deviation from a radial form is best discussed in terms of a possible rotation component, which was done in Section \ref{sec3.4}. A possible dependence of the radial velocity on $r$ would obviously affect the results. However, we have seen (Figures \ref{fig25} and \ref{fig28}) that the radial fluctuations of the mass-loss rate and of the density calculated assuming an $r$-independent wind velocity are small. As wind velocity and mass-loss rate are linearly related, fluctuations of one must be expected to be accompanied by fluctuations of the other. Namely, one might use for de-projection a wind velocity distribution that would display small fluctuations in $r$ mimicking those observed on the mass-loss rate; one would then obtain a new density that would be just as reliable as that displayed in Figure \ref{fig28}. Indeed, both wind velocity and density are likely to display related fluctuations in $r$; the important point is that both are expected to be small. The density enhancement observed at $r<\sim2$ arcsec is independent of the choice of inclination angle $\varphi$ and would imply a positive wind velocity gradient in this region if the flux of matter were to be constant.

Taking all above arguments into account, we estimate that, on average, temperatures are measured with a precision of $\sim\pm25$\% and densities with a precision of $\sim\pm15$\%. 

\section{Summary and discussion}\label{sec7}

\subsection{Summary}\label{sec7.1}

The present analysis of ALMA observations has confirmed the overall picture that had been previously drawn of the circumstellar envelope of EP Aqr, displaying approximate axi-symmetry about an axis making an angle of $\sim10$\dego\ with the line of sight and projecting on the sky plane some 20\dego\ west of north. The Doppler velocity spectra have been separated into two components that have been shown to be associated with distinct entities: a narrow central component associated with an equatorial region expanding at low velocity $V_{eq}\sim(0.33\pm0.03)/\sin\!\varphi$ \kms\ and a broad component associated with a bipolar outflow expanding at a velocity increasing with stellar latitude $\alpha$ up to 10 to 11 \kms\ near the poles. The presence of faint wings at approximate noise level possibly extending up to some $\pm16$ \kms\, as suggested by the observation of SiO emission by \citet{Homan2018}, cannot be excluded. An upper limit of 38\% of the expansion velocity has been placed on a possible rotation velocity of the equatorial outflow. Both components merge at an angular distance of $\sim2$ arcsec from the star ($\sim200$-250 au). A joint description of both the narrow and broad components, for both \mbox{CO(1-0)} and \mbox{CO(2-1)} emissions, has been presented, taking temperature and absorption effects in due account. The values taken by the temperature and density have been evaluated in the star meridian plane and the uncertainties attached to these have been discussed.

The flaring of the equatorial outflow was shown to be small from the narrow width of the line, at the scale of the inclination angle $\varphi$, which is estimated not to exceed $\sim20$\dego. Corrected for inclination of the star axis with respect to the line of sight, thermal broadening and instrumental resolution, line widths in the equatorial outflow reach $\sim1.2$ \kms\ FWHM, consistent with a flaring of $\pm17$\dego\ FWHM for $\varphi=10$\dego. Both \mbox{CO(1-0)} and \mbox{CO(2-1)} emissions of the equatorial outflow display remarkably similar intensity fluctuations, at the level of $\sim\pm36$\%, dominated by an eastern enhancement in the form of an arc and a south-western depression, together having the appearance of a spiral. Evidence was presented for a radial modulation of the Doppler velocity of very low amplitude, $\sim0.16$ \kms, and a period of $\sim3.8$ arcsec. Apart from a small anti-correlation of their amplitudes, no clear correlation has been found between intensity and velocity fluctuations of the equatorial outflow. The identification of a spiral in the intensity map \citep[]{Homan2018} has been thoroughly discussed.

The bipolar outflow is smoother than the equatorial outflow, with density fluctuations at the level of $\sim\pm26$\%, but reveals both an asymmetry between blue and red sides and significant differences between \mbox{CO(1-0)} and \mbox{CO(2-1)} emissions. Evidence has been presented for the emission to reach a maximum at intermediate star latitudes, with a significant depression near the poles. The reliability of this conclusion has been discussed. 
The distribution of temperature in space has been evaluated from the ratio between \mbox{CO(2-1)} and \mbox{CO(1-0)} emissions. A radiative transfer calculation has been performed to evaluate the absorption, which was found at the level of 20\% to 30\% on average, not exceeding $\sim40$\%.  Accounting for it, temperature $T^*$ and density $d^*$ averaged over stellar longitude have been evaluated in the meridian plane of the star as functions of stellar latitude $\alpha$ and distance $r$ to the star. Apart from fluctuations of the mass-loss history evaluated at the level of $\pm10$\%, the mass-loss rate is found approximately constant at the level of ($1.6\pm0.4$) 10$^{-7}$ M$_\odot$\,yr$^{-1}$ over the past 2500 years. Such fluctuations are described in terms of density in the present model because the expansion velocity has been assumed to be free of them but they could as well be blamed on velocity fluctuations with a smooth density and are most likely the effect of both density and velocity fluctuations of the flux of matter in the recent history of the star. The temperature decreases exponentially with $r$, approximately as $109\mbox{[K]}\exp(-r\mbox{[arcsec]}/3.1)$ or $106\mbox{[K]}/r\mbox{[arcsec]}$ and is maximal at intermediate latitudes. Apart from a $\sim30$\% excess close to the star ($r<\sim2$ arcsec), the density decreases slowly with distance as approximately expected from UV dissociation. Its equatorial value depends trivially on the inclination angle of the star axis, small inclinations implying large equatorial wind velocities. Uncertainties on the measurements of temperature and density are estimated at the levels of 25\% and 15\% respectively.

\subsection{State of the art}\label{sec7.2}

Our current understanding of the mass-loss mechanism of AGB stars is still relatively vague \citep[for a recent review see][]{Hofner2018}. The basic picture is of isotropic, uniform and constant slow winds, with mass-loss rates in the range of 10$^{-7\,\rm{to}\, -5}$ M$_\odot$\,yr$^{-1}$, triggered by pulsation-induced shock waves, creating favourable conditions for the formation of dust grains that absorb the stellar light and transfer the resulting momentum to the surrounding gas by colliding with it. Acceleration occurs at distances from the star not exceeding a few tens of stellar radii, resulting in wind velocities at the scale of a few 10 \kms.  However, in the case of oxygen-rich stars like EP Aqr, serious doubts have been voiced on the ability of such a mechanism to produce sufficient acceleration \citep[]{Woitke2006a}, the dust grains being formed lacking the absorption power necessary to produce it. It has instead been proposed that micron-size iron-free and transparent silicate grains close to the star are accelerated by photon scattering \citep[]{Hofner2008} and confirmation of this hypothesis has been subsequently obtained \citep[]{Norris2012}. Recently, \citet{McDonald2016} and \citet{McDonald2018} have proposed convincing arguments in favour of a scenario according to which the mass-loss episode would be triggered predominantly by pulsations rather than by radiation pressure on dust, in excellent agreement with the model predictions of \citet{Winters2000} (their
B-regime that applies to short period, low luminosity stars); such a scenario would particularly apply to EP Aqr, which, with its low pulsation period of 55 days, is, according to McDonald et al., just starting to lose mass.

In any case, while the standard picture provides a satisfactory zero order approximation to the mass-loss mechanism, the physics behind deviations from it, whether bipolar outflow, equatorial enhancement or inhomogeneity in the form of arcs and lumps, is poorly understood. If we exclude spectacular events such as the formation of a super-wind at the end of the AGB phase \citep[]{Schroder1999} or of thin spherical detached shells around some carbon-rich stars \citep[]{Olofsson2000, Schoier2005, Maercker2010, Maercker2016} which are apparently unrelated to the case of EP Aqr, two main candidates for breaking the symmetry of the zero order picture have been considered and studied in some detail: the presence of large convection zones in the star atmosphere and the attraction of a stellar companion or massive planet.

Major convection cells in the star cause the mass loss to occur in patches in its upper atmosphere, producing an initially clumpy wind \citep[]{Freytag2017}. Recent observations of the stellar atmospheres of AGB stars such as W Hya \citep[]{Vlemmings2017} or R Sculptoris \citep[]{Wittkovski2017} have triggered a surge of interest in this topic. To which extent such lumpiness is preserved over large time scales is unclear. Similarly, while the effect of the short period pulsations is expected to be washed away when looking at the morphology of the circumstellar envelope at large distances from the star, the role played by episodic violent events such as He flashes is also unclear \citep[]{Olofsson1990}. An extensively studied example of a circumstellar envelope displaying the imprint of its mass-loss history is Mira Ceti \citep[]{Ramstedt2014, Nhung2016}, with fragmented distinct lumps moving out radially from the star.

The co-existence in the circumstellar envelope of radial winds of different velocities can generate important inhomogeneity from their interaction. Such interacting winds have been studied extensively, in particular at the end of the AGB, before entering the Planetary Nebula phase \citep[]{Schroder1999}, but also in less extreme configurations \citep[]{Villaver2002}. In the superwind case, the strong temperature dependence of the micro-physics and dust-formation chemistry dominates the effect of gravity, suggesting that qualitatively similar but quantitatively more modest effects might be present in younger AGB stars.  Of possible relevance to EP Aqr is the interaction between the AGB slow wind and a polar jet emitted by a companion accreting matter from it \citep[]{Soker2000, Livio2001, GarciaArredondo2004}.

The main effect of the presence of a stellar companion or massive planet has been shown to be twofold \citep[]{Kim2012}: if the companion is massive enough for the reflex motion of the mass-losing star to be significant, the wind velocity is the sum of the orbital velocity of the star and the radial expansion velocity of the gas, producing helical spiral patterns having a pitch that decreases with stellar latitude; moreover, the companion attracts some gas toward the plane of its orbit, thereby producing an equatorial enhancement in its wake. How significant these effects may be depends on the parameters describing the companion and its movement, but the less massive the companion, the smaller they are. A signature of such effects is the presence of spiral, or spiral-like arc enhancements of the gas density \citep[]{Mastrodemos1998, Mastrodemos1999, Kim2012, Homan2015}. A few remarkable examples have been identified, all from carbon stars, leaving no doubt on their interpretation. Among these are LL Pegasi \citep[]{Kim2017}, R Sculptoris \citep[]{Maercker2012} or CIT6 \citep[]{Kim2013}. Apart from the case of EP Aqr, no convincing example has yet been found among M or S oxygen rich stars, although it is tempting, in the absence of other compelling explanation of the presence of arcs of enhanced emissivity, to identify some of these with spiral arms \citep[]{Ramstedt2014, Mayer2014}. In the present case \citep[]{Homan2018}, however, the evidence for the presence of a spiral is much weaker than it is for carbon stars.

In the case of carbon stars simple hydrodynamic models \citep[]{Woitke2006b} give evidence for instabilities resulting from the episodic formation of dust above the stellar surface sufficient to break the initial spherical symmetry and produce incomplete dust shells or dust arcs. At least qualitatively similar phenomena may be expected to occur in the envelope of M stars, the intrinsic instabilities inherent to astrophysical gases under the influence of a distant source of radiation being of a rather basic and general nature \citep[]{Woitke2000}.

The presence of a companion is not the only mechanism that has been invoked to explain the formation of an equatorial enhancement. The equatorial outflow around \mbox{EP Aqr}, keeping the imprint of the recent mass-loss history, might also be caused by the rotation of the star atmosphere, as described by \citet{Dorfi1996}. The observation by \citet{Homan2018} of the emission of SO$_2$ molecules gives indeed evidence for significant rotation at distances from the star well below 1 arcsec or so. Finally, we must mention the possible influence of magnetic fields \citep[]{Vlemmings2013}.

\subsection{Discussion}\label{sec7.3}

When discussing the bipolar outflow detected from EP Aqr, we must keep in mind that the departures from the zero order picture that we are talking about are relatively modest. In particular, the equatorial wind velocity, $\sim2$ \kms\ is well below typical terminal wind velocities and below escape velocity over a significant fraction of the explored range. While of no serious concern for the results of the de-projection and the conclusions summarized in Section \ref{sec7.1} above, this is of high relevance to the understanding of the mass loss dynamics. Not many oxygen-rich AGB stars have been studied to a same degree of detail as EP Aqr but for at least one of these, RS Cnc, a similar low equatorial expansion velocity has been observed \citep[]{Nhung2015b}. At variance with EP Aqr, RS Cnc has its axis making a large angle, $\sim45$\dego, with respect to the line of sight: the evaluation of the equatorial velocity, $\sim2$ \kms, is not significantly affected by uncertainties on its precise value. A meaningful description of such a low velocity wind implies a deeper understanding of the mechanism of acceleration, and particularly of the radial range over which it is efficient, than currently available. In particular, it requires including information from observations using other tracers, such as SiO and SO$_2$, that explore shorter distances from the star. Further comments are therefore postponed to a forthcoming publication (\citet{TuanAnh}, in preparation).

An isotropic mass-loss shaped by a bipolar wind would produce a density that is maximal at the equator, which is far from being the case. The present results imply therefore that, even before acceleration, the flux of matter is anisotropic and peaks at intermediate latitudes. Indeed, as elaborated in \citet{Nhung2018a}, whatever form is adopted for the wind velocity, a spherical flux of matter is incompatible with the present observations. This implies in turn complex mechanisms for both mass-loss and acceleration, none of the simple scenarios alluded to in the preceding paragraph producing such morpho-kinematics on its own. In particular, the pulsations that trigger the wind, corresponding to the radial first overtone \cite[][and references therein]{Wood2015} are expected to be mostly isotropic although some non-radial $l=1$ contributions are known to be often present.

We remark that qualitatively the latitudinal dependence of the wind velocity, temperature and density resembles that observed in the Red Rectangle \citep[]{Anh2015, Bujarrabal2016}: a biconical enhancement of the density and temperature, peaking at intermediate latitudes, an equatorial expansion velocity of 1.6 \kms\ together with a rotation velocity at the scale of 1 \kms\ and a bipolar outflow velocity reaching $\sim$7.6 \kms. An important difference is the temperature, typically 5 times higher in the Red Rectangle than in EP Aqr. Yet, the Red Rectangle is a proto-planetary nebula while EP Aqr is a rather young AGB star. The Red Rectangle is believed to have evolved from oxygen-rich to carbon-rich \citep[]{Waters1998} and is known to be a binary, with a probable White Dwarf companion. Binarity is generally held responsible for shaping the envelope of the Red Rectangle, the companion accreting from the main star in an equatorial disc and ejecting along the line of poles a pair of high velocity (a few 100 \kms), high temperature (at the scale of $\sim$1000 K)  jets. However, no agreement has yet been reached on the precise mechanism \citep[]{MartinezGonzalez2015}; in particular, the companion polar jets may be narrow in precession about the binary axis with a period of 17.6 years \citep[]{Velazquez2011} or broad along the binary axis \citep[]{Thomas2013}.

The similarities between the two stars suggest that binarity may be the main shaping mechanism of the EP Aqr envelope. If one takes seriously the evidence for a spiral enhancement claimed by \citet{Homan2018} and thoroughly discussed in Section \ref{sec4.2}, the separation of 3 arcsec between arms corresponds, for an expansion velocity of 2 \kms\ to 850 years, at strong variance with the Red Rectangle where the orbital period is at the scale of a single year \citep[]{Velazquez2011}. For a total mass (main star and companion) of 2 M$_\odot$, this means a distance between them of
$(850^2\times2)^{1/3}=112$ au or $\sim 1$ arcsec. The Red Rectangle hosts instead a close binary with a separation at the scale of only 1 au. In summary, the similarities remarked between the morpho-kinematic properties of the CO emission of the circumstellar envelopes of EP Aqr and the Red Rectangle contrast with the major quantitative differences between their binary properties and between their mass-loss rates, two orders of magnitude higher for the Red Rectangle than for EP Aqr: if a pair of jets were shaping the polar depressions of EP Aqr, their detection would probably be difficult \citep[]{Witt2008}.

To the extent that the fate of the expanding gas envelope is largely determined by what happens in the small central region, $r<\sim 1$ arcsec, a detailed knowledge of its morpho-kinematics, temperature and chemistry is obviously essential to understand the morpho-kinematics at larger distances. None of the models of the mass loss mechanism at stake in oxygen-rich AGB stars can be confirmed or denied by the results of the present work but such a thorough analysis of the properties of the circumstellar envelope at distances exceeding $\sim250$ au severely constrains their main features. To definitely choose with confidence between the currently proposed models requires at least including what can be learned of the dynamics at shorter distances from the emission of tracers such as SiO and SO$_2$. Even so, the under-determination of the problem of de-projection of radio observations restricts the reliability of the conclusions that can be drawn and imposes particularly critical and rigorous scrutiny.

\section*{ACKNOWLEDGEMENTS}

We are deeply grateful to the referee, Professor Albert Zijlstra, for very helpful comments on the manuscript. We thank Dr Iain McDonald for comments on pulsation-induced acceleration and Dr Pierre Lesaffre for his interest in our work and useful comments. This paper makes use of the following ALMA data: 2016.1.00026.S. ALMA is a partnership of ESO (representing its member states), NSF (USA) and NINS (Japan), together with NRC (Canada), NSC and ASIAA (Taiwan), and KASI (Republic of Korea), in cooperation with the Republic of Chile. The Joint ALMA Observatory is operated by ESO, AUI/NRAO and NAOJ. This work was supported by the Programme National Physique et Chimie du Milieu Interstellaire (PCMI) of CNRS/INSU with INC/INP co-funded by CEA and CNES. The Hanoi team acknowledges financial support from VNSC/VAST, the NAFOSTED funding agency, the World Laboratory, the Odon Vallet Foundation and the Rencontres du Viet Nam. This research is funded by Graduate University of Science and Technology under grant number GUST.STS.DT2017-VL01.\\

%%%%%%%%%%%%%%%%%%%%%%%%%%%%%%%%%%%%%%%%%%%%%%%%%%



%\bibliographystyle{mnras}


%%%%%%%%%%%%%%%%%%%%%%%%%%%%%%%%%%%%%%%%%%%%%%%%%%

%%%%%%%%%%%%%%%%% APPENDICES %%%%%%%%%%%%%%%%%%%%%

%%%%%%%%%%%%%%%%%%%%%%%%%%%%%%%%%%%%%%%%%%%%%%%%%%

\appendix
\section{} \label{app1}
\begin{figure*}
\centering
\includegraphics[height=6cm]{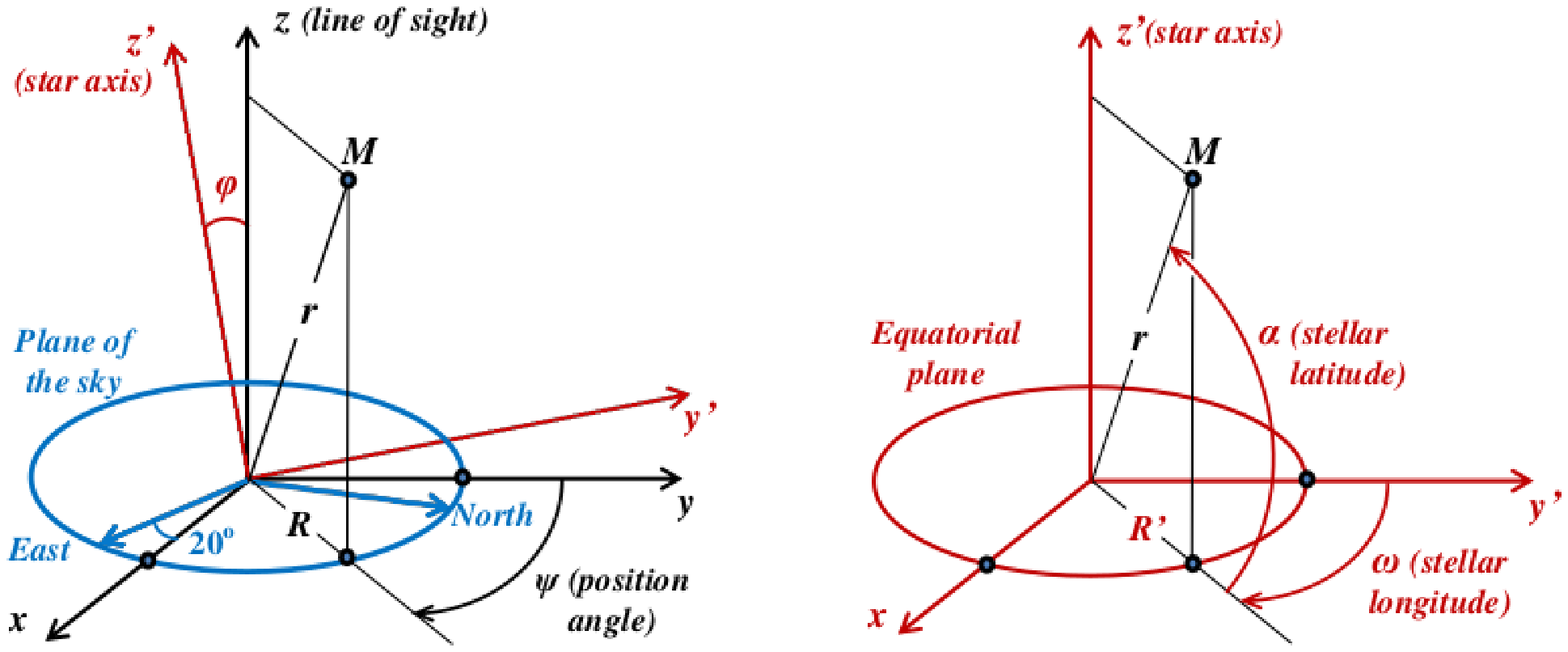}
\caption{Left: sky coordinates; $x$ is in the plane of the sky, 20\dego\ north of east; $y$ is in the plane of the sky, 20\dego\ west of north; $z$ is along the line of sight; a vector \textbf{\textit{r}} pointing to M projects as $R$ on the sky plane, $R^2 =x^2 +y^2$ ; the position angle $\psi$ measured counter-clockwise from north, is such that $y=R\cos\psi$ and $x=R\sin\psi$. The stellar frame of coordinate $(x,y',z')$ is obtained from the sky frame $(x,y,z)$ by rotation of angle $\varphi$ about the $x$ axis. Right: stellar coordinates; a vector \textbf{\textit{r}} pointing to M projects on the equatorial plane $(x,y')$ as $R'=\sqrt{x^2 +y'^2}$; the stellar longitude $\omega$ is such that $y'=R'\cos\omega$ and $x=R'\sin\omega$; the stellar latitude $\alpha$ is such that $R'=r\cos\alpha$ and $z'=r\sin\alpha$.}
\label{figA1}
\end{figure*}

Figure \ref{figA1} summarizes the variables used in relation to the sky and stellar systems of coordinates used in the article. The velocity vector is written \textbf{\textit{V}}, the Doppler velocity $V_z$ being its component on the $z$ axis.

Tables \ref{table4} to \ref{table7} provide additional information that is being referred to in the text.

\begin{table*}
  \centering
  \caption{Parameters of the fits of the form $I_1\exp(-R/R_1-k_1R^2)+ I_2\exp(-R/R_2-k_2R^2)$
    to the $R$-dependence of the integrated intensity for $R<8$ arcsec.}
  \begin{tabular}{|c|c|c|c|c|c|c|c|}
    \hline
    &Line&\makecell{$I_1$\\(Jy\,beam$^{-1}$\,\kms)}&\makecell{$I_2$\\(Jy\,beam$^{-1}$\,\kms)}&\makecell{$R_1$\\(arcsec)}& \makecell{$R_2$\\(arcsec)}&\makecell{$k_1$\\(arcsec$^{-2}$)}&\makecell{$k_2$\\(arcsec$^{-2}$)}\\
    \hline
    \multirow{2}{*}{Narrow component}&CO(1-0)&0.170&0.0593&2.69&24.4&0.458&0.00945\\
    \cline{2-8}
    &CO(2-1)&0.334&0.141&2.38&6.20&0.687&0.00895\\
    \hline
    \multirow{2}{*}{Broad component}&CO(1-0)&0.781&0.506&2.93&6.58&1.65&0.00400\\
    \cline{2-8}
    &CO(2-1)&4.00&0.875&0.347&3.99&0.138&0.0122\\
    \hline
  \end{tabular}
  \label{table4}
\end{table*}

\begin{table}
  \centering
  \caption{Narrow component line widths (Gaussian $\sigma$'s, \kms) measured after correction. Distances on the sky plane are measured in arcsec.}
  \label{table6}
  \begin{tabular}{|c|c|c|c|}
    \hline
    \multicolumn{2}{|c|}{Selection}&CO(1-0)&CO(2-1)\\
    \hline
    \multicolumn{2}{|c|}{All ($R$ \& $\psi$ corrected)}&0.51&0.55\\
    \hline
    \multicolumn{2}{|c|}{All (each px recentred)}&0.44&0.47\\
    \hline
    \multirow{8}{*}{$\psi\pm22.5^\circ$}&0\dego&0.42&0.45\\
    \cline{2-4}
    &45\dego&0.46&0.48\\
    \cline{2-4}
    &90\dego&0.43&0.49\\
    \cline{2-4}
    &135\dego&0.41&0.44\\
    \cline{2-4}
    &180\dego&0.39&0.42\\
    \cline{2-4}
    &225\dego&0.45&0.45\\
    \cline{2-4}
    &270\dego&0.43&0.46\\
    \cline{2-4}
    &315\dego&0.42&0.44\\
    \hline
    \multicolumn{2}{|c|}{$|R-R_{sp}|<h/4$, $R>1.5$}&0.51&0.54\\
    \hline
    \multicolumn{2}{|c|}{$|R-R_{sp}|>h/4$, $R>1.5$}&0.56&0.61\\
    \hline
    \multicolumn{2}{|c|}{$|R-R_{circle}|<0.95$, $R>1.5$}&0.56&0.61\\
    \hline
    \multicolumn{2}{|c|}{$|R-R_{circle}|>0.95$, $R>1.5$}&0.50&0.54\\
    \hline
    \multicolumn{2}{|c|}{$|R-R_S|<0.7$}&0.40&0.42\\
    \hline
  \end{tabular}
\end{table}

\begin{table}
  \centering
  \caption{Parameters of the reference distributions used for the broad component. They are defined as $F_1(r)$ being a fit to $\rho r^2$ of the form $\exp(a+b\,r+c\,r^2+d\,r^3+e\,r^4)$ and $F_2(\sin\!\alpha)$ being a fit to $\rho r^2/F_1(r)$ of the form $a+b\sin\!\alpha+c\sin^2\!\alpha+d\sin^4\!\alpha$.}
  \label{table7}
  \begin{tabular}{|c|c|c|c|c|}
  \hline
  \multirow{2}{*}{}&\multirow{2}{*}{Parameter}&\multicolumn{2}{c|}{CO(1-0)}&\multirow{2}{*}{CO(2-1)}\\
  \cline{3-4}
  &&$r<4$ arcsec&$r>4$ arcsec&\\
  \hline
  \multirow{5}{*}{$F_1(r)$}&$a$&$-$1.918&$-$1.632&0.514\\
  &$b$&$-$0.0405&0.206&0.00398\\
  &$c$&0.149&$-$0.0118&0.105\\

  &$d$&$-$0.0206&0&$-$0.0229\\
  &$e$&0&0&0.00125\\
  \hline
  \multirow{4}{*}{$F_2(\sin\!\alpha)$}&a&\multicolumn{2}{c|}{0.370}&0.304\\
  &b&\multicolumn{2}{c|}{0.164}&0.278\\
  &c&\multicolumn{2}{c|}{4.755}&5.117\\
  &d&\multicolumn{2}{c|}{$-$3.882}&$-$3.919\\
  \hline
  \end{tabular}
\end{table}

% Don't change these lines
\bsp	% typesetting comment
\label{lastpage}
\end{document}